%% file: main.tex
\documentclass[traditabstract]{aa}

\usepackage{graphicx}
\usepackage{txfonts}
\usepackage{color}
\usepackage[draft]{hyperref}
\usepackage{natbib}
\bibpunct{(}{)}{;}{a}{}{,} 

\usepackage[english]{babel}
\newcommand{\beqa}{\begin{eqnarray}} 

\newcommand{\eeqa}{\end{eqnarray}}

\newcommand{\bsub}{\begin{subequations}}
\newcommand{\esub}{\end{subequations}}
\newcommand{\beal}{\begin{align}}
\newcommand{\ealn}{\end{align}}

\newcommand{\kms}{$\mathrm{km\;s^{-1}}$}
\newcommand{\kmpc}{$\mathrm{km\; s^{-1}\; Mpc^{-1}}$}
\newcommand{\ho}{$H_0$\,}

\newcommand{\avg}[1]{\left\langle{#1}\right\rangle}

\def\gsim{\mathrel{\rlap{\lower 4pt \hbox{\hskip 1pt $\sim$}}\raise 1pt \hbox {$>$}}}
\def\lsim{\mathrel{\rlap{\lower 4pt \hbox{\hskip 1pt $\sim$}}\raise 1pt \hbox {$<$}}}

\makeatletter
\renewcommand*\aa@pageof{, page \thepage{} of \pageref*{LastPage}}
\makeatother

\begin{document}
\title{Measuring the Hubble constant with
Type Ia supernovae \\ as near-infrared standard candles}

\titlerunning{$H_0$ with SN~Ia as NIR standard candles}
\authorrunning{Dhawan, Jha, \& Leibundgut}
\author{\textbf{Suhail Dhawan\inst{1,2,3,4}
    \and Saurabh W.~Jha\inst{5}
 	\and Bruno Leibundgut\inst{1,2}
    }}

\institute{European Southern Observatory, Karl-Schwarzschild-Strasse 2,
D-85748 Garching bei M\"unchen, Germany \\
\email{suhail.dhawan@fysik.su.se}
\and  Excellence Cluster Universe, Technische Universit\"at M\"unchen,
Boltzmannstrasse 2, D-85748, Garching, Germany\\
\and Physik Department, Technische Universit\"at M\"unchen,
James-Franck-Strasse 1, D-85748 Garching bei M\"unchen, Germany \\
\and Oskar Klein Centre, Department of Physics, Stockholm University, 
SE 106 91 Stockholm, Sweden \\
\and Department of Physics and Astronomy, Rutgers, the State University of
New Jersey, 136 Frelinghuysen Road, Piscataway, NJ 08854, USA
} 

\date{Received; accepted }

\offprints{S.~Dhawan}

\abstract{
The most precise local measurements of $H_0$ rely on observations of
Type Ia supernovae (SNe~Ia) coupled with Cepheid distances to SN~Ia
host galaxies. Recent results have shown tension comparing $H_0$ to
the value inferred from CMB observations assuming $\Lambda$CDM, making
it important to check for potential systematic uncertainties in either
approach. To date, precise local $H_0$ measurements have used SN~Ia
distances based on optical photometry, with corrections for light
curve shape and colour. Here, we analyse SNe~Ia as standard candles in
the near-infrared (NIR), where luminosity variations in the supernovae
and extinction by dust are both reduced relative to the optical. From
a combined fit to 9 nearby calibrator SNe with host Cepheid distances
from \citet{Riess2016} and 27 SNe in the Hubble flow, we estimate
the absolute peak $J$ magnitude $M_J = -18.524\;\pm\;0.041$ mag and
$H_0 = 72.8\;\pm\;1.6$ (statistical) $\pm$ 2.7 (systematic) km
s$^{-1}$ Mpc$^{-1}$. The 2.2\% statistical uncertainty demonstrates
that the NIR provides a compelling avenue to measuring SN~Ia
distances, and for our sample the intrinsic (unmodeled) peak $J$
magnitude scatter is just $\sim$0.10 mag, even without light curve
shape or colour corrections. Our results do not vary significantly
with different sample selection criteria, though photometric
calibration in the NIR may be a dominant systematic uncertainty. Our
findings suggest that tension in the competing $H_0$ distance ladders
is likely not a result of supernova systematics that could be expected
to vary between optical and NIR wavelengths, like dust extinction. We
anticipate further improvements in $H_0$ with a larger calibrator
sample of SNe~Ia with Cepheid distances, more Hubble flow SNe~Ia with
NIR light curves, and better use of the full NIR photometric data set
beyond simply the peak $J$-band magnitude.
}

\maketitle
\section{Introduction}
\label{sec:intro}

The Hubble constant ($H_0$) can be measured locally and also derived from the
sound horizon observed from the cosmic microwave background (CMB), providing
two absolute distance scales at opposite ends of the visible expansion history
of the Universe. The ``reverse'' distance ladder, in which $H_0$ is inferred
from CMB and other high-redshift observations, requires a cosmological model.
Thus, a comparison between this inference and a direct local measurement of
$H_0$ becomes a stringent test of the standard cosmological model and its
parameters. For instance, a precise local determination of $H_0$, combined
with high-$z$ Type Ia supernova \citep[SN~Ia;][]{Betoule2014}, baryon acoustic
oscillation \citep[BAO;][]{Alam2016} and cosmic microwave background
\citep[CMB;][]{Planck2015} can offer key insights into the dark energy
equation of state \citep{Freedman2012,Riess2011,Riess2016}.

Improved measurements (3--5$\%$ precision) of $H_0$ at low redshifts \citep[$z \lesssim 0.5$; e.g.,][]{Riess2009,Riess2011,Freedman2012,Suyu2013,Bonvin2017}, along
with recent progress in CMB measurements
\citep{Bennett2013,Hinshaw2013,Planck2015} hint at mild tension
(2--2.5$\sigma$) between the different determinations. 
The most precise estimates of the distances to local SNe~Ia come from
observations of Cepheid variables in host galaxies of 19 SNe~Ia from the SH0ES
program \citep{Riess2016}. Combined with a large SN~Ia sample going to 
$z\approx0.4$ this calibration results in an uncertainty in $H_0$ of 
just 2.4\%, which appears to be in tension with the CMB inference from
\emph{Planck} at the 3.4$\sigma$ level and with WMAP+SPT+ACT+BAO
at a reduced, 2.1$\sigma$ level. Moreover, recent re-analyses of the \citet{Riess2016} data confirm a high value of $H_0$ \citep{Feeney2017,Follin2017}. If this effect is corroborated by more future data and analyses, it would imply ``new physics'', including possibilities like
additional species of relativistic particles, non-zero curvature, dark
radiation or even a modification of the equations of general relativity
\citep[e.g., see][]{Verde2017,DiValentino2017,Sasankan2017,Dhawan2017b,Renk2017}. 
However, in
order to be confident that we are uncovering a genuine shortcoming in our
standard cosmological model, we need to ensure that systematic uncertainties
are correctly estimated for both the local and distant probes.

The local distance ladder measurement of $H_0$ with SNe~Ia has been simplified
to three main steps \citep{Riess2016}: 1. calibrating the Leavitt law
\citep{Leavitt1912} for Cepheids with geometric anchor distances (Milky Way
parallaxes, LMC eclipsing binaries, or the Keplerian motion of masers in the
nucleus of NGC 4258); 2. calibrating the luminosity of SNe~Ia with Cepheid
observations in supernova hosts; and 3. applying this calibration to SNe~Ia in
the smooth Hubble flow. Numerous instrumental and astrophysical systematics
can apply in each of these three steps, and the increase in precision in the
SN~Ia $H_0$ measurement can largely be attributed to mitigating these
systematics, for example, by tying all the Cepheid observations to the same
Hubble Space Telescope photometric system. The Cepheid measurements are
further improved by calibrating them in the near-infrared ($H$-band using
WFC3/IR) rather than the optical. In the NIR, Cepheids have a lower
variability amplitude, reduced sensitivity to metallicity and uncertainties 
from dust extinction are minimized, 
leading to a Leavitt law with less scatter and thus more precise distances
\citep{Macri2015,Wielgorski2017}.

However, unlike for the Cepheids, \citet{Riess2016} used SN~Ia distances
standardized from optical light curves, comprising
the vast majority of SN~Ia photometric data for which light-curve fitter 
and distance models have been developed \citep{Guy2007,Jha2007,Scolnic2015}. 
To mitigate
against extinction, \citet{Riess2016} restrict their SN~Ia samples to 
objects with low reddening ($A_V \leq 0.5$ mag). In this paper
we use near-infrared (NIR) observations of nearby SNe~Ia as an alternate route
to measure $H_0$. Like for Cepheids, NIR distances to SNe~Ia have a few
advantages: SNe observed in the NIR have a lower intrinsic scatter 
than those observed in the optical
\citep{Meikle2000,Krisciunas2004,Wood-Vasey2008,Barone-Nugent2012,Dhawan2015}, the
corrections to the peak magnitude from the light curve shape and colour are
smaller \citep{Mandel2009,Mandel2011,Kattner2012}, and the effects of dust are
mitigated. In particular, the $J$-band extinction is a factor of $\sim$4 lower 
than in $V$ for typical dust. By calibrating the NIR luminosity of 
SN~Ia with the NIR Cepheid distances from \citet{Riess2016},
we can test whether the locally measured $H_0$ is consistent with the one
derived from optical SN~Ia light curves, and consequently determine whether
the observed tension is robust to potential systematic uncertainties in the SN
data that could be expected to vary with wavelength (e.g., colour/extinction 
corrections).

Because SNe~Ia are nearly standard candles in the near-infrared \citep[e.g.,][]{Barone-Nugent2012}, in this work
we take a much simpler approach to measuring SN~Ia distances than traditional
optical light-curve parametrisation and fitting. Here we derive distances
based \emph{only} on the peak magnitudes in the $J$-band, with no corrections
for light curve shape or colour, and find this truly ``standard'' candle
approach competitive with SN~Ia distances derived with optical light curves.

In Section~\ref{sec:data} we describe the SN sample. We then present the
analysis method and results in Sections~\ref{sec:analysis} and
\ref{sec:results}. We discuss our findings and conclude in
Section~\ref{sec:dis}.

\section{Data} 
\label{sec:data}

For our analysis, we use NIR photometry for the calibrator and Hubble-flow
samples from the literature. Our method (see Section~\ref{sec:analysis} below)
requires SNe with at least 3 $J$-band points between $-$6 and $+$10 days (relative
to the time of $B$-band maximum) with at least one of them before the $J$-band
maximum. Of the 19 Cepheid-calibrated SN~Ia in \citet{Riess2016}, 12 of them
have published $J$-band data, and of these, 9 have the requisite sampling to
precisely determine a peak magnitude. A summary of the sources for the
calibrator sample data is provided in Table~\ref{tab:calib}.

We compile our Hubble-flow sample from SNe with NIR photometry available in
the literature. This includes the Carnegie Supernova Project \citep
[CSP;][]{Contreras2010,Stritzinger2011} data releases 1 and 2, the Center for
Astrophysics (CfA) IR program \citep{Wood-Vasey2008,Friedman2015}, and a
Palomar Transient Factory (PTF) NIR follow-up program
\citep{Barone-Nugent2012,Barone-Nugent2013}. Table 3 of \citet{Friedman2015}
presents a snapshot of published NIR SN~Ia photometry. Of the 213 SN in the
table, 149 have $z > 0.01$, which we take to be the lower limit of the Hubble
flow\footnote{\citet{Riess2016} also explore a limit of $z > 0.0233$ for the 
Hubble flow; only 99 of the 213 SNe~Ia would pass this redshift cut.}. However, 
only 30 meet our light curve sampling criterion. The vast
majority of objects are observed only after maximum and/or are sparsely
sampled, and hence cannot be included in our current analysis. In particular,
we are not able to use any objects from the large sample of the
\emph{Sweetspot} survey \cite{Weyant2014,Weyant2017}. Our full Hubble-flow
data set is shown in Table~\ref{tab:hflow}.

\begin{table*}
\caption{The calibrator sample of Cepheid-calibrated SN~Ia, with
tabulated absolute magnitudes $M_J$ and uncertainties $\sigma_M$. The peak $J$
magnitudes $m_J$ are listed along with the uncertainty from the Gaussian
process fit. The host galaxy Cepheid distances $\mu_{\rm Ceph}$ and
uncertainties $\sigma_{\rm Ceph}$ are from \citet{Riess2016}. A correction for
dust extinction from the Milky Way (MW $A_J$) has been applied. $K$-corrections 
have been applied to the photometry; a representative value at the time of 
$J$ maximum for each supernova is tabulated as $K_J$.}
\centering
\small
\input{table_calibrators}
\label{tab:calib}
\end{table*}

\begin{table*}
\caption{CMB frame redshifts, with and without corrections for local flows,
peak $J$-band magnitudes and uncertainties for SNe in the Hubble flow. A
correction for dust extinction from the Milky Way (MW $A_J$) has been
applied. $K$-corrections have been applied to the photometry; a representative 
value at the time of $J$ maximum for each supernova is tabulated as $K_J$.}
\begin{minipage}{70mm}
\centering
\small
\input{table_hflow}
\tablefootwide{
\tablefoottext{a}{These fast-declining objects are excluded from our fiducial sample.}
}
\end{minipage}
\label{tab:hflow}
\end{table*}

\section{Analysis}
\label{sec:analysis}

Our approach is simple: we take SNe~Ia to be \emph{standard} candles in their
peak $J$-band magnitudes, which we derive directly from the observations. We
take the published $J$-band photometry, correct for Milky Way extinction using
the dust maps of \citet{Schlafly2011}, and apply a $K$-correction using the
SED sequence of \citet{Hsiao2007} calculated with the \texttt{SNooPy} package
\citep{Burns2011}. 

Because we have constructed our sample to contain only objects with sufficient
observations near peak (including pre-maximum data), we can estimate the peak
$J$ magnitude $m_J$ through straightforward Gaussian process interpolation,
using the Python \texttt{pymc} package, as implemented in 
\texttt{SNooPy} \citep{Burns2011}. This routine uses a
uniform mean function and a Matern covariance function with 3 parameters:
the time-scale at which the function varies (taken as 10 observer
frame days, except as below), the amplitude by which the function varies on
these scales (estimated from the photometric data), and the degree of
differentiability or ``smoothness'' (taken to be 3). Our light curve fits are
shown in Appendix~\ref{sec:gpfits}.

Note that we do not make the standard corrections used to measure SN~Ia
distances with optical light curves. We do not correct for light-curve shape,
supernova colour, or host-galaxy extinction. We have opted for this approach
because it is simple, and as seen below, effective. We describe shortcomings
in our approach and potential improvements in Section~\ref{sec:dis}.

The derived peak magnitudes $m_J$, with uncertainties estimated in the fit
$\sigma_{\rm fit}$, for the nearby calibrator sample are presented in Table
\ref{tab:calib}. For these objects, we adopt Cepheid distances and
uncertainties $(\mu_{\rm Ceph}, \sigma_{\rm Ceph})$ as reported by
\citet{Riess2016} to calculate absolute magnitudes. These distances are
``approximate'' in the sense that they are an intermediate step in the global
model presented by \citet{Riess2016}. Here, we take the Cepheid distances as
independent in deriving statistical uncertainties, and treat their covariances
(e.g., from the anchor distances or the form of the Leavitt Law) as
separately-estimated systematic uncertainties (see Section
\ref{sec:systematics}). In that case the SN~Ia $J$-band absolute magnitudes
are given simply by $M_J = m_J -
\mu_{\rm Ceph}$ and the calibrator absolute magnitude uncertainty is the
quadrature sum $\sigma_M^2 = \sigma_{\rm fit}^2 + \sigma_{\rm Ceph}^2$.

\begin{figure}
\centering \includegraphics[width=0.5\textwidth]{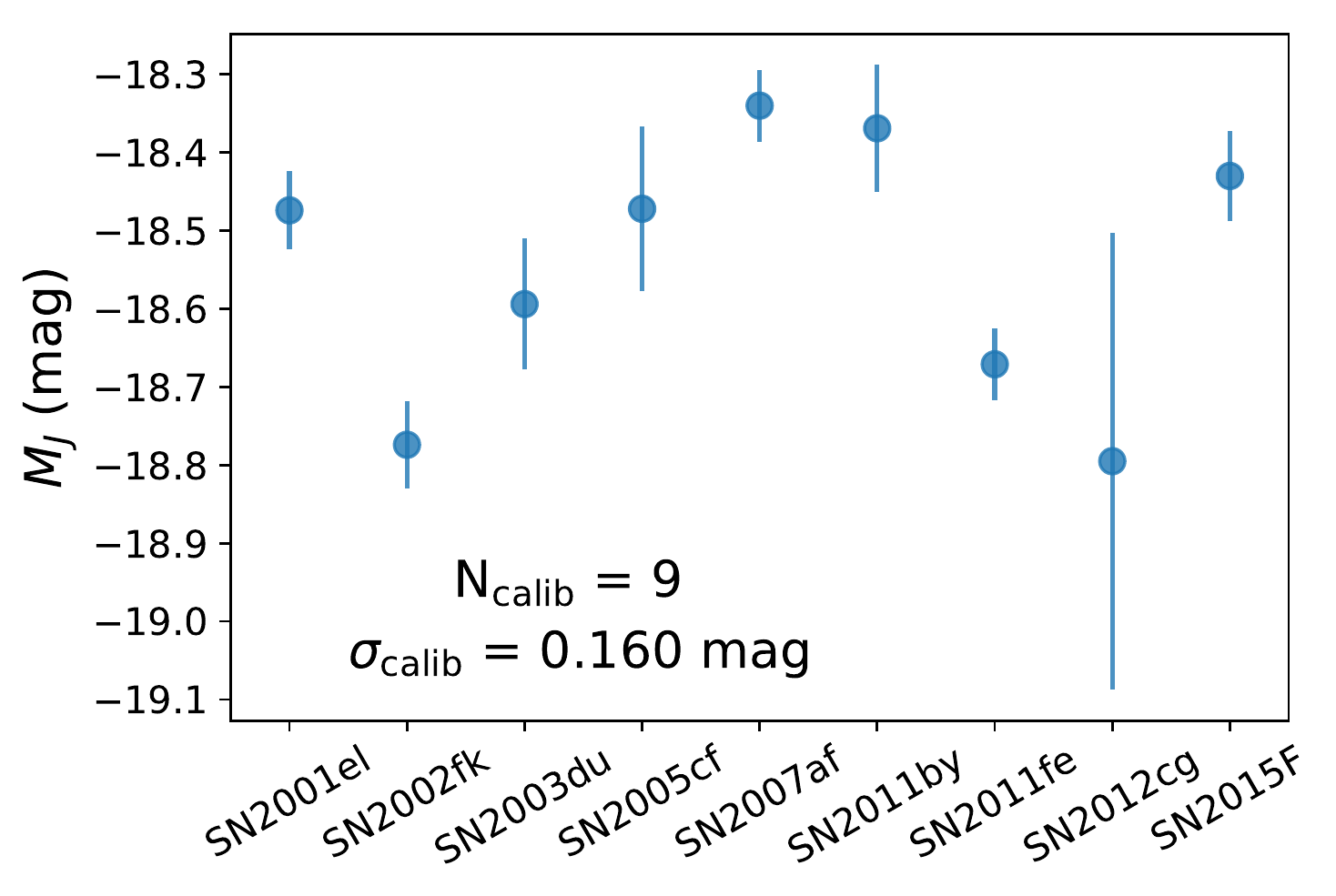}
\caption{The peak $J$ absolute magnitude distribution for the calibrator SN~Ia
sample, based on the Cepheid distances of \citet{Riess2016}. The data have
been corrected for Milky Way extinction and K-corrections, but no further
light curve shape or colour correction is applied.}
\label{fig:calib}
\end{figure}

The 9 calibrator SN absolute magnitudes are listed in Table~\ref{tab:calib}
and displayed in Figure~\ref{fig:calib}. The calibrator absolute magnitudes
show a dispersion of just $\sigma_{\rm calib} = 0.160$ mag. This
scatter, derived from treating the SN~Ia as \emph{standard} candles in $J$, is
comparable to the typical scatter in SN~Ia distances from optical light curves
\emph{after} light-curve-shape and colour corrections. The NIR dispersion,
though small, is nevertheless larger than can be accounted for by the formal
uncertainties $\sigma_M$, with $\chi^2 = 55.2$ for 8 degrees of freedom. This
suggests that the SN~Ia have an additional intrinsic (or more precisely,
unmodeled) scatter, $\sigma_{\rm int}$, that we need to include in our
analysis. Such a term is also routinely used in optical SN~Ia distances. If we
neglect the intrinsic scatter term, the weighted mean peak absolute magnitude
is $\avg{M_J} = -18.524 \pm 0.021$ mag.

In the same way as for the calibrator sample, we derive peak apparent
magnitudes $m_J$ and uncertainties $\sigma_{\rm fit}$ for the Hubble-flow
sample from the Gaussian process interpolation, applying Milky Way extinction
and $K$-corrections as above. For the $K$-corrections, we warp the
model SED to match the SN colour \citep[using the \texttt{mangle} option in \texttt{SNooPy};][]{Burns2011}. The difference between  colour-matching and not
colour-matching the SED is small ($\lesssim 0.01$ mag) for all SNe. For SNe that have $Y$ and $H$ band observations we use the $Y$, $J$, $H$ filters to match colours, otherwise we use $J$, $H$, $K$ filters. The former approach is more reliable, as pointed out by \citet{Boldt2014}; nonetheless these differences are small for SNe in our sample ($\lesssim 0.01$ mag). For four
objects (SN 2005eq, SN 2006lf, PTF10mwb, and PTF10ufj), the default Gaussian 
process covariance function parameters do not produce a satisfactory fit. 
For these objects, we
reduce the scale and increase the amplitude of the covariance function (the
parameters are specified in the fitting module code). Our final results are
listed in Table \ref{tab:hflow}.

We retrieved CMB-frame redshifts for the Hubble-flow host galaxies from
NED\footnote{\url{http://ned.ipac.caltech.edu}}. These are largely consistent
with values previously reported in the literature, except for NGC~2930, host
of SN~2005M, which has a previously erroneous redshift now corrected in NED.
Four of our Hubble-flow host galaxies are cluster members: for these we take
the cosmological redshift as the cluster redshift reported in NED rather than the
specific host galaxy redshift to avoid large peculiar velocities from the
cluster velocity dispersion. These objects are noted in Table \ref{tab:hflow}.
Following the analysis of \citet{Riess2016}, we also tabulate redshifts
corrected for coherent flows derived from a model based on visible large scale
structure \citep{Carrick2015}. Along with the reported redshift uncertainties
$\sigma_z$, we adopt an additional peculiar velocity uncertainty of
$\sigma_{\rm pec} =$ 150 \kms (for all SNe except PTF10ufj the redshift 
uncertainty is sub-dominant compared to the peculiar velocity uncertainty).

\begin{figure*}{}
\centering
\includegraphics[width=0.6\textwidth]{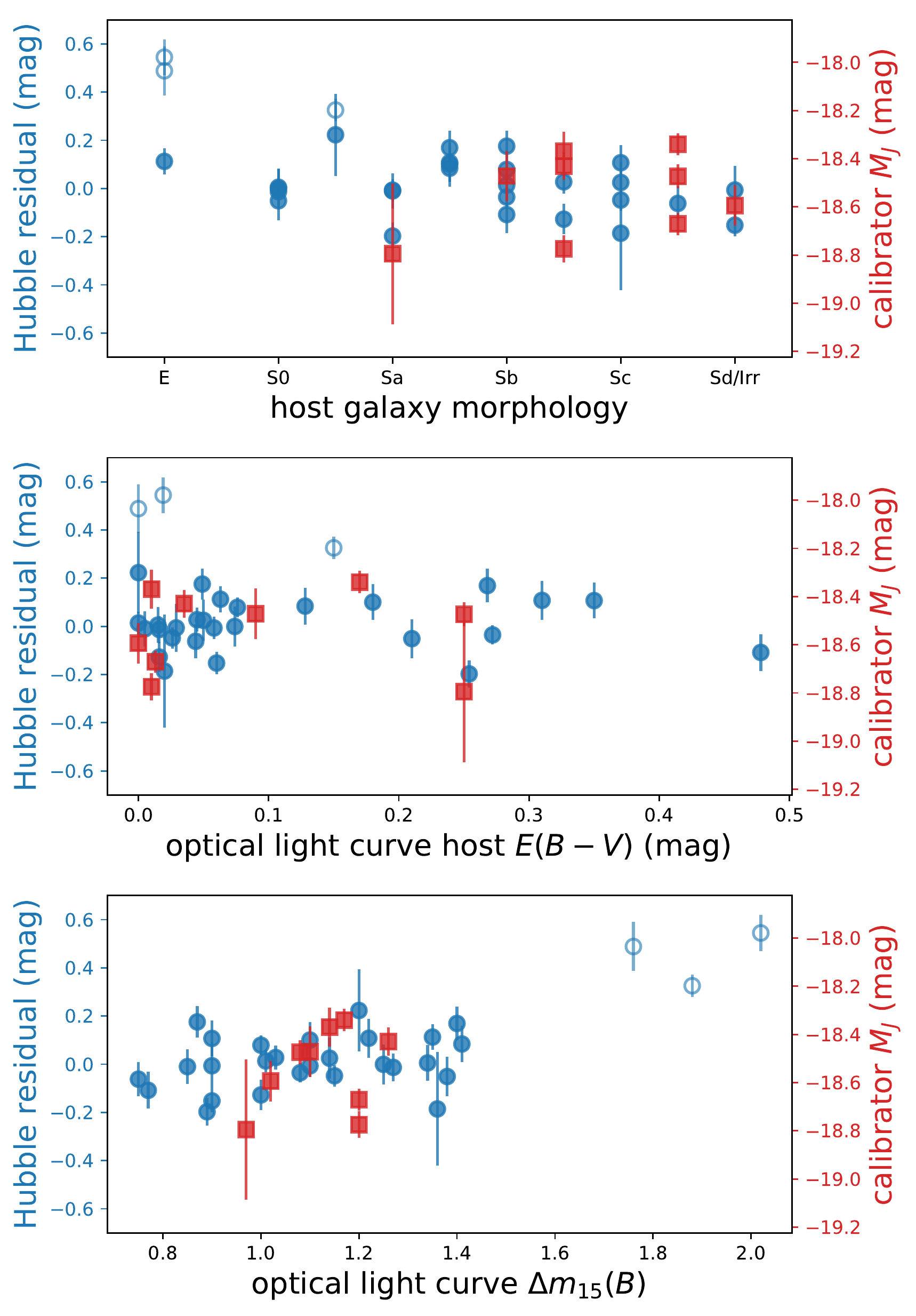}
\caption{A comparison of the calibrator and Hubble-flow samples in
host-galaxy morphology, host-galaxy reddening, and optical light-curve decline
rate. Blue circles show the Hubble-flow sample $J$-band Hubble-diagram
residuals (left axis), while red squares show the calibrator absolute $J$
magnitudes (right axis). The open circles indicate three fast-declining SN~Ia
that are excluded from our fiducial sample as outliers. These plots are used
to define sample cuts only. Distances are based on the $J$-band photometry
alone, with no corrections from these diagnostic parameters.}
\label{fig:diagnostic}
\end{figure*}

The high precision of modern SN~Ia \ho\ measurements
\citep{Riess2009,Riess2011,Riess2016} is due in part to selecting an ``ideal''
set of calibrator SN~Ia, with low extinction and typical light curve shapes.
The Hubble-flow SN~Ia are a much more heterogeneous set than these ideal
calibrators. Given that we are not applying colour or light curve shape
corrections and treating the SN~Ia as standard candles in their peak $J$
magnitude, it is important to ensure that our Hubble-flow objects are on the
whole similar to the calibrators. In Figure~\ref{fig:diagnostic}, we plot the
Hubble-flow Hubble diagram residuals and calibrator absolute magnitudes on the
same scale, as a function of host-galaxy morphology and two parameters
estimated from the optical light curves of these SN~Ia: host galaxy reddening
$E(B-V)$ and light-curve decline rate $\Delta m_{15}(B)$
\citep{Phillips93,Hamuy96_morph}. These quantities are taken from the
literature and tabulated in Table \ref{tab:diagnostic}; we have not attempted
to derive them in a uniform way. Rather, we are interested in comparing the
Hubble-flow and calibrator SN~Ia to suggest sample cuts. Beyond that, we do
not use the optical photometry in any way in our results.

\begin{table*}
\centering
\caption{Host-galaxy reddening, light-curve decline rate, and host-galaxy 
morphology for the calibrator and Hubble-flow SN~Ia, compiled from the
literature. $E(B-V)_{\rm host}$ and $\Delta m_{15}(B)$ are based on optical
data and used as diagnostics for sample cuts, but do not directly affect our
distance estimates. The morphology is mainly taken from NED, with a numerical
code given by: E=0, S0=1, Sa=2, Sb=3, Sc=4, and Sd/Irr=5.}
\small
\input{table_diagnostic}

\label{tab:diagnostic}
\tablebib{ The host-galaxy reddening and light-curve decline rate for the
calibrator sample are taken from \citet{Krisciunas2004}, \citet{Cartier2014}, 
\citet{Stanishev2007}, \citet{Wang2009}, \citet{Contreras2010}, \citet{Friedman2015}, 
\citet{Matheson2012}, \citet{Marion2016}, \citet{Cartier2017}. For the Hubble flow SNe 
observed by the CSP and CfA, they are derived from data presented in 
\citet{Contreras2010}, \citet{Stritzinger2011}, and \citet{Hicken2009,Hicken2012}, 
while for PTF10mwb and PTF10ufj the parameters are from \citet{Maguire2012}. }
\end{table*}

Figure \ref{fig:diagnostic} shows that the Hubble-flow SN~Ia span a broader
range of the displayed diagnostic parameters than the calibrators. This is to
be expected. For example, the calibrator galaxies are chosen to host Cepheids,
excluding early-type galaxies. Similarly the ``ideal'' calibrators have low
host reddening and normal decline rates. Nonetheless, the visual impression
from Figure \ref{fig:diagnostic} is that the broader Hubble-flow sample does
not show obvious trends with the parameters, except for the three fast-declining 
($\Delta m_{15}(B) > 1.7$)
SN~Ia, which are clear outliers (open circles). Indeed,
\citet{Krisciunas2009}, \citet{Kattner2012}, and \citet{Dhawan2017} have
demonstrated that the NIR absolute magnitudes of fast-declining SN~Ia diverge
considerably from their more normal counterparts (similar to the behaviour in
optical bands). We define a fiducial sample for analysis excluding these three
SN~Ia, and explore further sample cuts in Section~\ref{sec:samples}.

\begin{figure*}
\centering
\includegraphics[width=0.6\textwidth]{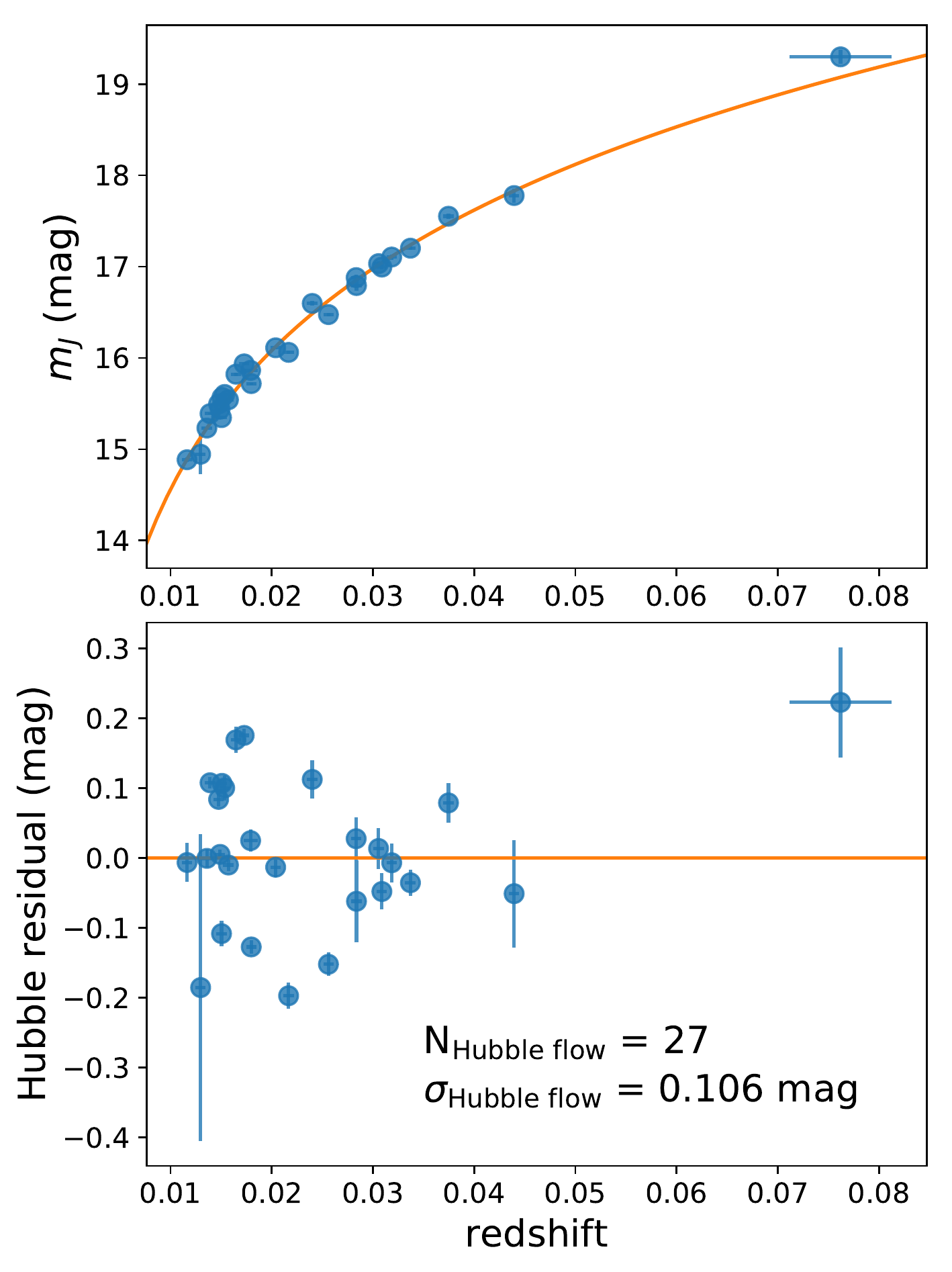}
\caption{Hubble diagram for our fiducial sample of 27 Hubble-flow SN~Ia.}
\label{fig:hflow}
\end{figure*}

The Hubble diagram of our fiducial sample, with 27 objects in the Hubble flow,
is shown in Figure \ref{fig:hflow}. The standard deviation of the
residuals is just $\sigma_{\rm Hflow} = 0.106$ mag in our simple 
standard-candle approach. Very few optical SN~Ia samples have such a low scatter,
even after light curve shape and colour correction. This scatter
\emph{includes} known components like the photometry and redshift measurement
uncertainties and peculiar velocities. We convert the redshift/velocity
uncertainties to magnitudes, with
\begin{equation}
\sigma_{z{\rm ,mag}} \approx \frac{5}{\ln 10}\frac{\sigma_z}{z} 
\quad \mbox{and} \quad
\sigma_{\rm pec,mag} \approx \frac{5}{\ln 10}\frac{\sigma_{\rm pec}}{cz}
\label{eq:pecv}
\end{equation}
and adopt $\sigma_{\rm pec}$ = 150 \kms\ as noted above. The individual
Hubble-flow object uncertainty is then the quadrature sum of these terms and
the Gaussian process fit peak magnitude uncertainty, $\sigma_m^2 = \sigma_{\rm
fit}^2 + \sigma_{z{\rm ,mag}}^2 + \sigma_{\rm pec,mag}^2$.

As for the calibrators, though to a lesser extent, the Hubble-flow sample
shows more scatter than can be explained by the formal uncertainties, here
with $\chi^2 = 62.8$ for 26 degrees of freedom. Again, this points to the need
for an additional intrinsic scatter component to explain the variance in the
data.

Combining the calibrator sample and the Hubble-flow sample yields our estimate
of $H_0$, with 
\begin{equation}
m_J - M_J = 5 \log d_L + 25
\label{eq:distmod}
\end{equation} 
for the luminosity distance $d_L$ measured in Mpc. Following \citet{Riess2016}
we use a kinematic expression for the luminosity distance-redshift relation,
with
\begin{equation}
d_L(z) = \frac{cz}{H_0}\left[1+\frac{(1-q_0)z}{2} - \frac{(1 - q_0 - 3
q^2_0 + j_0)z^2}{6} + \mathcal{O}(z^3)\right]
\label{eq:klumdist}
\end{equation}
and we fix $q_0 = -0.55$ and $j_0 = 1$. We have also explored the dynamic
parametrisation of the luminosity distance in a flat, $\Omega_M +
\Omega_\Lambda = 1$, Universe \citep[see,
e.g.,][]{Jha2007},
\begin{equation}
d_L(z) = \frac{c(1+z)}{H_0}\int_0^z \left[\Omega_M(1 + z')^3 +
\Omega_\Lambda\right]^{-1/2} dz'.
\end{equation}
Because our Hubble-flow sample is at quite low redshift, we find no
significant differences in our results with either approach, nor when
varying cosmological parameters within their observational limits.

In estimating $H_0$ from SN~Ia it is traditional to rewrite
equations~\ref{eq:distmod}
and \ref{eq:klumdist} as
\begin{equation}
\mathrm{log}\, H_0 = \frac{M_J + 5a_J + 25}{5}.
\label{eq:h0}
\end{equation}
where $M_J$ is constrained by the calibrator sample, and $a_J$ is the
``intercept of the ridge line'' that can be determined separately from the
Hubble-flow sample. Ignoring higher order terms, the intercept is given by
\begin{equation}
a_J = \log cz + \log \left[1+\frac{(1-q_0)z}{2} - \frac{(1 - q_0 - 3
q^2_0 + j_0)z^2}{6}\right] - 0.2 m_J.
\label{eq:intercept}
\end{equation}

We vary the traditional analysis slightly to account for the necessary
intrinsic scatter parameter, $\sigma_{\rm int}$, that we interpret as
supernova to supernova variance in the peak $J$ luminosity. We introduce
$\sigma_{\rm int}$ as a nuisance parameter that is to be constrained by the
data and marginalized over. We assume that the intrinsic scatter is a property
of the supernovae, independent of whether an object is in the calibrator
sample or the Hubble-flow sample (and test this assumption in
Section~\ref{sec:intscatter}). In this case the full uncertainty for a given
calibrator object $i$ is
\begin{equation}
\sigma_{M,i}^2 = \sigma_{{\rm fit},i}^2 + \sigma_{{\rm Ceph,}i}^2 +
\sigma_{\rm int}^2
\label{eq:calibunc}
\end{equation}
and the total uncertainty for a Hubble-flow object $k$ is
\begin{equation}
\sigma_{m,k}^2 = \sigma_{{\rm fit},k}^2 + \sigma_{z{\rm ,mag,}k}^2 +
\sigma_{{\rm pec,mag},k}^2 + \sigma_{\rm int}^2.
\end{equation}

Because the same intrinsic scatter affects the relative weights of both
calibrator and Hubble-flow objects, we cannot solve for $M_J$ and $a_J$
independently. Instead we fit a joint Bayesian model to the combined data set,
with MCMC sampling of the posterior distribution using the \texttt{emcee}
package \citep{FM13code,FM13}. In principle we have four fit parameters:
$M_J$, $a_J$, $\sigma_{\rm int}$, and $H_0$, but we can simplify this to just
three using equation \ref{eq:h0}. We choose $H_0$, $M_J$, and $\sigma_{\rm
int}$ as our parameterisation, and simply calculate $a_J$ for each MCMC sample
given $H_0$ and $M_J$. The results would be identical if we had fit for $a_J$
and calculated $H_0$. For convenience, rather than $a_J$, we tabulate $-5 a_J$
which can be expressed in units of magnitudes and interpreted in the same
sense as the Hubble-flow peak magnitudes $m_J$. In our Bayesian analysis we
take uninformative priors: uniform on $H_0 > 0$ and $M_J$, and scale-free on $\sigma_{\rm int} > 0$, with $p(\sigma_{\rm int}) =
1/\sigma_{\rm int}$. Our full analysis code, including notebooks that produce
Figures \ref{fig:calib}, \ref{fig:diagnostic}, \ref{fig:hflow}, and
\ref{fig:corner}, is available at \url{https://github.com/sdhawan21/irh0}.

\section{Results}
\label{sec:results}

\begin{figure*}
\centering
\includegraphics[width=0.9\textwidth]{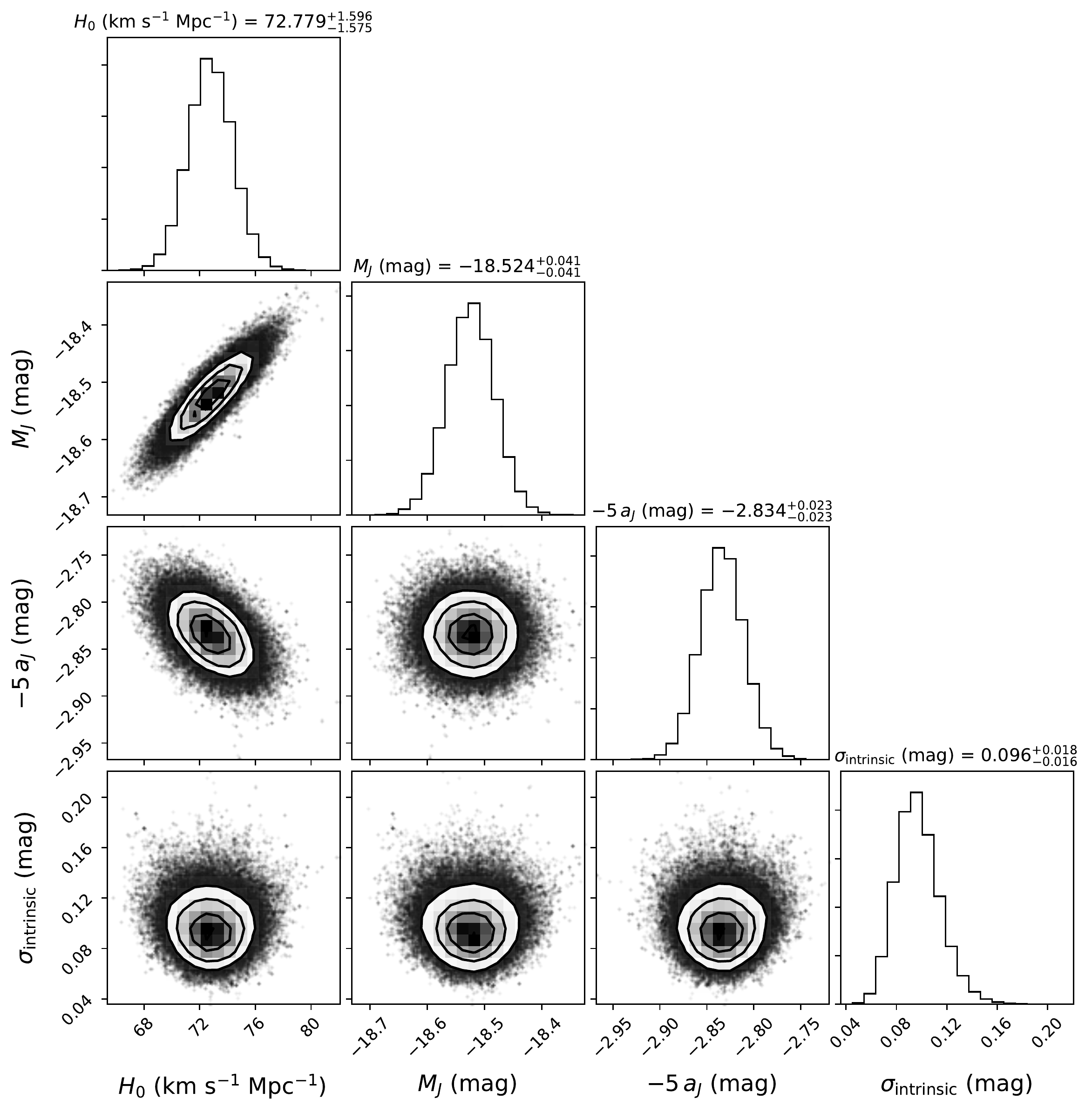}
\caption{Distribution and covariances of the model parameters for our
fiducial sample. The uncertainties are \emph{statistical only}, with the
median value and 16th and 84th percentile differences listed. As discussed in
the text, $a_J$ is not a fit parameter; it is calculated from the other
parameters for each sample. This plot uses the \texttt{corner} package by
\citet{FM17}.}
\label{fig:corner}
\end{figure*}

Our fiducial sample consists of the 9 calibrator SN~Ia and 27 Hubble-flow
SN~Ia (i.e., excluding the three fast-declining outliers). We use the NED
redshifts and uncertainties (columns 3 and 4 of Table \ref{tab:hflow}) for the
Hubble-flow objects. The results from $2 \times 10^5$ posterior samples of our
model are shown in Figure \ref{fig:corner} and tabulated in Table
\ref{tab:results}. We find a sample median $H_0 = 72.78_{-1.57}^{+1.60}$
\kmpc, where the uncertainty is statistical only, and is measured down (up) to
the 16th (84th) percentile\footnote{As seen in Figure~\ref{fig:corner}, the
marginal distributions are largely symmetric, so using the medians or means
give similar results.}. The 2.2\% statistical uncertainty is impressive given
the small sample size. The results show that the median calibrator peak
magnitude ($M_J = -18.524 \pm 0.041$) contributes approximately 2\%
uncertainty to \ho, whereas the Hubble flow sample contributes about 1\% ($-5
a_J = -2.834 \pm 0.023$ mag), in line with the numbers of supernovae in each
category.

We also see the intrinsic scatter parameter is estimated clearly to be
non-zero: $\sigma_{\rm int} = 0.096_{-0.016}^{+0.018}$. This has the effect of
increasing the uncertainties in the other parameters, for instance, roughly
doubling the uncertainty on the peak absolute magnitude $M_J$ compared to the
straight weighted mean calculated in Section~\ref{sec:analysis}. Though our
analysis method was developed to allow the intrinsic scatter parameter to
connect to both the calibrator and Hubble-flow samples, we further see in
Figure \ref{fig:corner} that $M_J$ and $-5 a_J$ do not have much correlation,
reflecting the fact they are largely being constrained separately by the
calibrators and Hubble-flow objects, respectively.

\begin{table*}
\caption{Results with varying sample choices. Sample median values of the fit
parameters are given, with 16th and 84th percentile differences
(\emph{statistical uncertainties only}). }
\begin{minipage}{70mm}
\centering
\small
\renewcommand{\arraystretch}{1.3}
\input{table_results}
\end{minipage}
\label{tab:results}
\end{table*}

\subsection{SN sample choices}
\label{sec:samples}

To explore the sensitivity of our derived \ho, in Table~\ref{tab:results} we
present a number of different sample choices. First, we find that adopting the
flow corrections to the CMB frame redshifts from Table~\ref{tab:hflow} has
only a small effect, raising \ho\ by 0.5\% (0.4 \kmpc), though also slightly
increasing the Hubble-flow scatter from 0.106 mag to 0.115 mag. Because of the
increased scatter, we do not adopt these flow corrections for our fiducial
sample.

Similarly, restricting the Hubble-flow sample to low host reddening, low Milky
Way extinction, or spiral galaxies only has little effect on the derived \ho\
or intrinsic scatter. Combining these three cuts does decrease the Hubble flow
scatter slightly to 0.094 mag at the expense of eliminating nearly half of the
Hubble flow sample; this combination yields \ho\ that is 1\% higher than the
fiducial sample.

Limiting the redshift range of the Hubble-flow sample similarly has little
effect as shown in Table \ref{tab:results}, 1.6\% higher at most if the sample
is reduced to just the 12 objects with $0.02 \le z \le 0.05$. If we make an
even more extreme cut, identifying both calibrator and Hubble flow objects
that overlap in their diagnostics from Figure \ref{fig:diagnostic}, we retain
only 7 calibrators and 8 Hubble-flow objects. Even so, the derived value of
\ho\ in this ``strictest overlap'' sample shows no significant deviation than
from the larger, fiducial sample.

On the other hand, if we include the three fast-declining ``outlier'' SN~Ia,
these fainter objects \citep{Krisciunas2009,Dhawan2017} pull the value of \ho\
down to $71.3 \pm 2.1$ \kmpc\ (a 2.0\% decrease), while increasing the Hubble
flow scatter significantly, from 0.106 to 0.170 mag. As seen in Figure
\ref{fig:diagnostic}, these fast-declining Hubble-flow objects clearly have no
calibrator analogues and should be excluded.

\subsection{Intrinsic scatter}
\label{sec:intscatter}

The derived value of the intrinsic scatter for the fiducial sample,
$\sigma_{\rm int} = 0.096_{-0.016}^{+0.018}$ mag seems reasonable compared to
optical SN~Ia distances after standardization. Nevertheless, adopting a single
intrinsic scatter may serve to
obscure systematic uncertainties. In particular, we note that the dispersion
of the residuals of the fiducial Hubble flow sample is 0.106 mag (which
includes the measured photometric and redshift uncertainties as well as
peculiar velocity uncertainties), substantially less than the scatter in the
calibrators, 0.160 mag (including photometric and Cepheid distance
uncertainties). We can test these for consistency by including \emph{two}
separate intrinsic scatter parameters, one for the calibrators and one for the
Hubble flow. In that case, we find $\sigma_{\rm int, calib} =
0.147_{-0.039}^{+0.056}$ mag and $\sigma_{\rm int, Hflow} =
0.073_{-0.017}^{+0.020}$ mag. These are only consistent at the $\sim$2$\sigma$
level. Marginalizing over both of these parameters, our fiducial value of \ho\
is not significantly changed, but has slightly higher uncertainty, $H_0 =
72.81_{-2.04}^{+2.03}$ \kmpc, corresponding to 2.8\% precision rather than
2.2\%. This is mainly due to the higher intrinsic scatter in the calibrators,
for which the peak absolute magnitude is now measured to lower precision: $M_J
= -18.524 \pm 0.057$ mag.

Separating the intrinsic scatter in this manner leads to some other
complications. In principle the Hubble-flow intrinsic scatter parameter, which
is independent of redshift, should be separable from the adopted peculiar
velocity uncertainty (which for a fixed velocity uncertainty, corresponds to
larger magnitude uncertainty at lower redshifts; equation \ref{eq:pecv}).
However, because the redshift range of our Hubble-flow sample is small, in
practice the Hubble-flow intrinsic scatter is largely degenerate with the
adopted peculiar velocity uncertainty. Raising $\sigma_{\rm pec}$ to 250 \kms,
as used in \citet{Riess2016}, completely accounts for \emph{all} the Hubble
flow scatter, and yields $\sigma_{\rm int, Hflow} = 0.000_{-0.000}^{+0.007}$
mag, clearly inconsistent with the scatter in the calibrators. Conversely, if
we unrealistically assume $\sigma_{\rm pec} = 0$, the intrinsic scatter
parameter increases to explain the observed scatter, $\sigma_{\rm int, Hflow}
= 0.096_{-0.013}^{+0.017}$ mag. Regardless of the exact choice, the
marginalized result for \ho\ is largely unaffected, differing only by $\pm$0.1
\kmpc\ relative to our fiducial choice of $\sigma_{\rm pec} = 150$ \kms. Our
choice is plausible \citep{RS2004,Turnbull2012,Feindt2013} and previous
studies have also used this value for the peculiar velocity uncertainty
\citep{Mandel2009,Mandel2011,Barone-Nugent2012}. Nevertheless, the higher
scatter in the calibrator sample compared to the larger Hubble flow sample
is a concern that should be noted; any
systematic differences between the calibrators and the Hubble-flow objects
could create a bias in \ho. Augmenting both of these samples in the future
should clarify the situation.

\subsection{Systematic uncertainties}
\label{sec:systematics}

In this study, we assume that SNe~Ia are standard candles in the $J$-band; we
do not correct for the light curve shape or host galaxy reddening. From
Figure~\ref{fig:diagnostic} we see that these parameters are not significantly
different for the calibrators and the Hubble flow objects. For example, the
average difference (fiducial Hubble flow minus calibrators) in $E(B-V)_{\rm
host}$ is just 0.023 $\pm$ 0.043 mag. For $A_J \approx 0.8\,E(B-V)$, this
corresponds to a potential 1 $\pm$ 2 \% correction to \ho. Curiously, in the
middle panel of Figure \ref{fig:diagnostic}, we do not see any evidence that
objects with larger $E(B-V)_{\rm host}$ are observed to be fainter (positive
residuals). Given the data in Table 3 show a mild anti-correlation between 
$E(B-V)_{\rm host}$ and $\Delta m_{15}(B)$, there could be a fortuitous cancellation
between a colour correction and a light-curve shape dependence like that found by \citep{Kattner2012}. In Table \ref{tab:results} we note that restricting the samples to
low $E(B-V)_{\rm host} \leq 0.3$ mag does not change \ho\ significantly.
Similarly, Figure~\ref{fig:diagnostic} shows that for the fiducial sample
(excluding the fast-declining objects), there are no strong trends in residual
with host galaxy morphology or optical light curve shape. Were we to regress
these parameters and make corrections, our derived \ho\ would not
significantly change. Based on this analysis, we adopt a $\pm$1\% systematic
uncertainty from these potential sample differences.

One source of uncertainty that is systematically different between the
calibrators and the Hubble-flow sample is $K$-corrections. There are not as
many near-infrared spectra of SNe~Ia as in the optical, and ground-based
observations are complicated by atmospheric absorption. Nevertheless, the
median redshift of our Hubble-flow sample is only $z_{\rm med} = 0.018$, where
the $K$-correction uncertainties are $\sim$0.015 mag 
\citep{Boldt2014,Stanishev2015},
corresponding to only a $\sim$0.7\% uncertainty in \ho.

Our Hubble-flow sample is drawn largely from two surveys: CSP
\citep{Contreras2010,Stritzinger2011} and CfA
\citep{Wood-Vasey2008,Friedman2015}. \citet{Friedman2015} do an extensive 
comparison of CfA and CSP NIR photometry for 18 SNe~Ia in common, and find 
a mean offset in $J$ of just $\avg{\Delta m_J ({\rm CSP}-{\rm CfA})} = 
-0.004 \pm 0.004$  mag. Two of these objects, SN~2005el and SN~2008hv, 
have the requisite light-curve coverage to qualify for our Hubble-flow 
sample. Table~\ref{tab:csp_cfa} 
shows the excellent agreement in peak $m_J$ for these objects when comparing
the two surveys and justifies our combining the photometry for 
these two objects in our fiducial sample.

Despite these extensive cross-checks between the surveys, 
Table \ref{tab:results} shows the intercept of the
ridge line (in the form of $-5a_J$) differs by 0.078 mag between the 
CSP and CfA Hubble-flow samples,
translating into a $\sim$3.7\% difference in \ho. Unfortunately six of the
nine calibrators have their $J$-band photometry from sources other than these
surveys, so it is difficult to simply restrict our analysis to one system or
the other. Table~\ref{tab:results} presents results for the CSP Hubble-flow
sample with the one calibrator with CSP photometry (SN 2007af) and similarly
the CfA Hubble-flow sample with two CfA calibrators (SN~2011by and SN~2012cg),
but these calibrator samples are too small to meaningfully estimate an
intrinsic scatter and determine a secure value for \ho. The variety of
photometric systems for the calibrator sample may help explain its higher
scatter compared to the Hubble flow sample. Filter corrections
\citep[$S$-corrections;][]{Stritzinger2002,Friedman2015} are expected to be
modest in $J$-band at peak (the supernova near-infrared color is within the
stellar locus), so zero-point differences may play the largest role. Here we
adopt a $\pm$3\% systematic uncertainty on \ho\ from the average difference
between the calibrators and Hubble-flow sample based on the different
photometric systems. Our information is too limited here to better quantify
this uncertainty, but this estimate makes it the largest component in the
systematic error budget and a ripe target for future improvement.

\begin{table}
\caption{Best fit $J$-band peak magnitudes for the two Hubble-flow SNe observed by both CSP and CfA.}
\centering
\renewcommand{\arraystretch}{1.2}
\input{table_csp_cfa}
\label{tab:csp_cfa}
\end{table}

Combining these effects our near-infrared ``supernova'' standard candle
systematic uncertainty amounts to 3.2\%. Our estimate of \ho\ also relies on
the Cepheid distances of \citet{Riess2016} and thus we adopt their systematic
uncertainties (see their Table 7) for the lower rungs of the distance ladder
including the primary anchor distance, the mean Leavitt Law in the anchors,
the mean Leavitt Law in the calibrator host galaxies \citep[corrected for the
fact we only use 9 calibrators rather than 19 as in][]{Riess2016}, Cepheid
reddening and metallicity corrections, and other Leavitt Law uncertainties.
This gives a ``Cepheid+anchor'' systematic uncertainty of 1.8\%.

To check this we have analysed our sample using the alternate Cepheid
distances from \citet{Cardona2017}, who introduce hyper-parameters to account
for outliers and other potential systematic uncertainties in the data set. For
our calibrators, the differences based on this reanalysis are minor, mainly
somewhat increased uncertainties in a few of the Cepheid distances. The
biggest change is for NGC~4424 (host of SN~2012cg), for which
\citet{Cardona2017} find $\mu_{\rm Ceph} = 30.82 \pm 0.19$ mag compared to
$31.08 \pm 0.29$ mag from \citet{Riess2016}. In fact this reduces the scatter
for the 9 calibrators from 0.160 mag to 0.133 mag, as shown in Table
\ref{tab:results}, and increases \ho\ by 1.4\% compared to our fiducial
analysis, well within the 1.8\% Cepheid+anchor systematic uncertainty.

Summing our supernova (3.2\%) plus Cepheid+anchor (1.8\%) systematic
uncertainties in quadrature yields our total systematic uncertainty of 3.7\%,
and gives our final estimate of $H_0 = 72.8 \pm 1.6$ (statistical) $\pm\;2.7$
(systematic) \kmpc. Our result is completely consistent with
\citet{Riess2016}. Because the Cepheid and anchor data we adopt is from that
analysis, those uncertainties are in common and our near-infrared cross-check
of their result is actually more precise. Our result can be written $H_0 =
72.8 \pm 1.6$ (statistical) $\pm\;2.3$ (separate systematics) $\pm\;1.3$ (in
common systematics) \kmpc. Leaving out the in-common systematics we can
compare our result $72.8 \pm 2.8$ \kmpc\ with the \citet{Riess2016} result
(also leaving out systematics in common, which dominate their error budget),
$73.2 \pm 0.8$ \kmpc\ and find excellent agreement\footnote{Because 
\citet{Riess2016} used flow corrections for the Hubble flow sample, perhaps 
the best comparison is our flow-corrected redshifts value (Table 
\ref{tab:results}), which coincidentally matches their value.}.

\section{Discussion and Conclusion}
\label{sec:dis}

Our main conclusion is that replacing optical light curve standardized
distances of SNe~Ia with $J$-band standard candle distances gives a wholly
consistent (at lower precision) distance ladder and measurement of the Hubble
constant. This suggests that supernova systematic uncertainties that could be
expected to vary with wavelength (e.g., dust extinction or colour correction)
are not likely to play a dominant role in ``explaining" the tension between
the local measurement of \ho\ and its inference from CMB data in a standard
cosmological model. Studies have sought to determine whether or not SN~Ia 
luminosity variations relate to local environments in nearby samples 
\citep[$z \lesssim 0.1$;][]{Rigault2013,Rigault2015,Kelly2015,Jones2015,Roman2017}. 
To play a dominant role, such environmental factors must affect both the 
optical and $J$-band light curves in common. 

Our final result has lower precision than the \citet{Riess2016}, with total
(statistical+systematic) uncertainty of 4.3\%: 72.8 $\pm\;3.1$ \kmpc. We can
still compare this with the reverse distance ladder estimate from the 2016 
Planck intermediate results: 66.93 $\pm\;0.62$ \kmpc\  \citep{Planck2016}, 
and find a 1.8$\sigma$ ``tension''. The significance of this would be increased 
had we posited \emph{a priori} that we were only checking for a local value that 
was higher than the CMB value (i.e., a one-tailed test).

A number of aspects of our analysis can be improved, both in terms of
statistical and systematic uncertainty. Our calibrator sample size is less
than half that of \citet{Riess2016}, and our Hubble-flow sample nearly an
order of magnitude smaller (and at typically lower redshift, more susceptible
to peculiar velocities). Our statistical uncertainty would be improved by more
objects in both sets: more Cepheid-calibrated SN~Ia and importantly, more
well-sampled near-infrared light curves of nearby and Hubble-flow SN~Ia. Our
limited sample is due in part to our stringent requirement for NIR data before
$J$-band maximum light. This peak typically occurs a few days \emph{before}
$B$-band maximum light, and has been difficult to measure. New surveys that
discover nearby SN~Ia earlier, combined with rapid NIR follow-up will
certainly help.

Of course, we do not need to use only the $J$-band peak magnitude. For
example, $H$-band is even less sensitive to dust and may provide an even
better standard candle
\citep{Kasen2006,Wood-Vasey2008,Mandel2009,Mandel2011,Weyant2014}. However, 
the $H$-band
``peak'' is much broader in time and not as well-defined as in $J$, making it
difficult to measure in our approach. Fitting template NIR light-curves
\citep[e.g.][]{Wood-Vasey2008,Folatelli2010,Burns2011,Kattner2012} would allow
for additional filters and sparser light-curve coverage. It will be important 
to test whether data from later epochs has increased scatter relative to the 
$J$-band peak; for instance, in the redder optical bands, the second maximum 
does show more variation among SNe \citep{Hamuy96_morph,Jha2007,Dhawan2015}. 
In addition, here we measure the peak $J$ magnitude \emph{at the time of $J$ 
maximum}; previous studies have often measured the ``peak'' magnitude at the 
time of $B$ maximum. These times of maxima can show systematic variations
\citep{Krisciunas2009,Kattner2012} that could lead to a different magnitude 
scatter between the two approaches. 

Systematic uncertainties can also be mitigated. We have ascribed our dominant
systematic uncertainty to photometric calibration of the $J$-band data, as
evidenced by the difference in Hubble-flow intercepts from CSP and CfA, and
perhaps the increased scatter in the calibrator sample (which has more
heterogeneous photometric sources). Further in-depth analysis of the
photometry \citep[already discussed extensively in][]{Friedman2015}
could in principle significantly reduce this uncertainty. Augmented NIR
spectroscopic templates could better quantify $K$- and $S$-correction
uncertainties. Finally, we performed our analysis unblinded, raising the
possibility of confirmation bias in our results. Future analyses can be
designed with a blinded methodology, as recently applied to this problem by
\citet{Zhang2017}.

Perhaps the most remarkable of our results is how well a purely standard
candle approach works, with intrinsic (unmodeled) scatter comparable to
optical light curves after correction. Measuring and applying
corrections to the NIR light curves (based on light-curve shape, colour, host
galaxy properties, local environments, etc.) should only serve to increase the
precision. NIR observations of SNe~Ia may thus play a key role in
a distance ladder that makes the best future measurements of the
local value of \ho, an extremely valuable cosmological constraint.

\bibliographystyle{aa} 
\bibliography{main}


\begin{acknowledgements}
We are grateful to Adam Riess and Dan Scolnic for helpful comments. We thank 
Michael Foley for supplying the flow model velocity corrections. We appreciate 
useful discussions with Arturo Avelino, Anupam Bhardwaj, Chris Burns, 
Regis Cartier, Andrew Friedman, Kate Maguire, Kaisey Mandel, and Michael
Wood-Vasey.  We thank the anonymous referee for helpful suggestions. 
B.L. acknowledges support for this work by
the Deutsche Forschungsgemeinschaft through TRR33, The Dark Universe. This
research was supported by the Munich Institute for Astro- and Particle Physics
(MIAPP) of the DFG cluster of excellence "Origin and Structure of the
Universe". S.W.J.~acknowledges support from US Department of Energy grant
DE-SC0011636 and valuable discussion and collaboration opportunity during the
MIAPP program ``The Physics of Supernovae''.
\end{acknowledgements}

\onecolumn
\appendix

\section{Gaussian Process Light Curve Fitting} 
\label{sec:gpfits}
\begin{figure}[h]
\centering
\includegraphics[width=.3\linewidth]{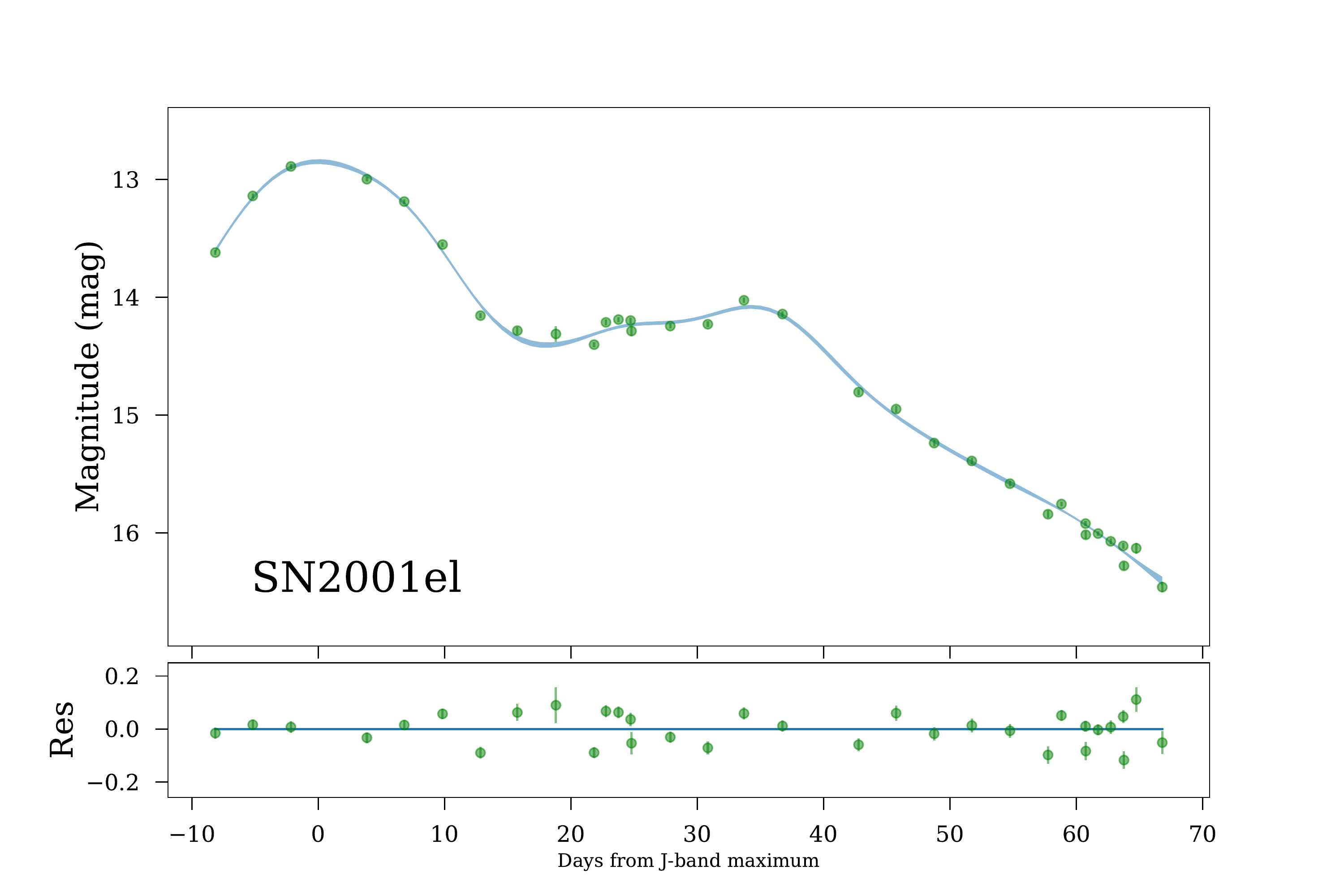}
\includegraphics[width=.3\linewidth]{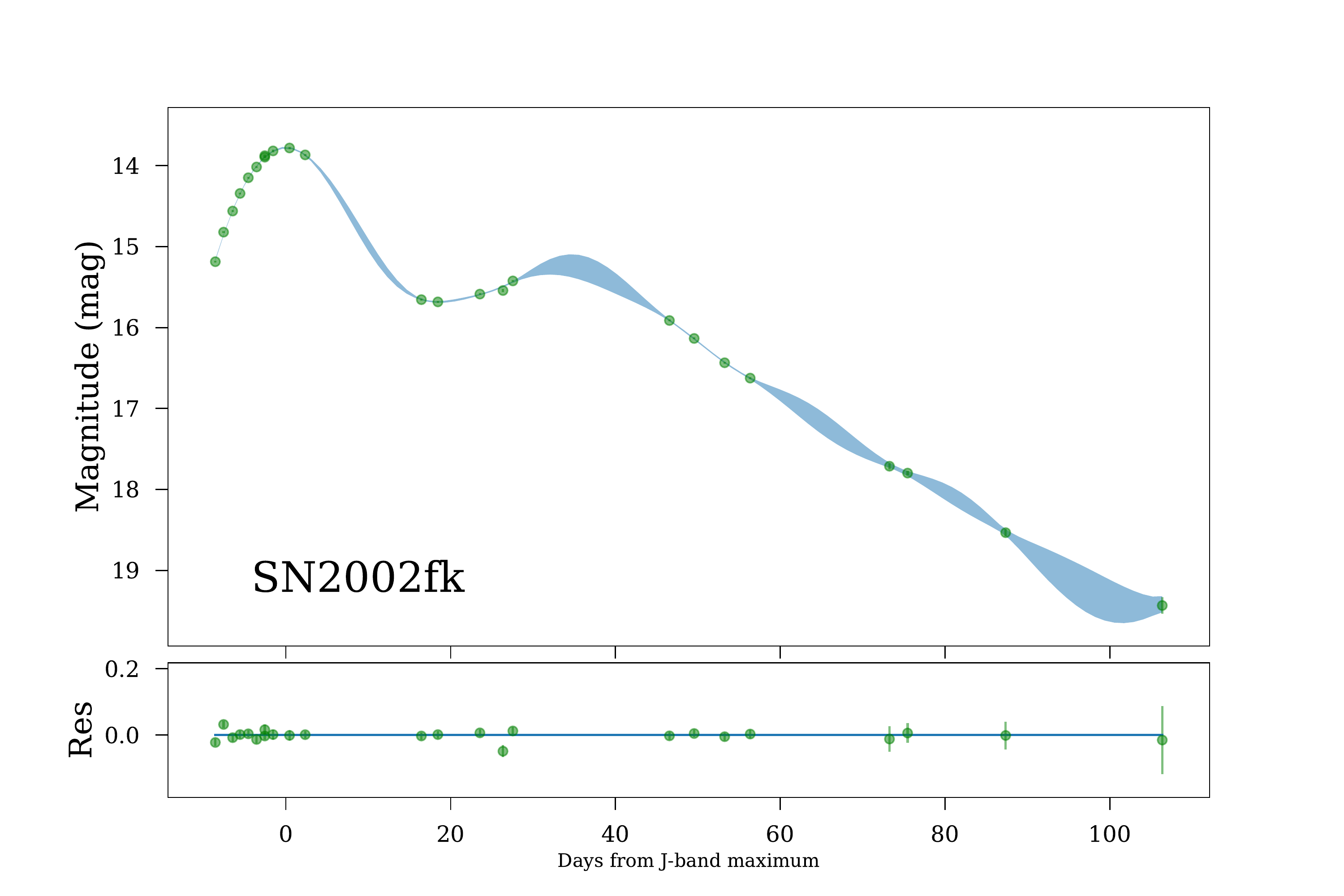}
\includegraphics[width=.3\linewidth]{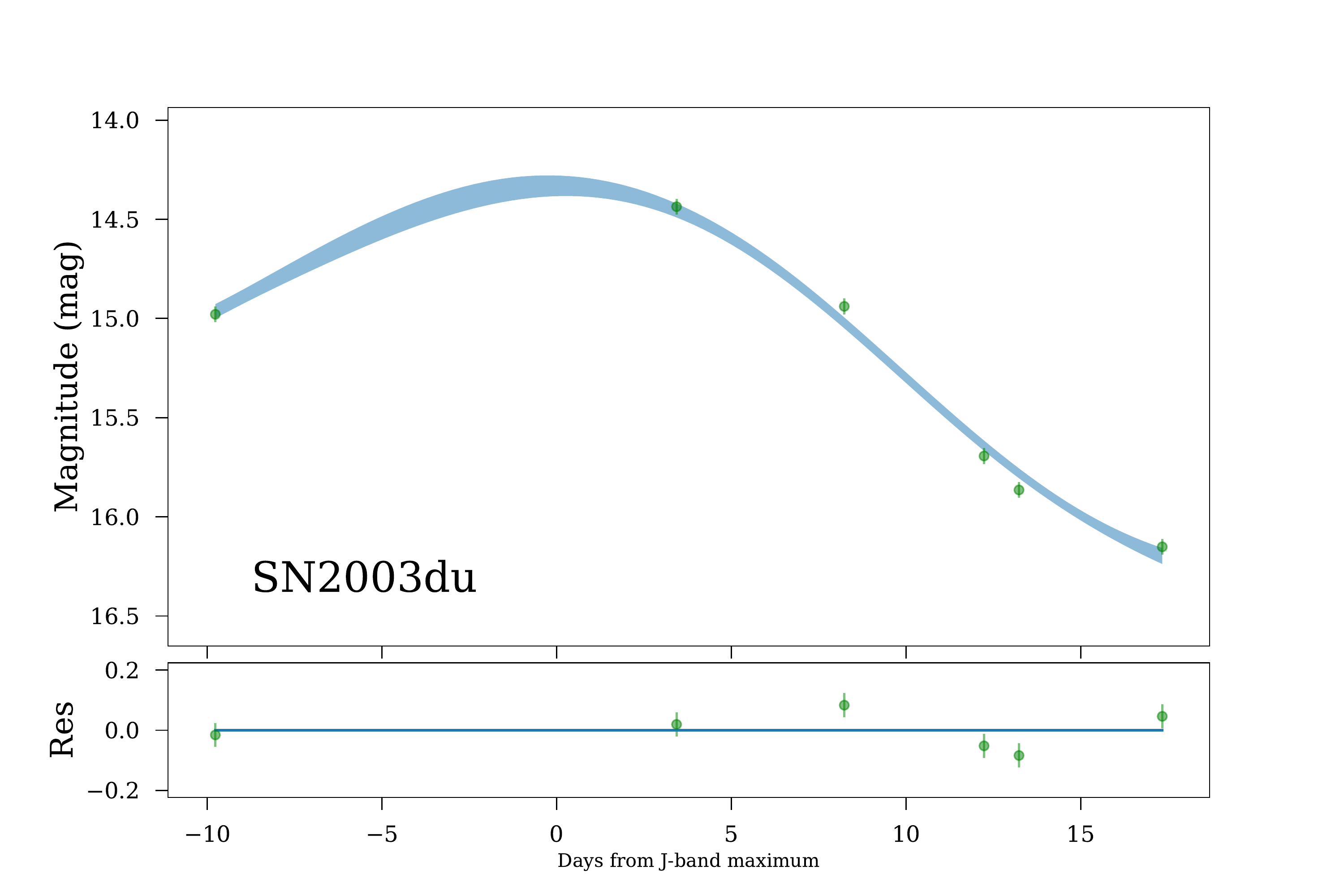}
\includegraphics[width=.3\linewidth]{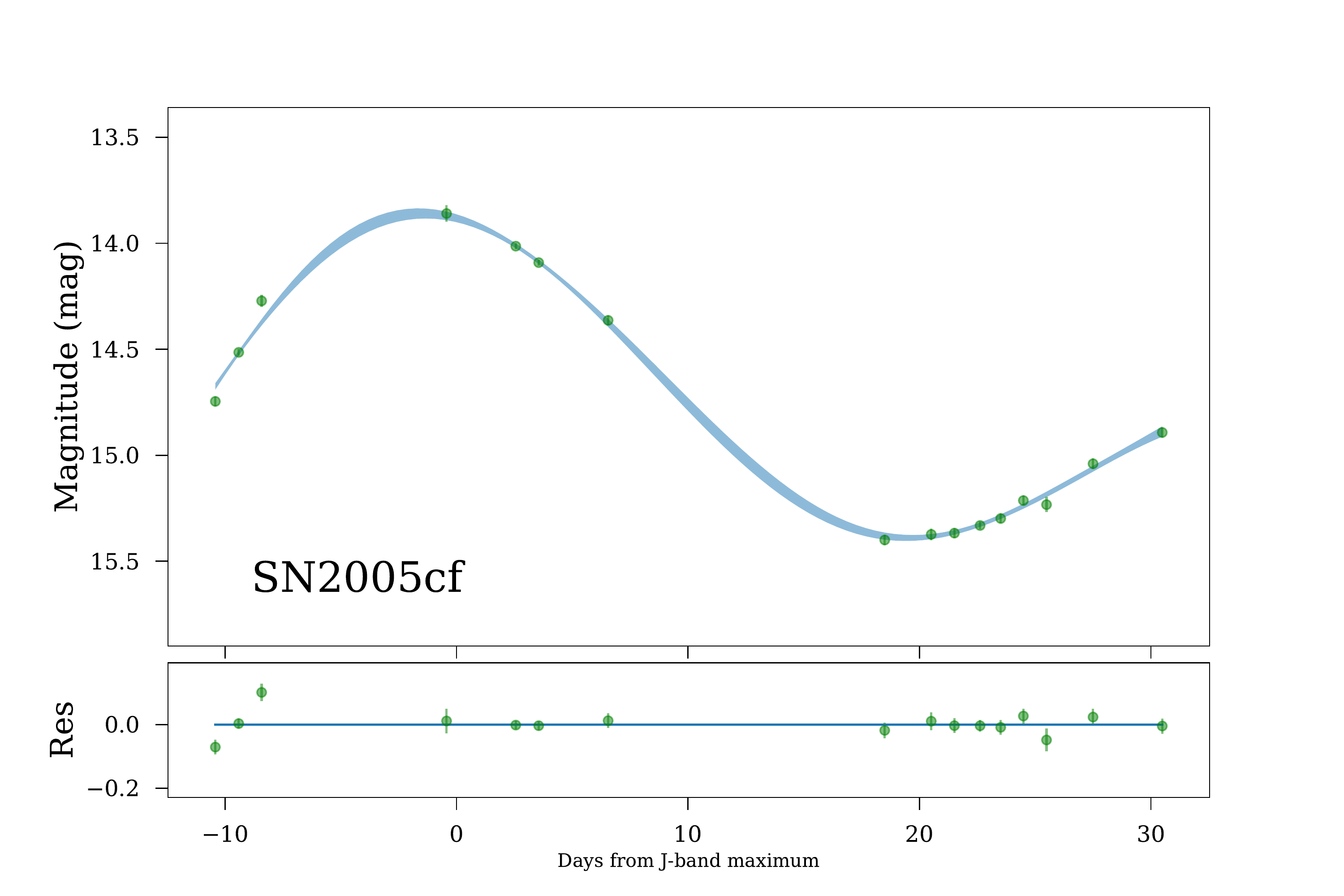}
\includegraphics[width=.3\linewidth]{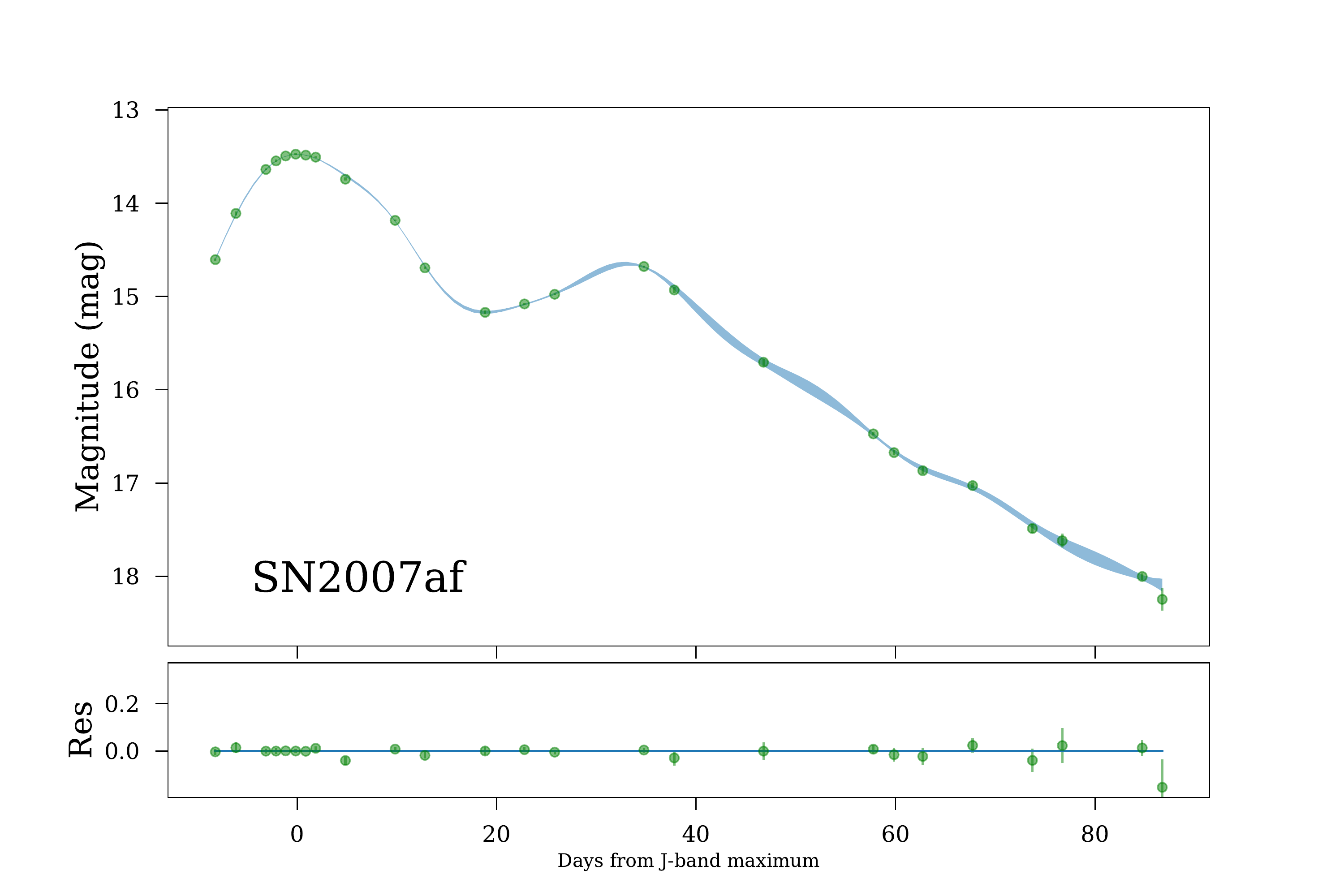}
\includegraphics[width=.3\linewidth]{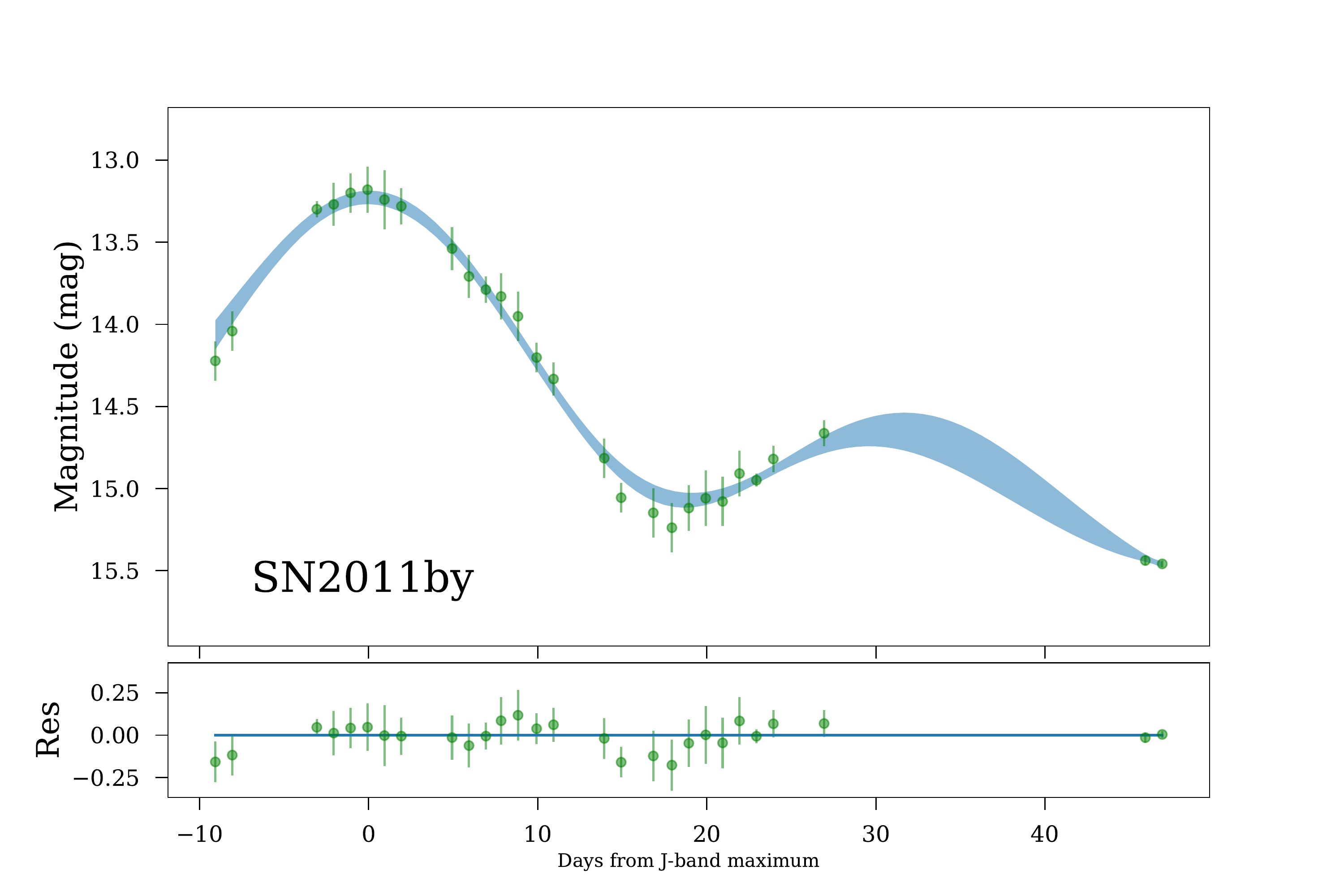}
\includegraphics[width=.3\linewidth]{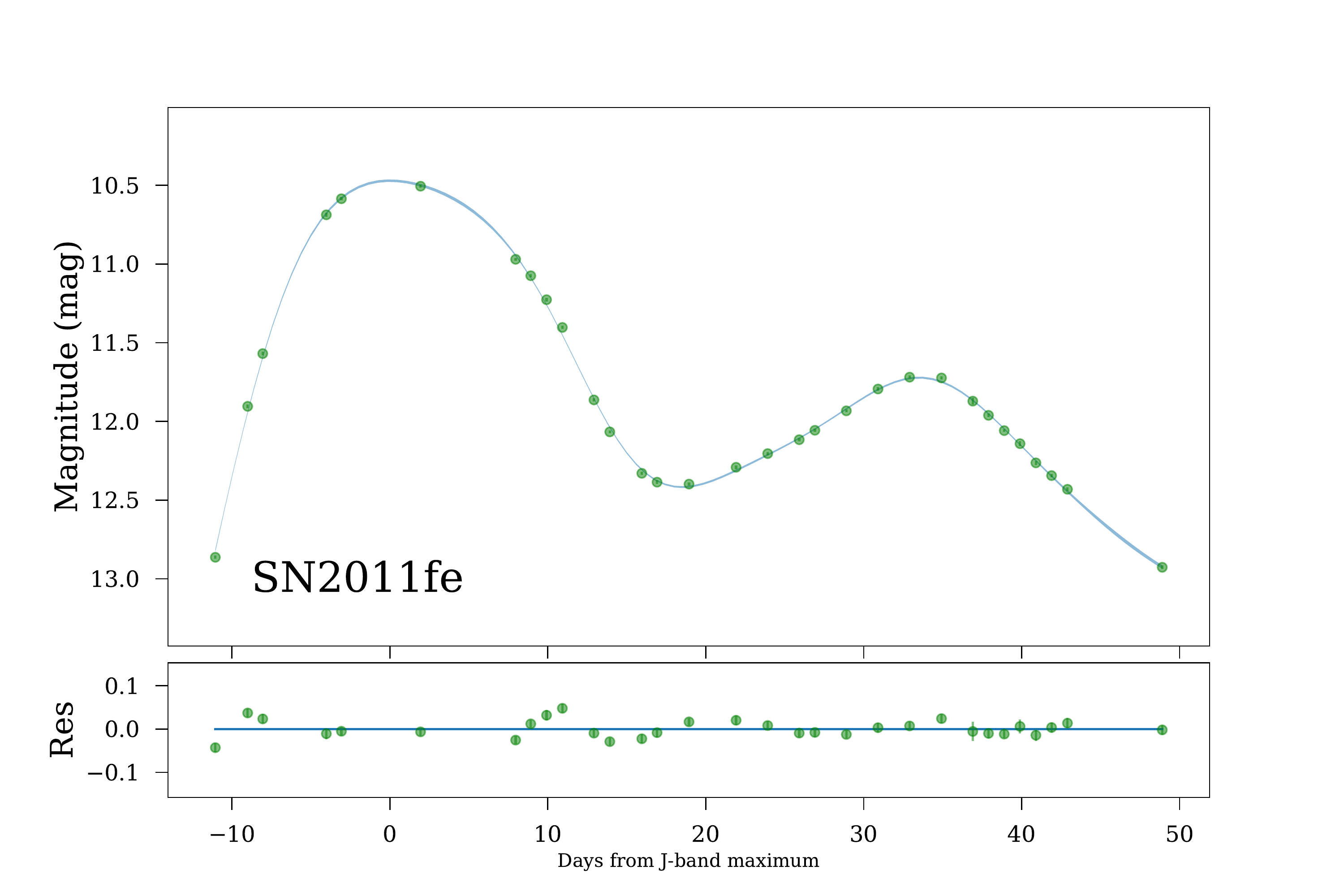}
\includegraphics[width=.3\linewidth]{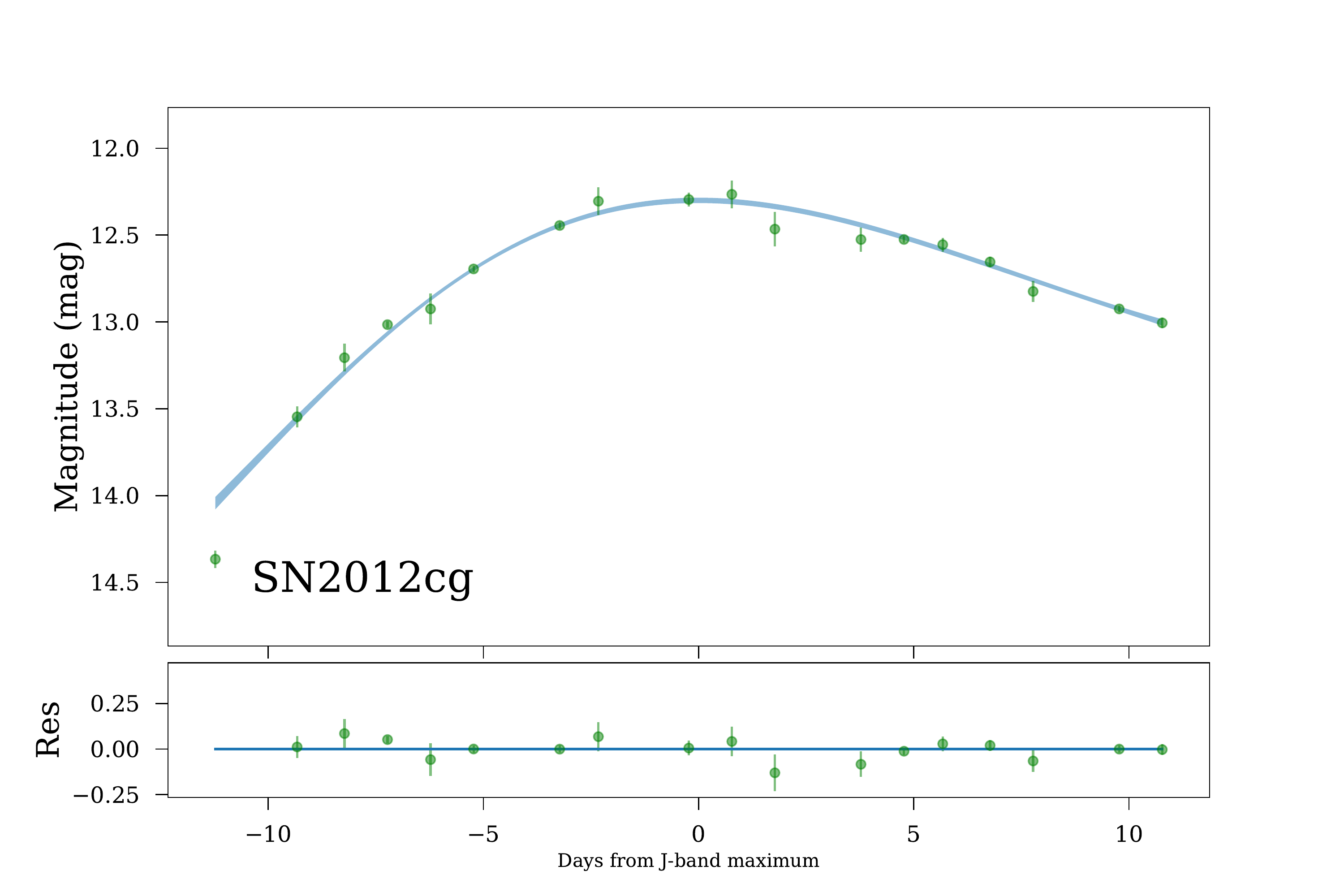}
\includegraphics[width=.3\textwidth]{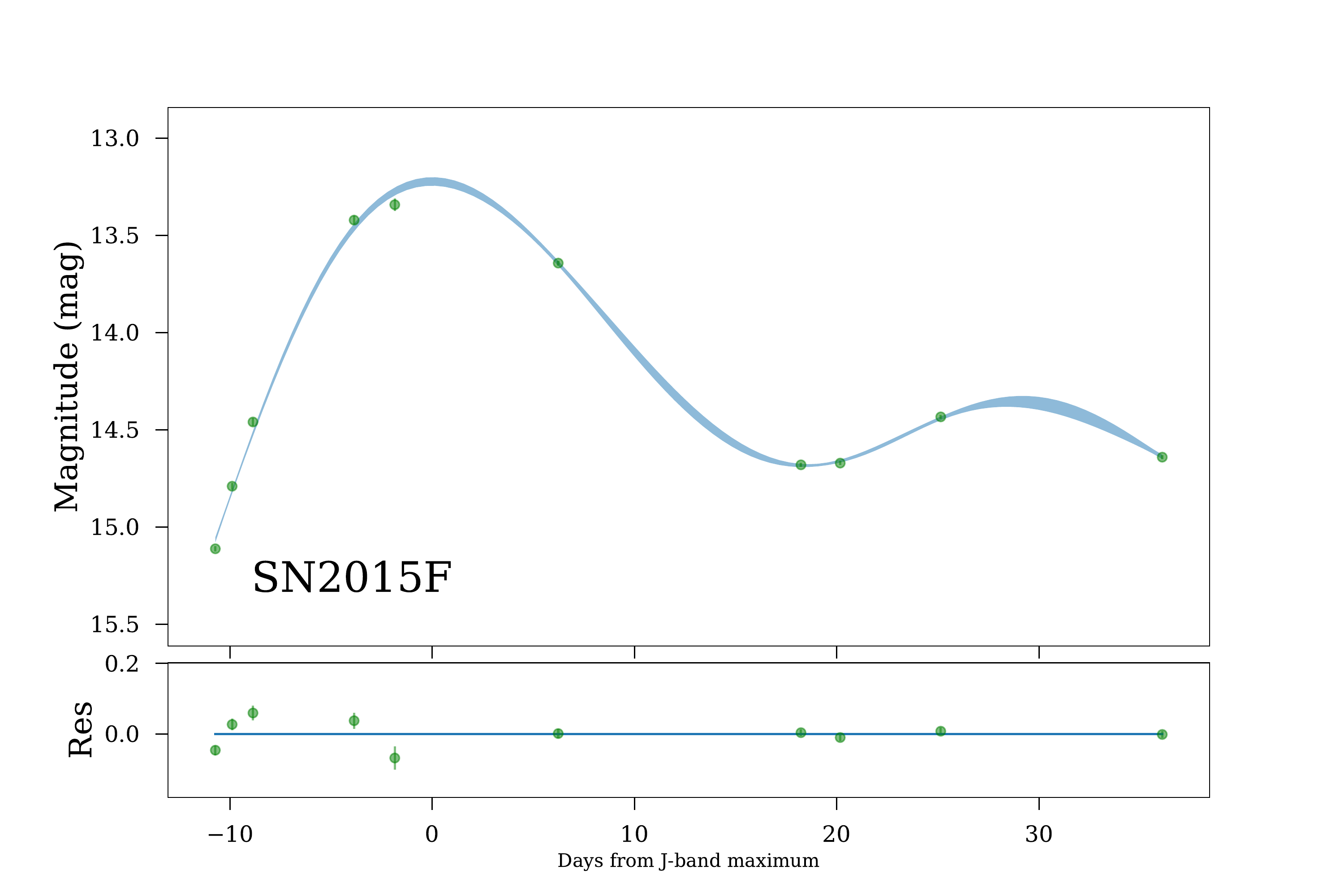}
\caption{Gaussian Process Fits for SNe in the calibration sample. The
errorbars are smaller than the point sizes in most cases. On the x-axis, the days from $J$-band maximum are in the observer frame.}
\label{fig:gpfits_calib}
\end{figure}

\begin{figure}[h]
\centering
\includegraphics[width=.2\linewidth]{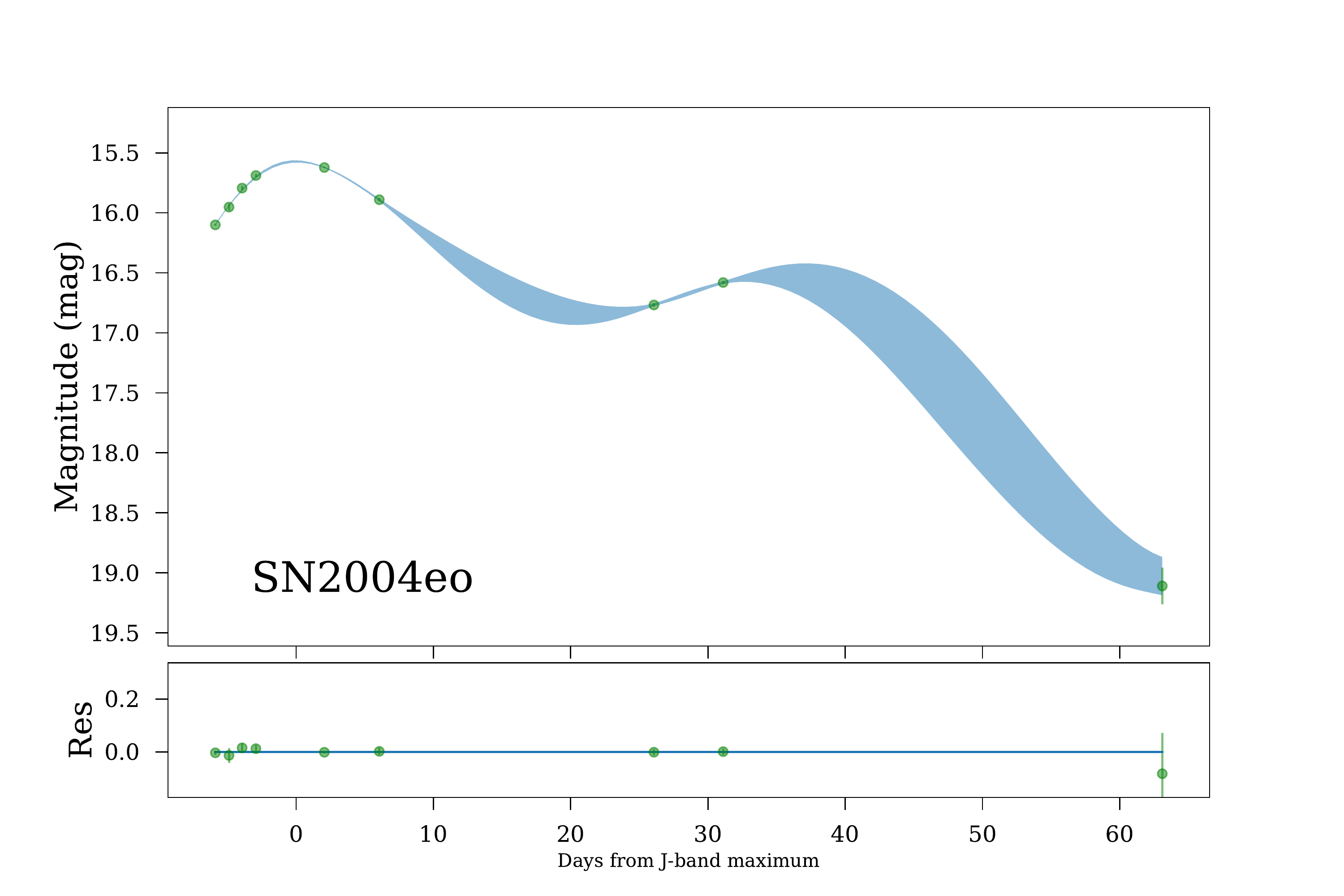}
\includegraphics[width=.2\linewidth]{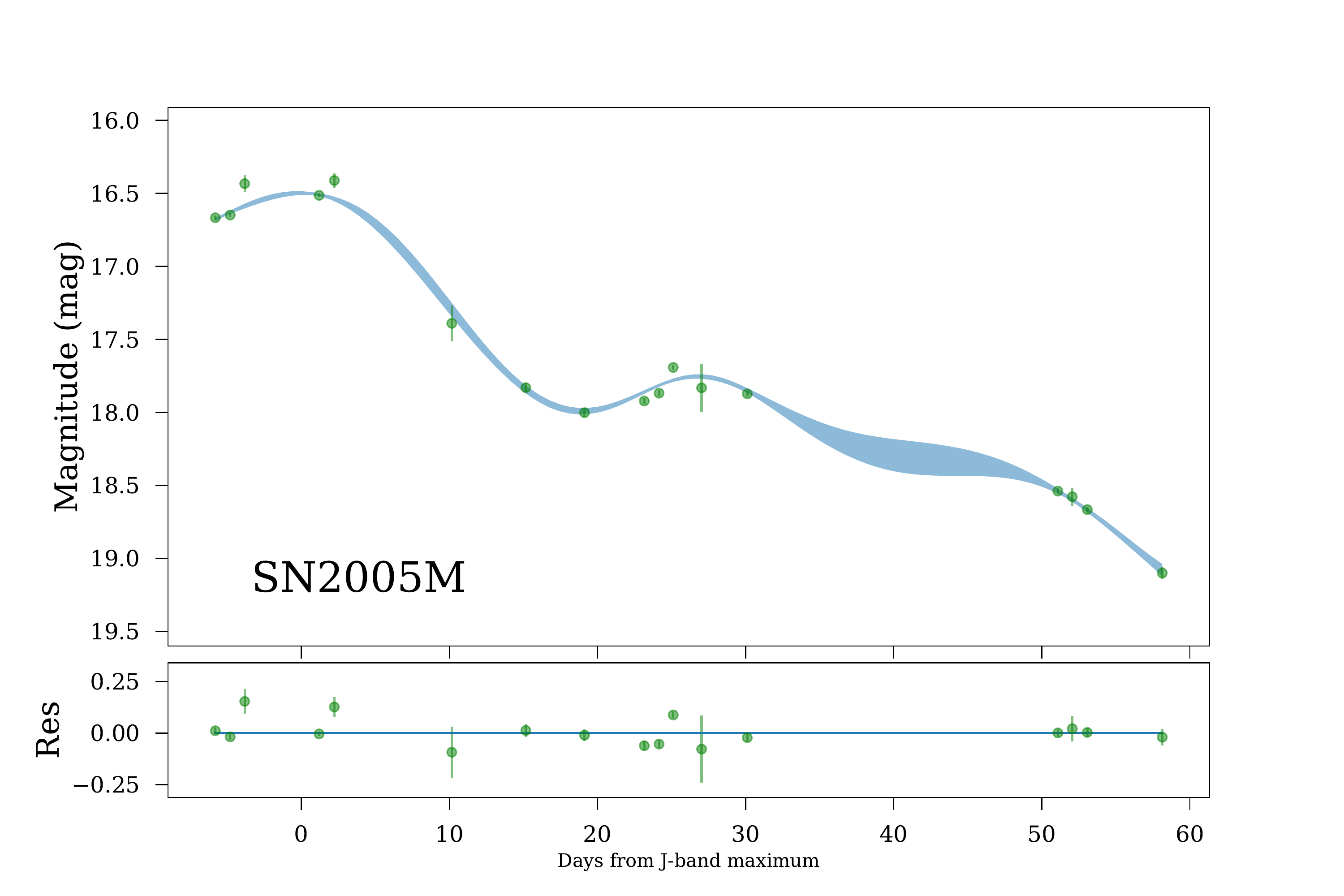}
\includegraphics[width=.2\linewidth]{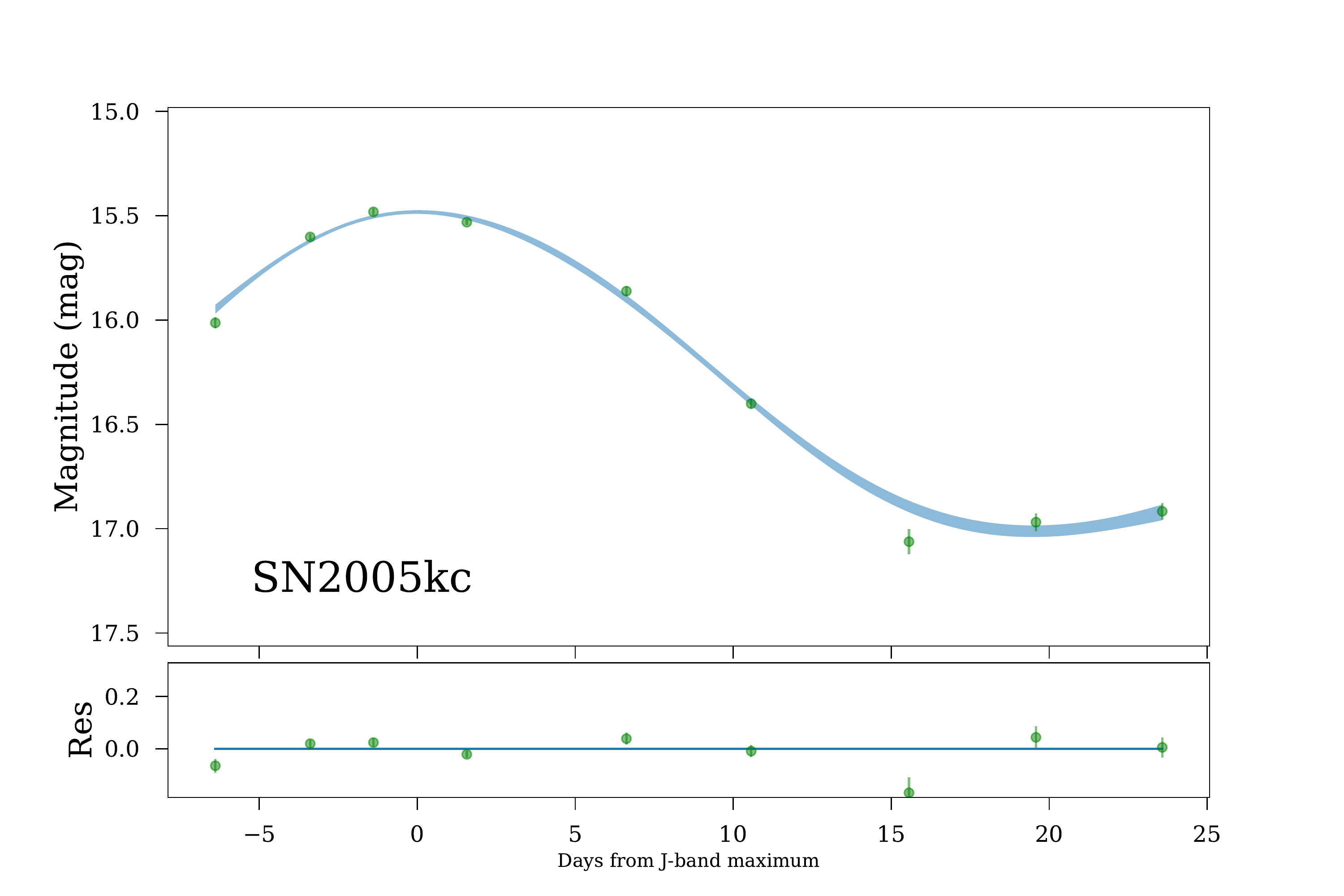}
\includegraphics[width=.2\linewidth]{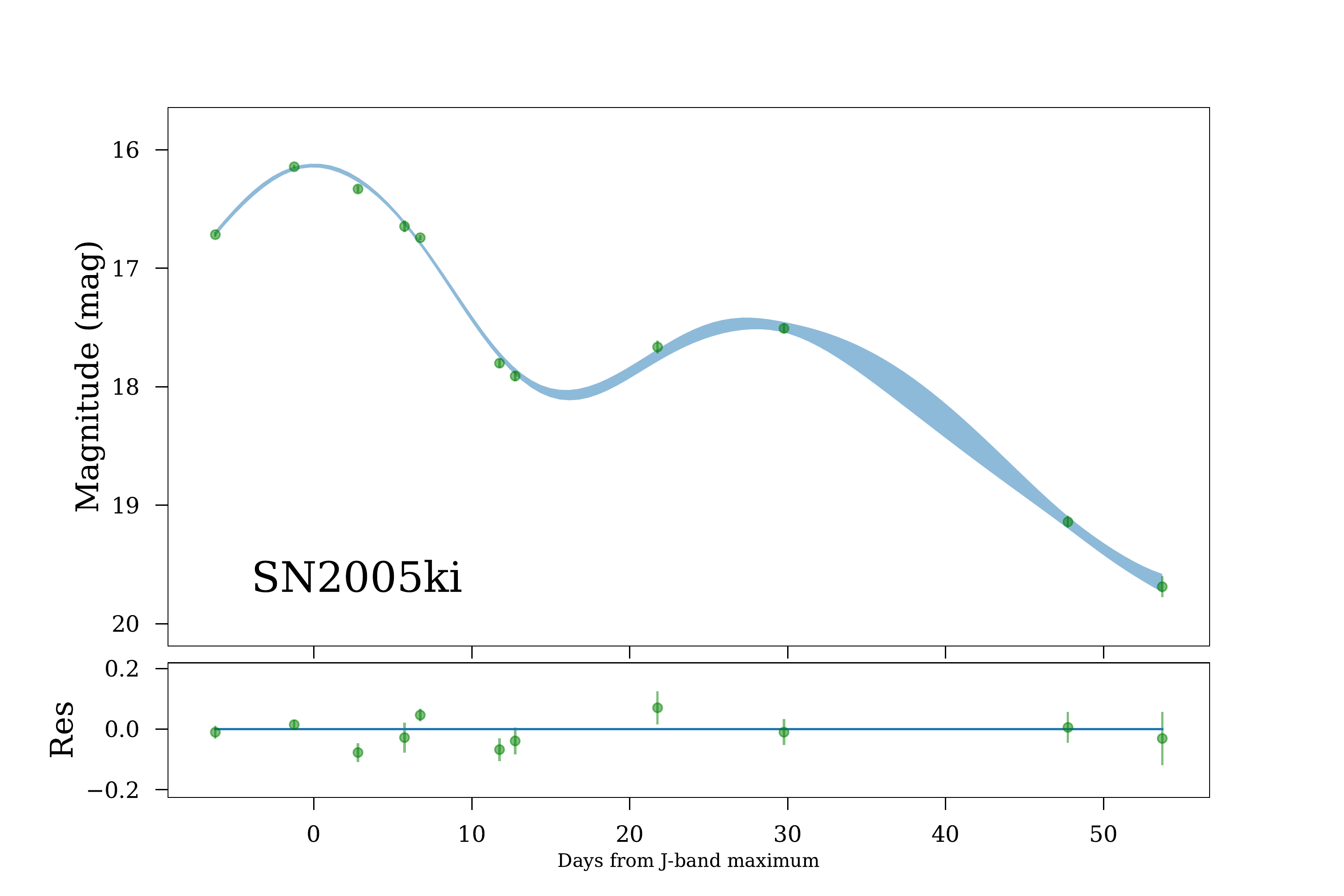}
\includegraphics[width=.2\linewidth]{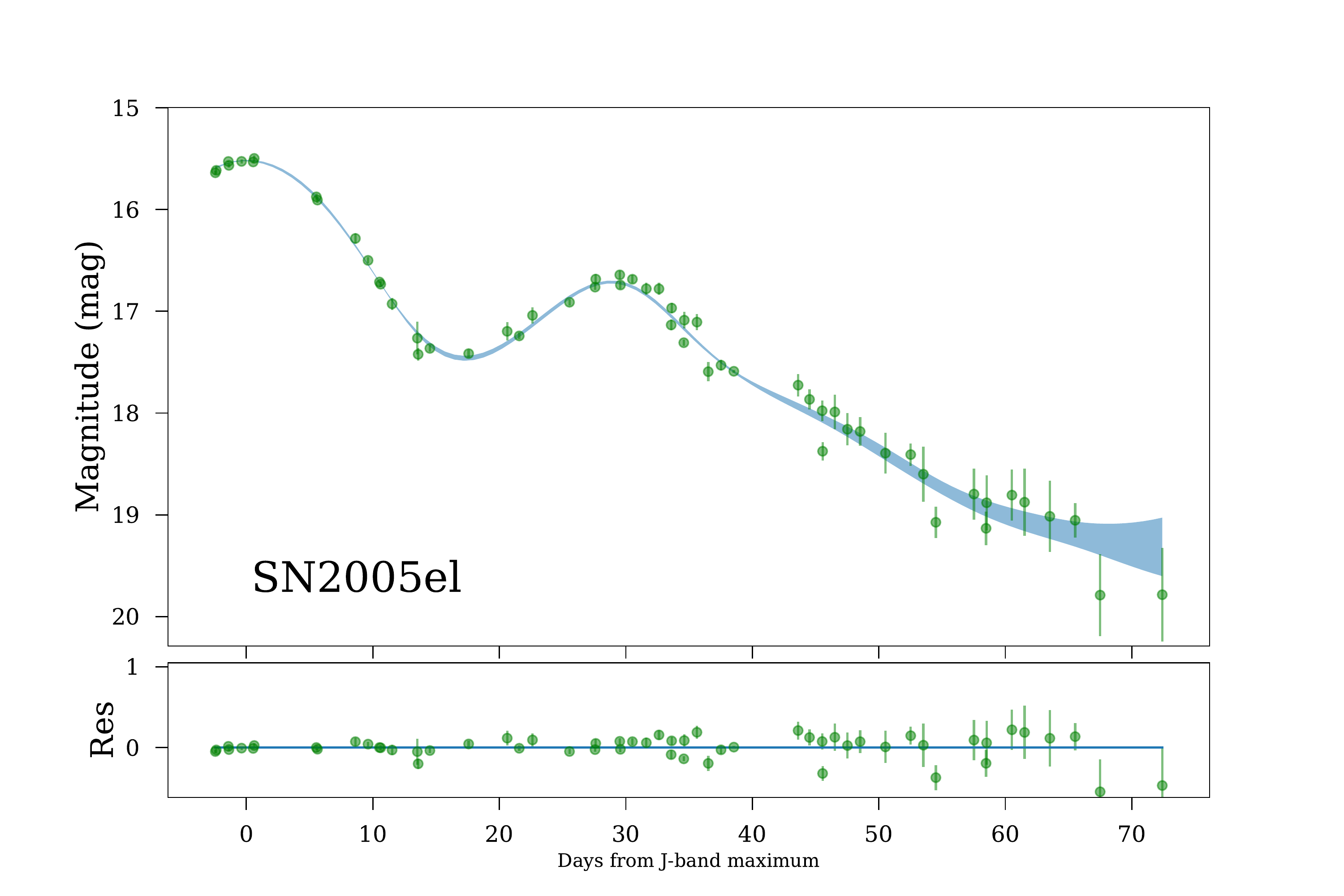}
\includegraphics[width=.2\linewidth]{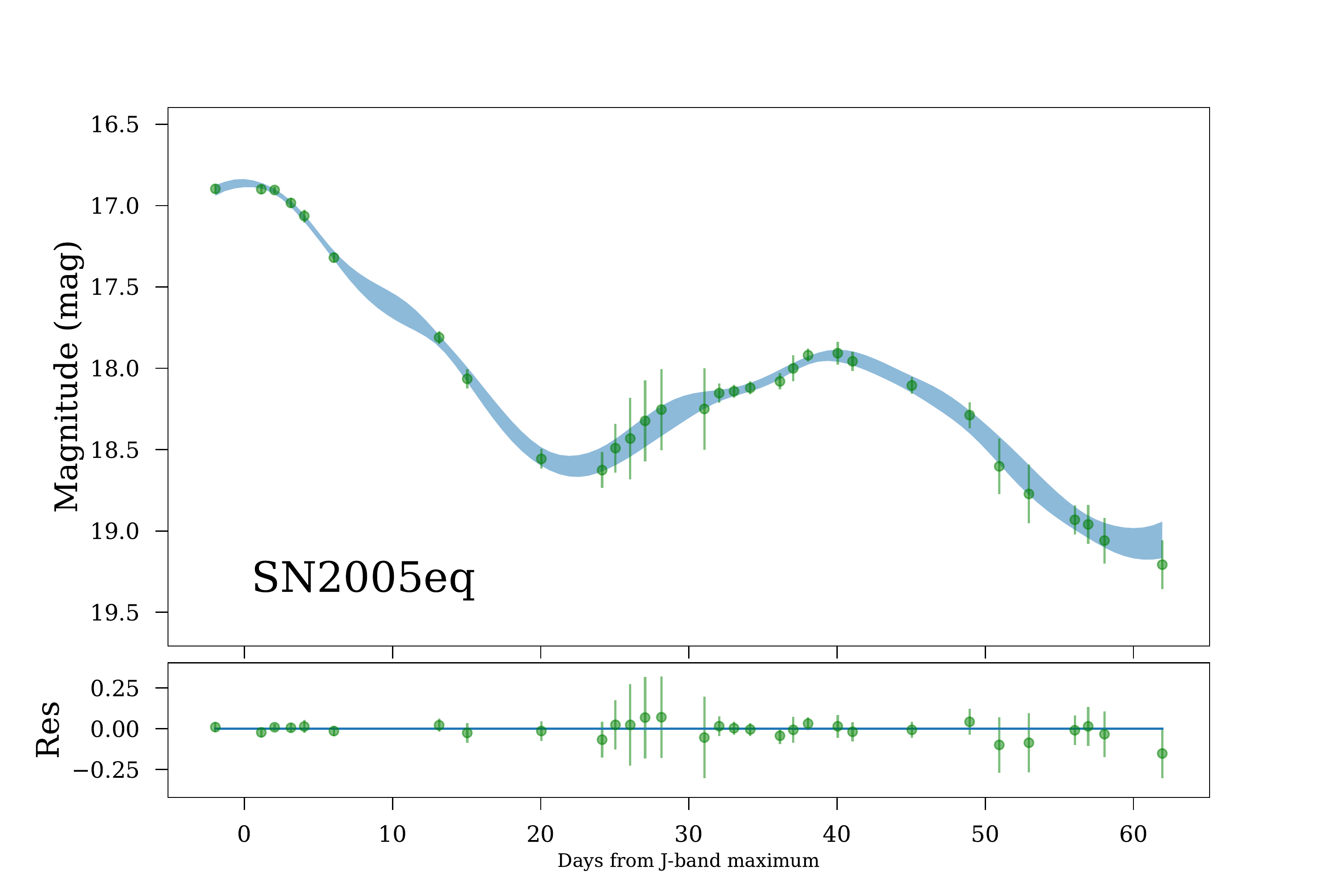}
\includegraphics[width=.2\linewidth]{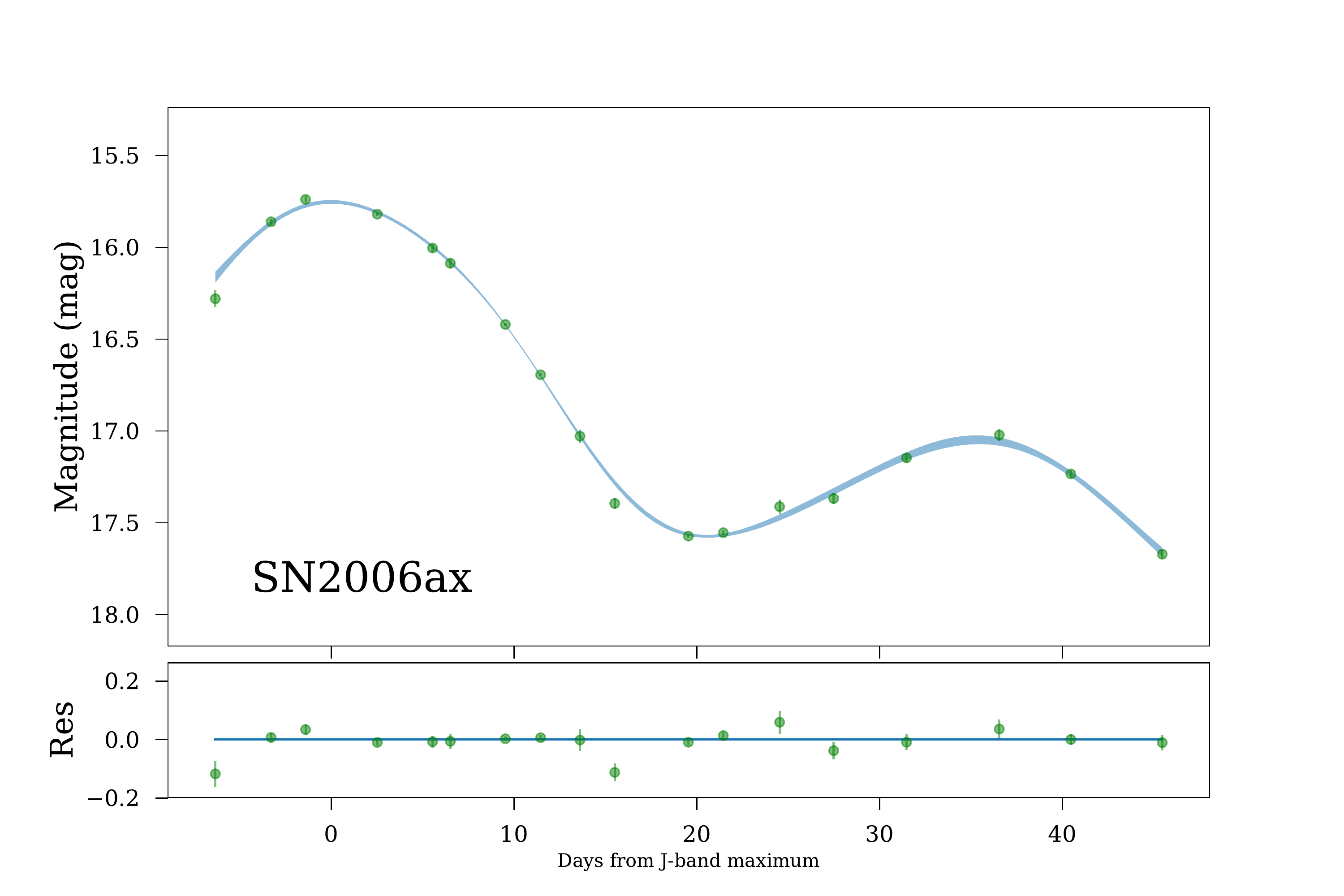}
\includegraphics[width=.2\linewidth]{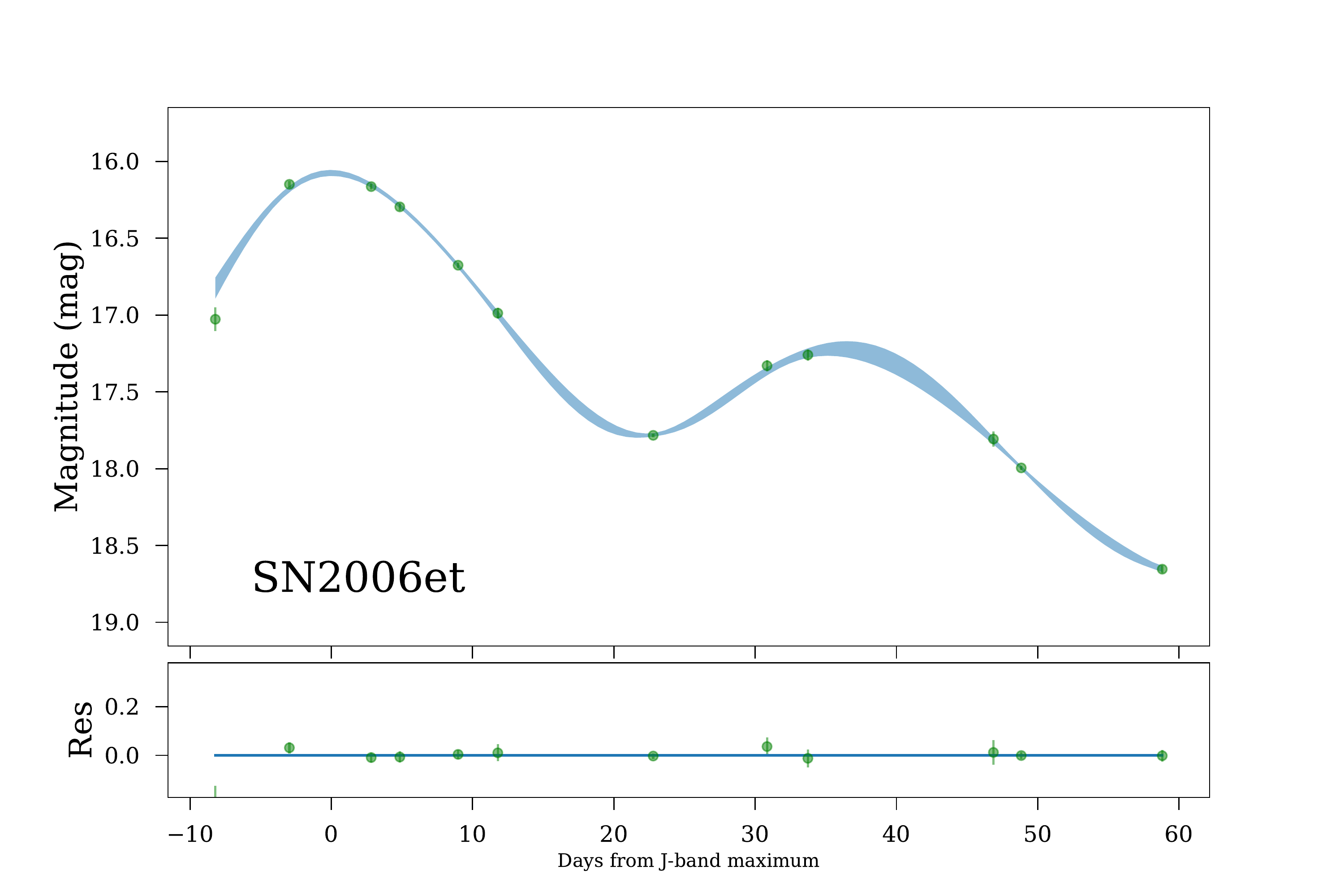}
\includegraphics[width=.2\linewidth]{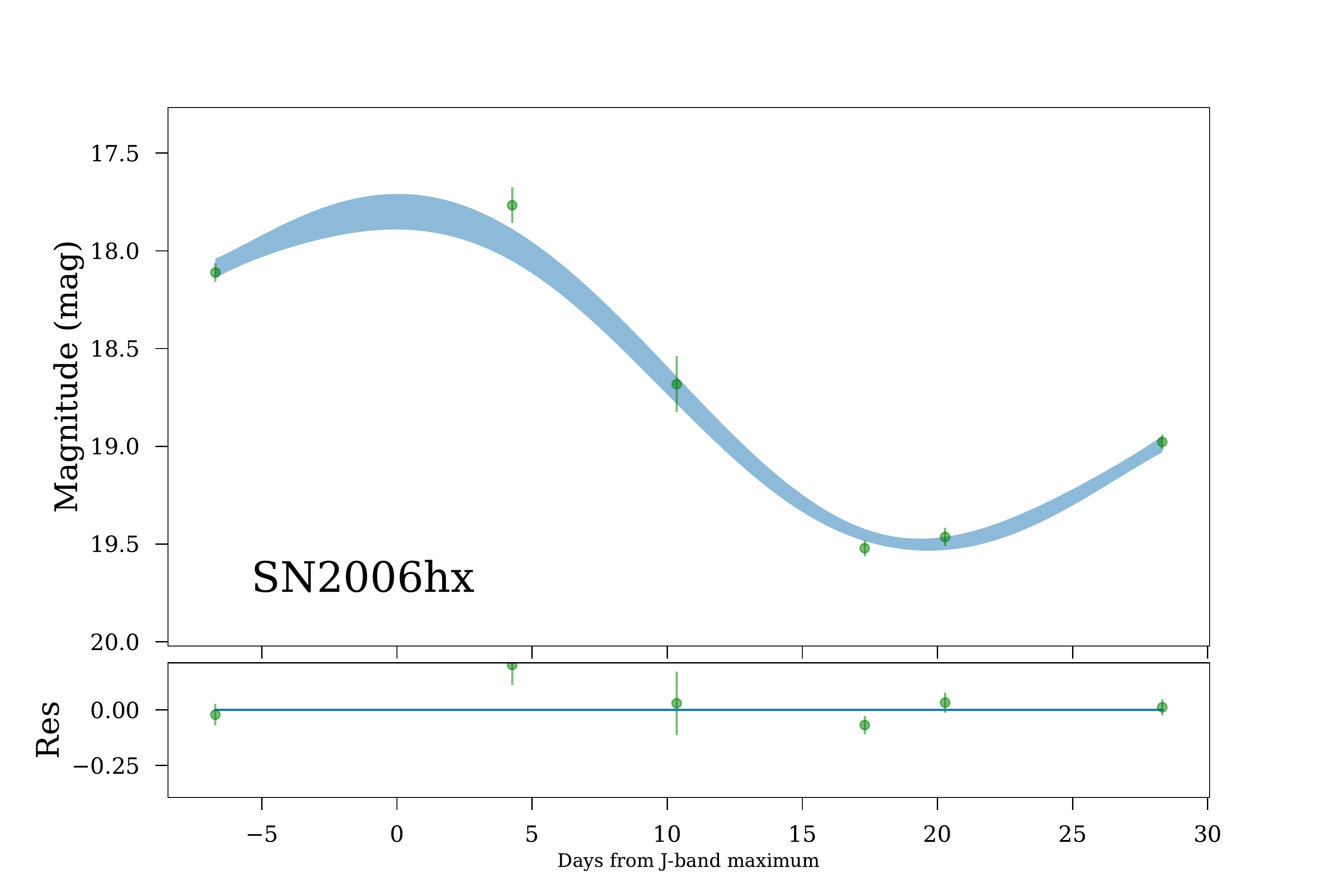}
\includegraphics[width=.2\linewidth]{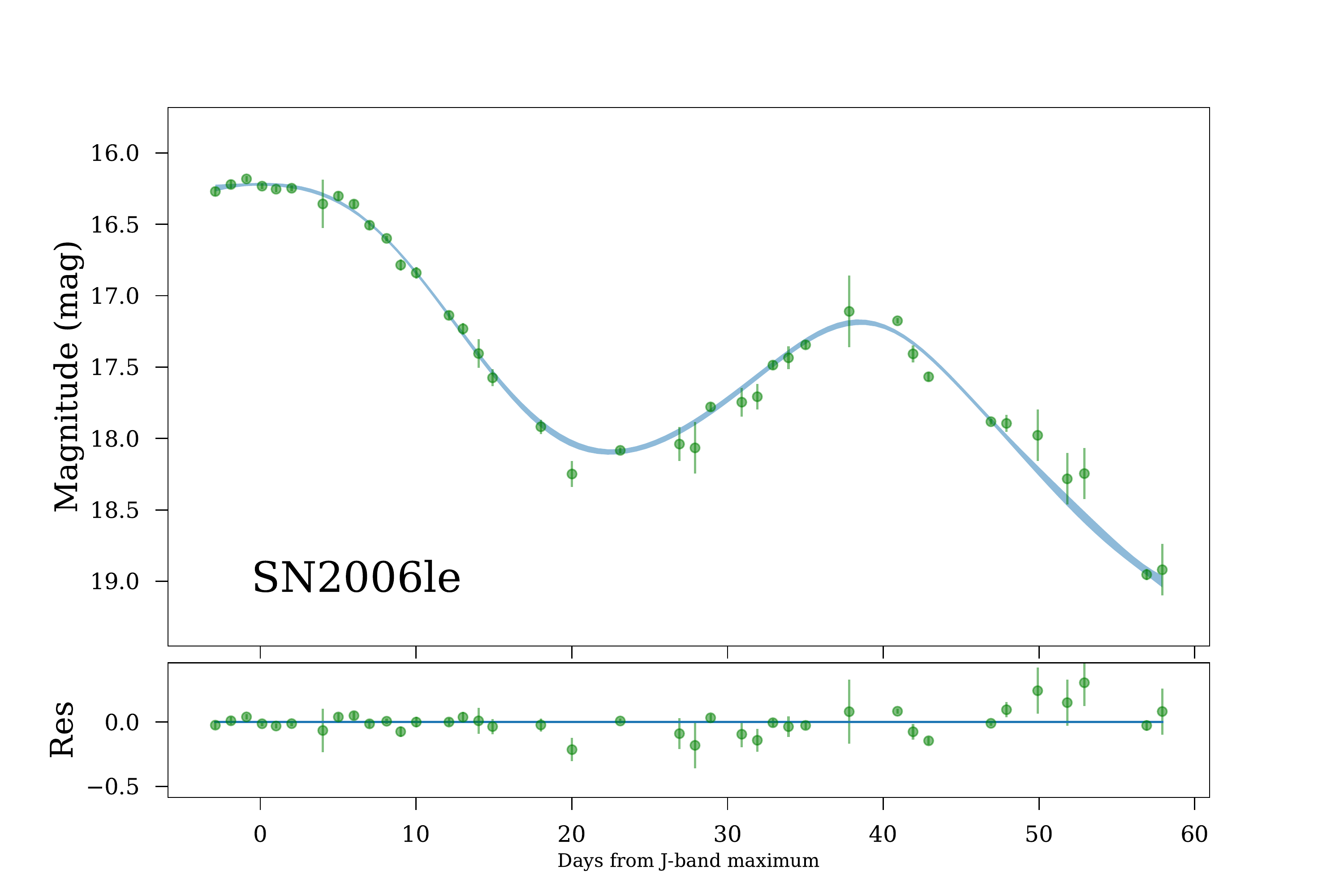}
\includegraphics[width=.2\linewidth]{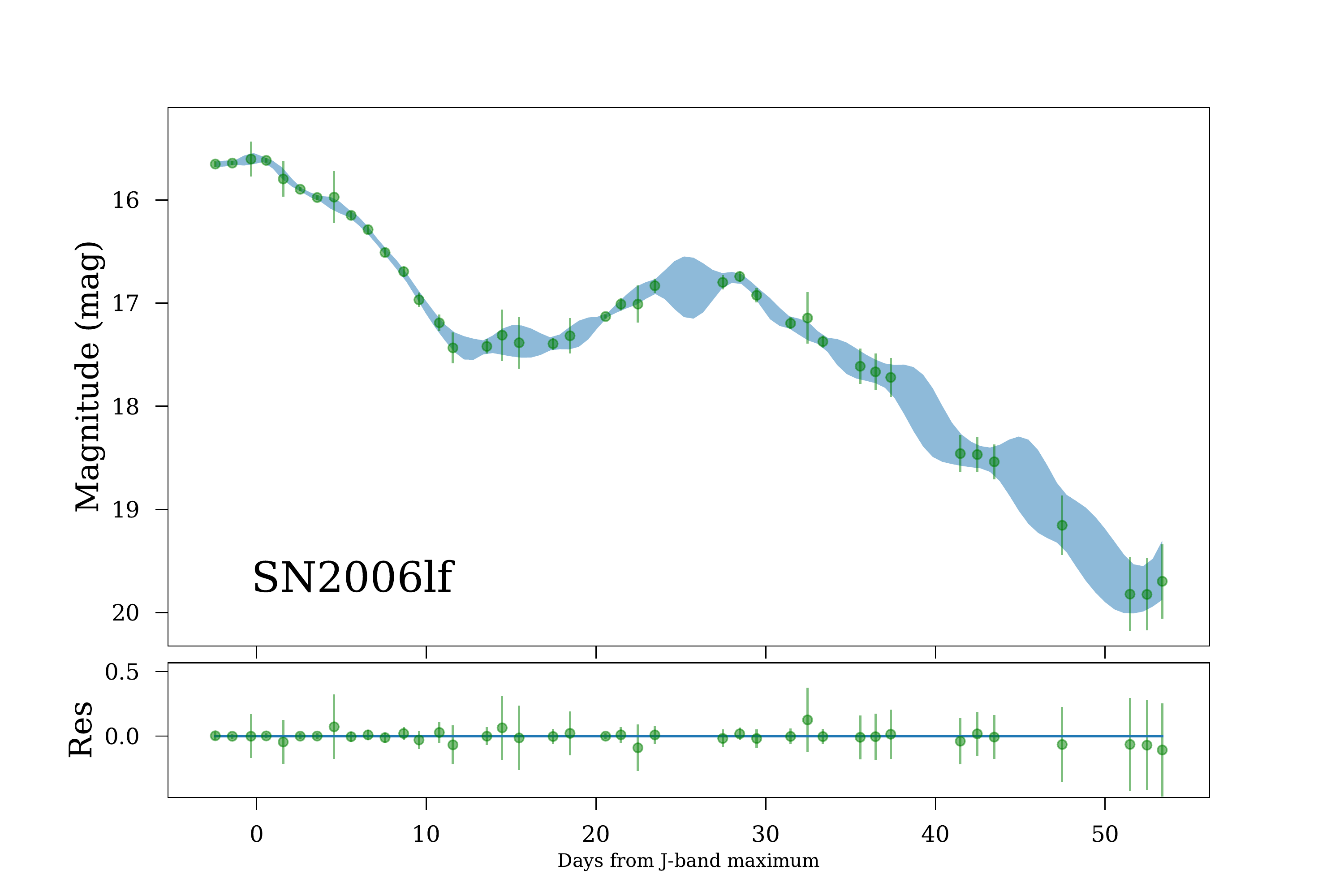}
\includegraphics[width=.2\textwidth]{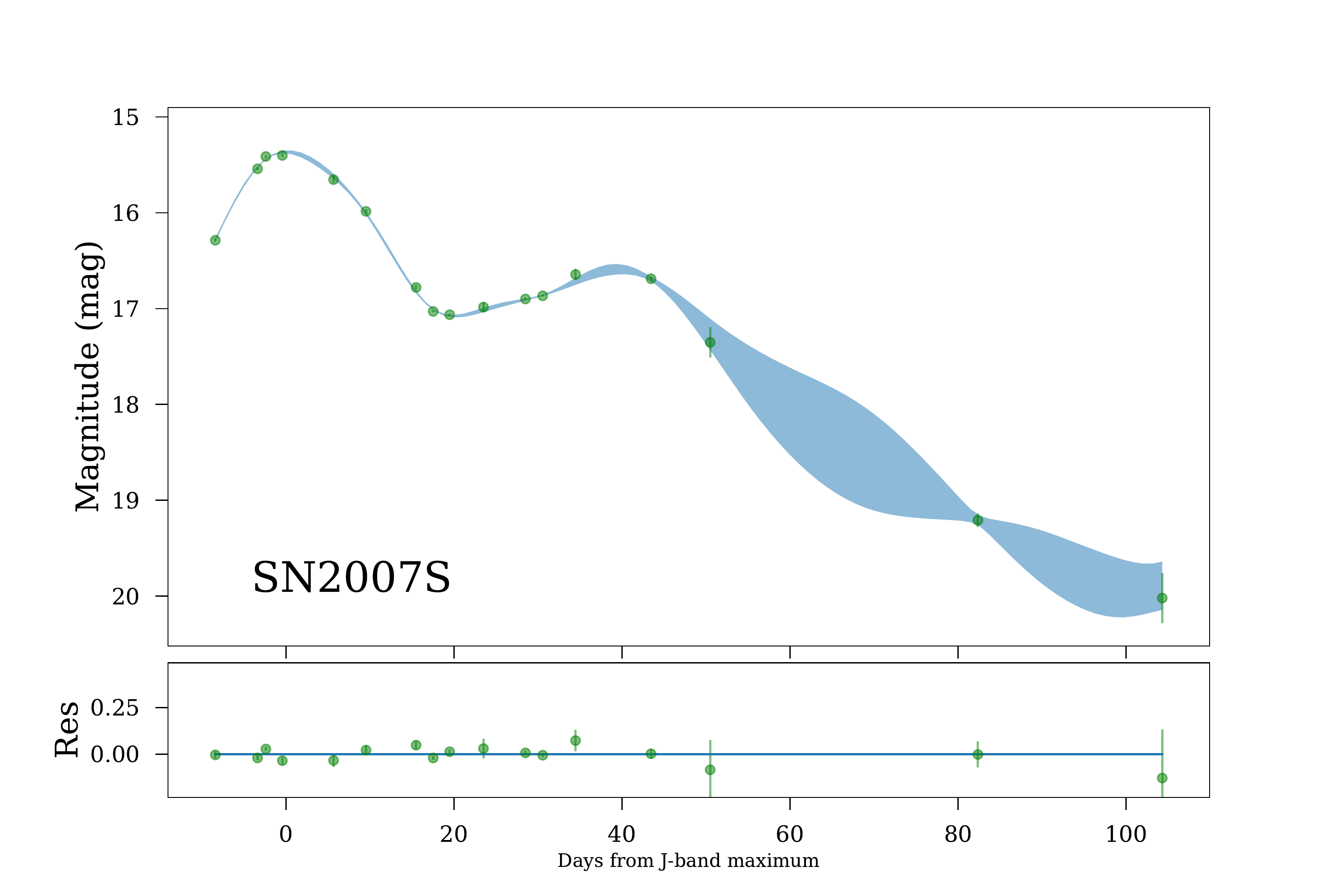}
\includegraphics[width=.2\textwidth]{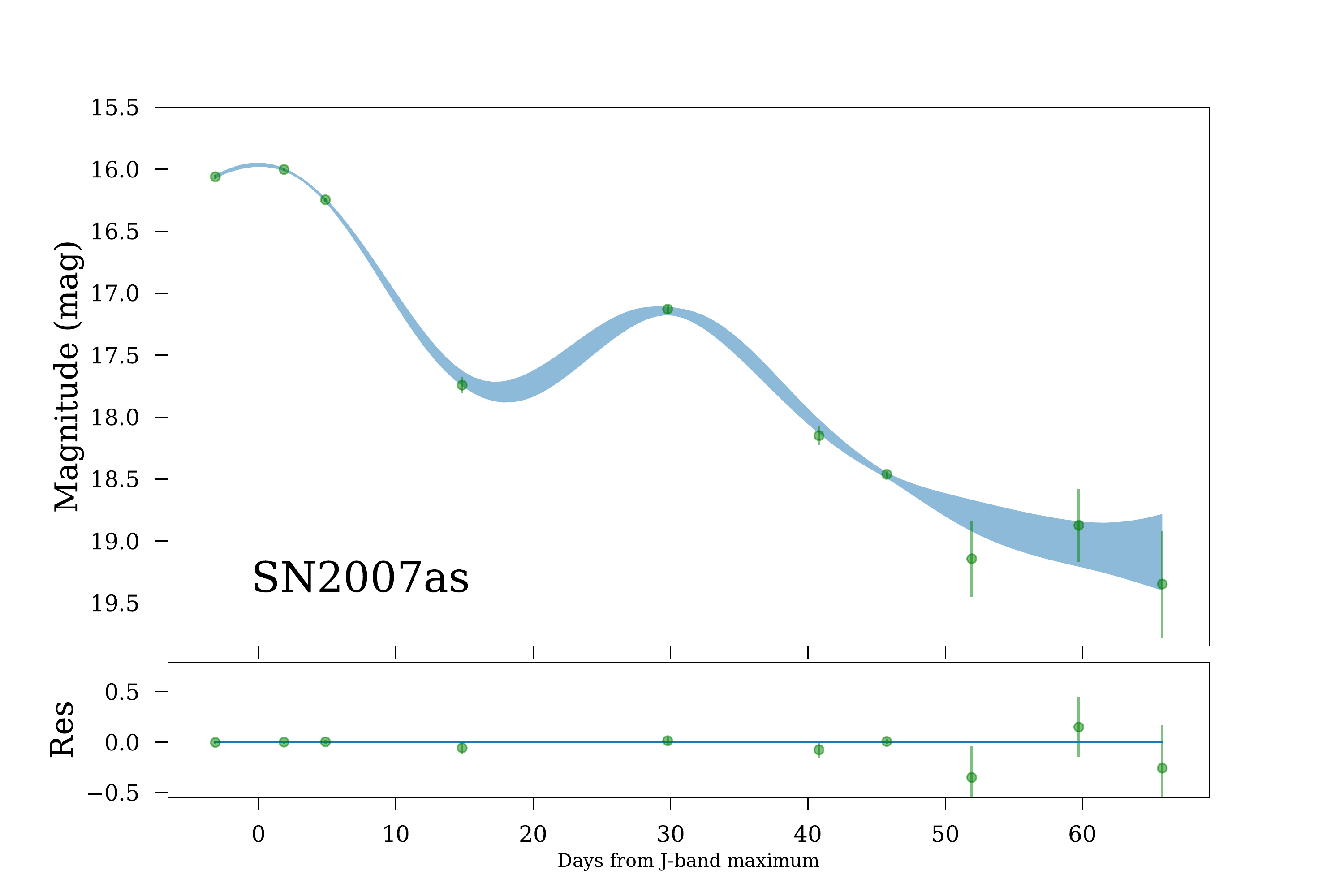}
\includegraphics[width=.2\textwidth]{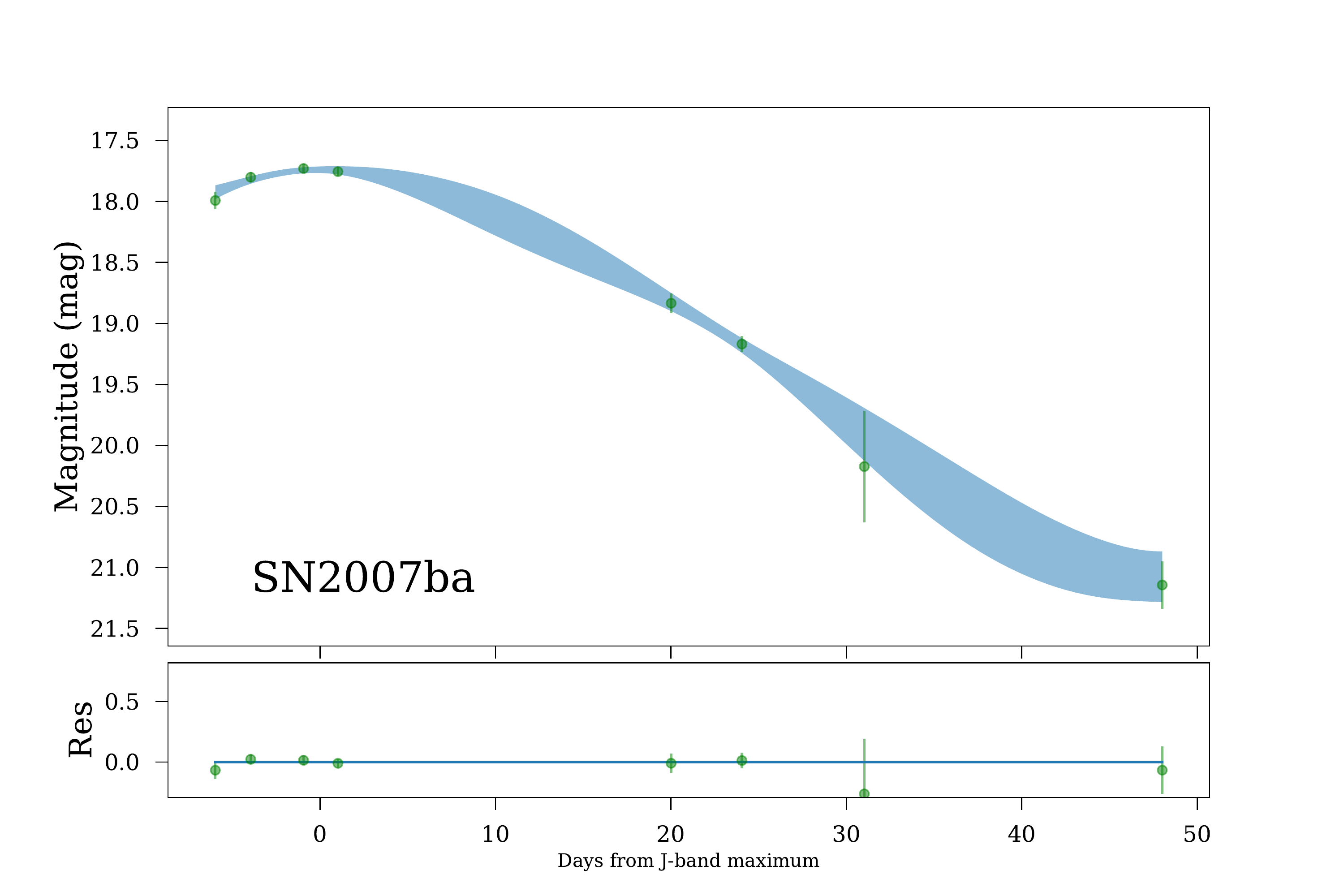}
\includegraphics[width=.2\textwidth]{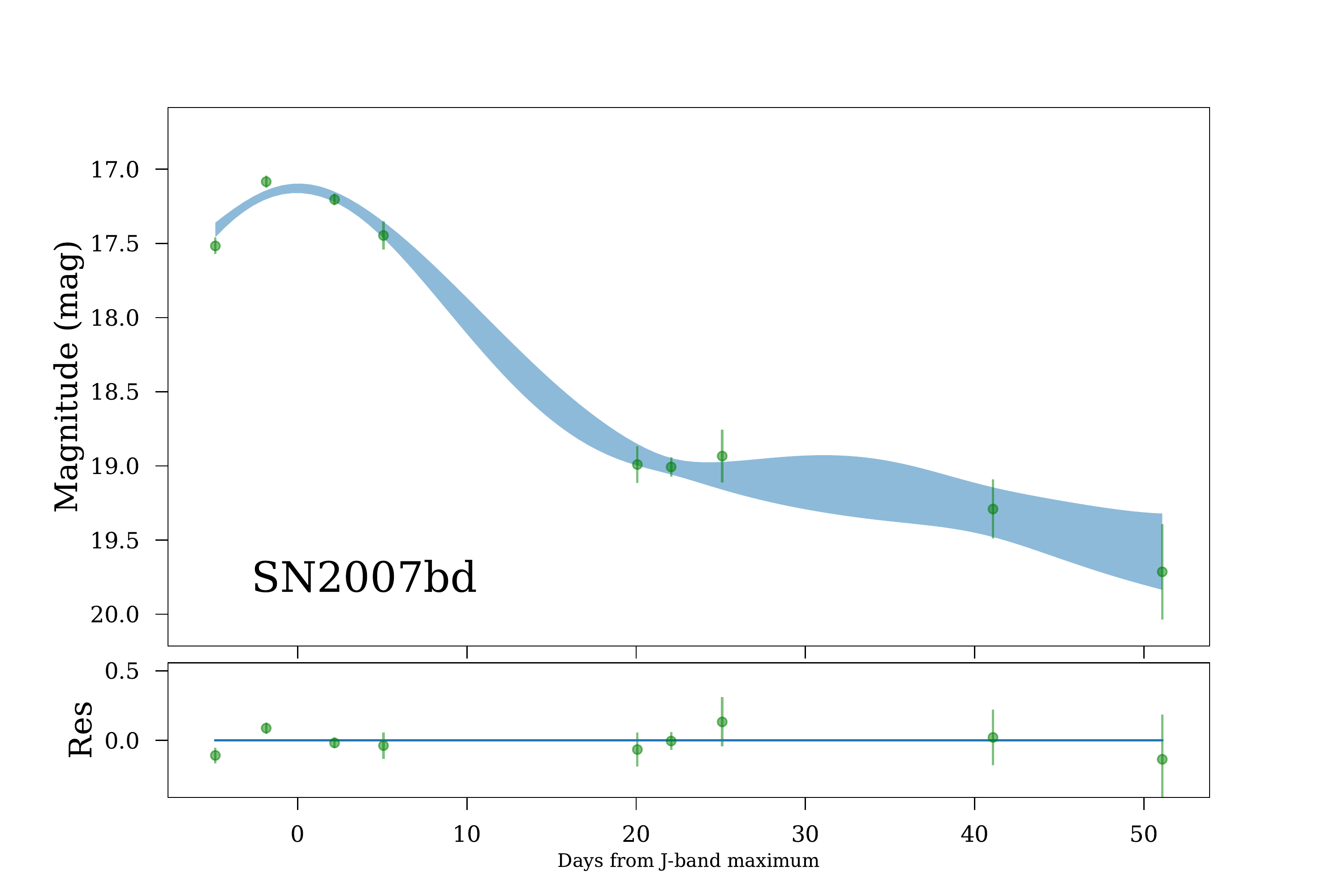}
\includegraphics[width=.2\textwidth]{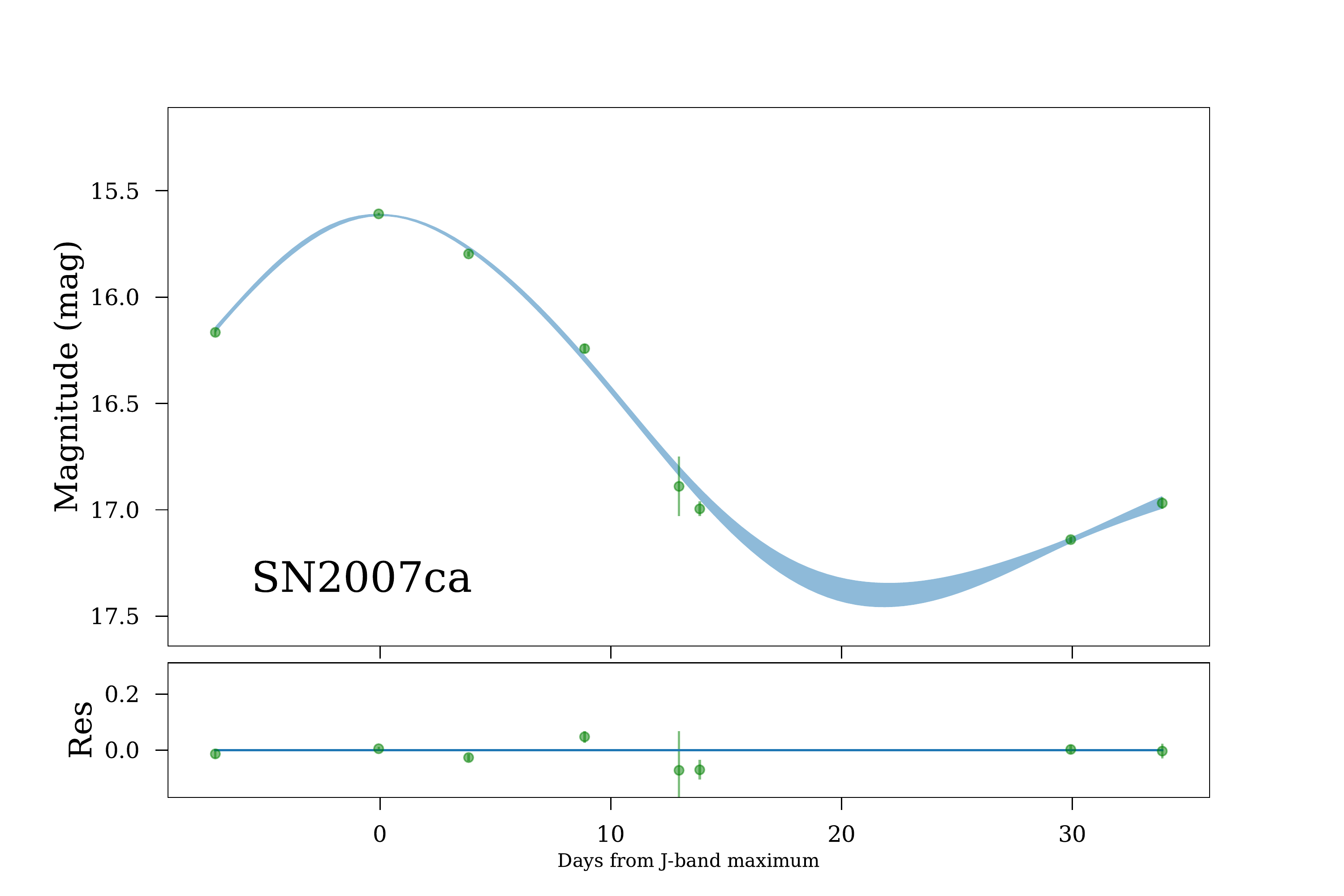}
\includegraphics[width=.2\textwidth]{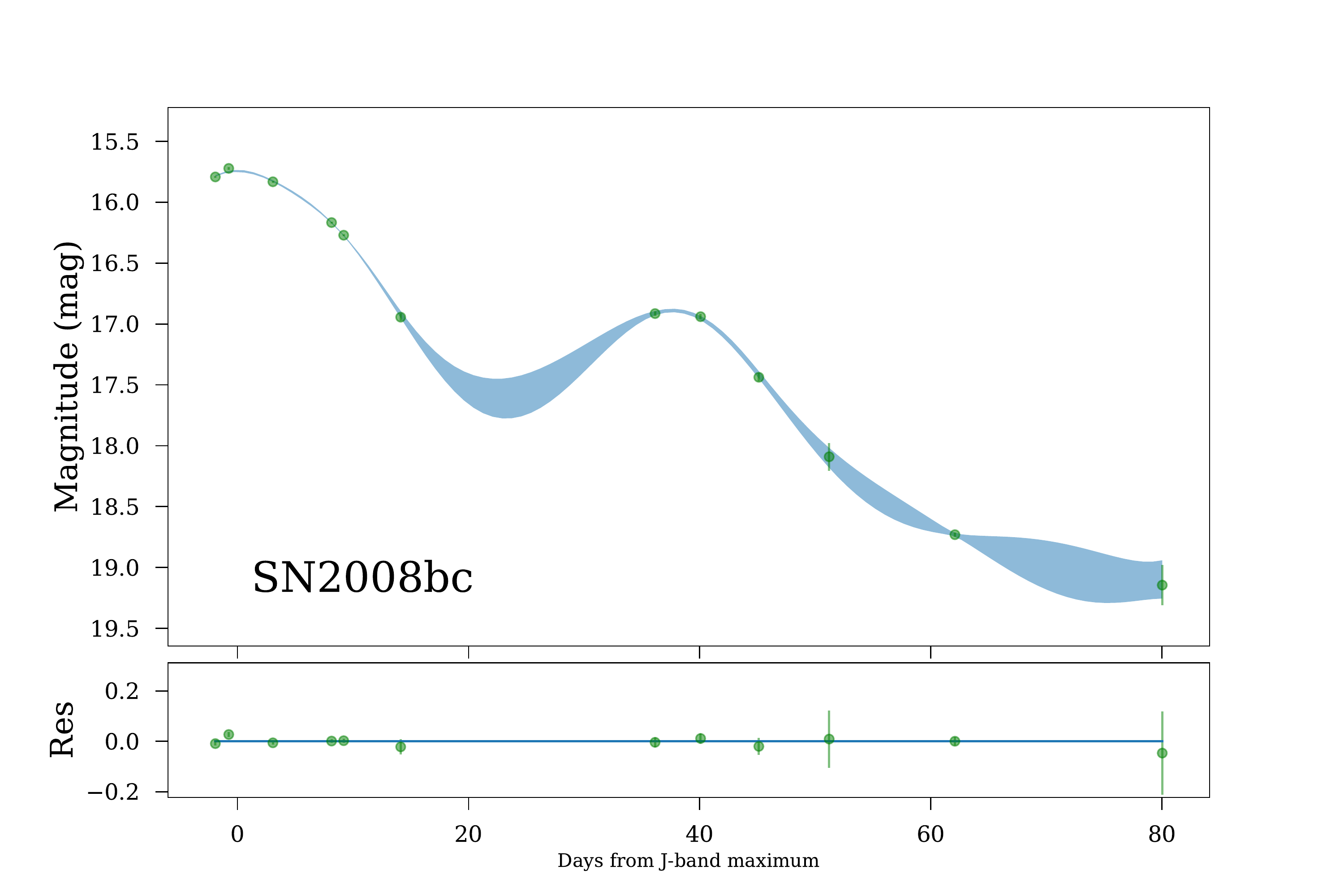}
\includegraphics[width=.2\textwidth]{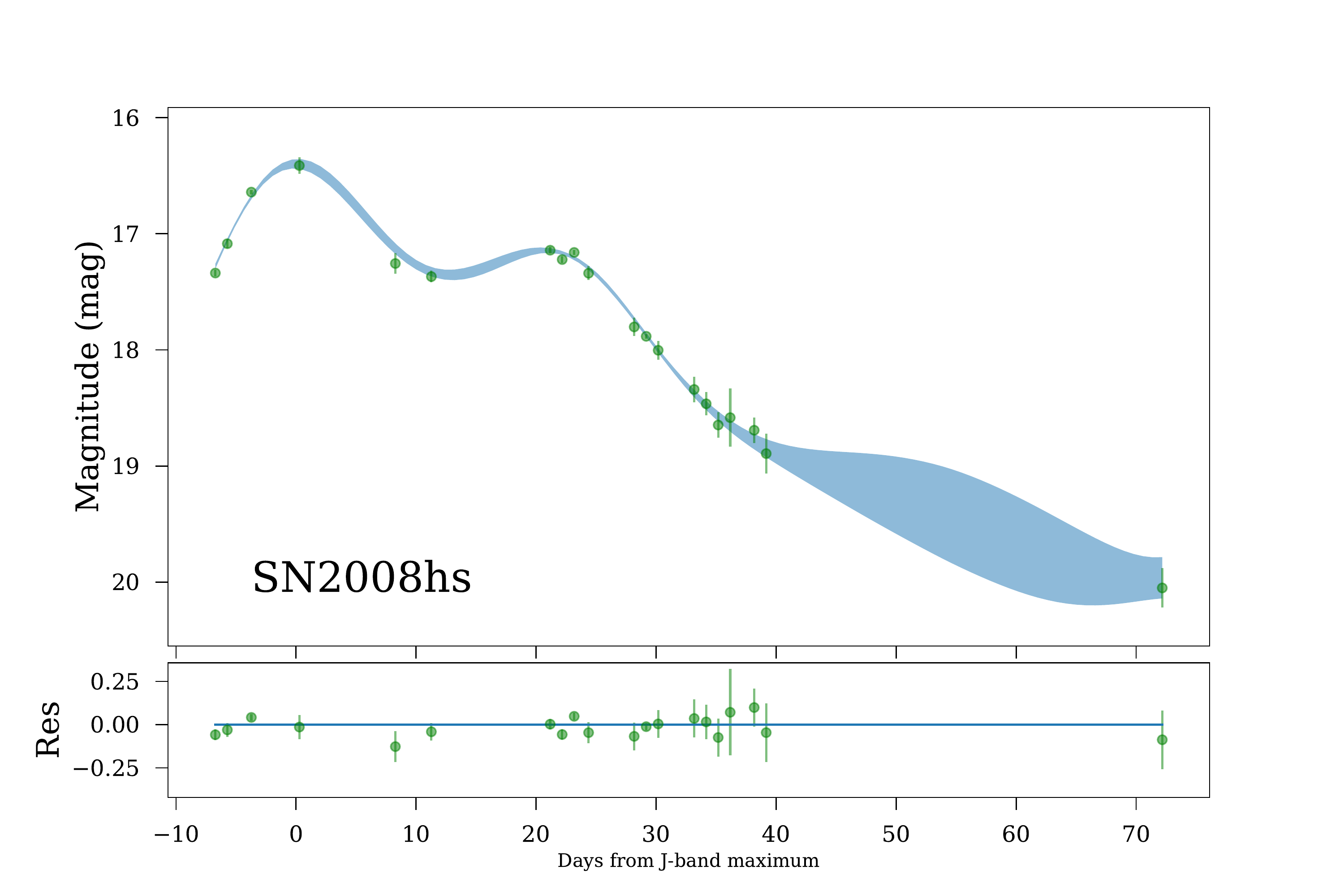}
\includegraphics[width=.2\textwidth]{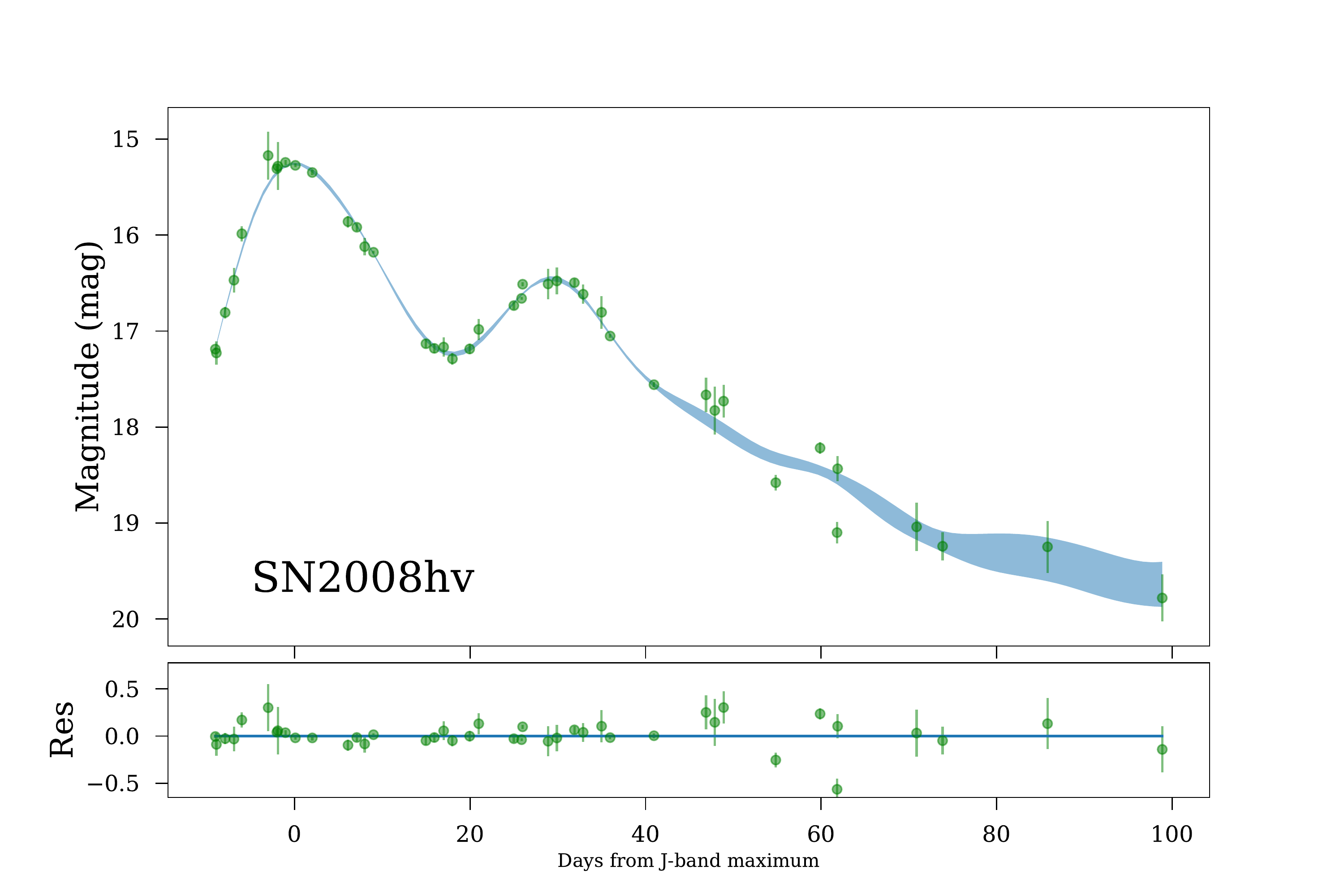}
\includegraphics[width=.2\textwidth]{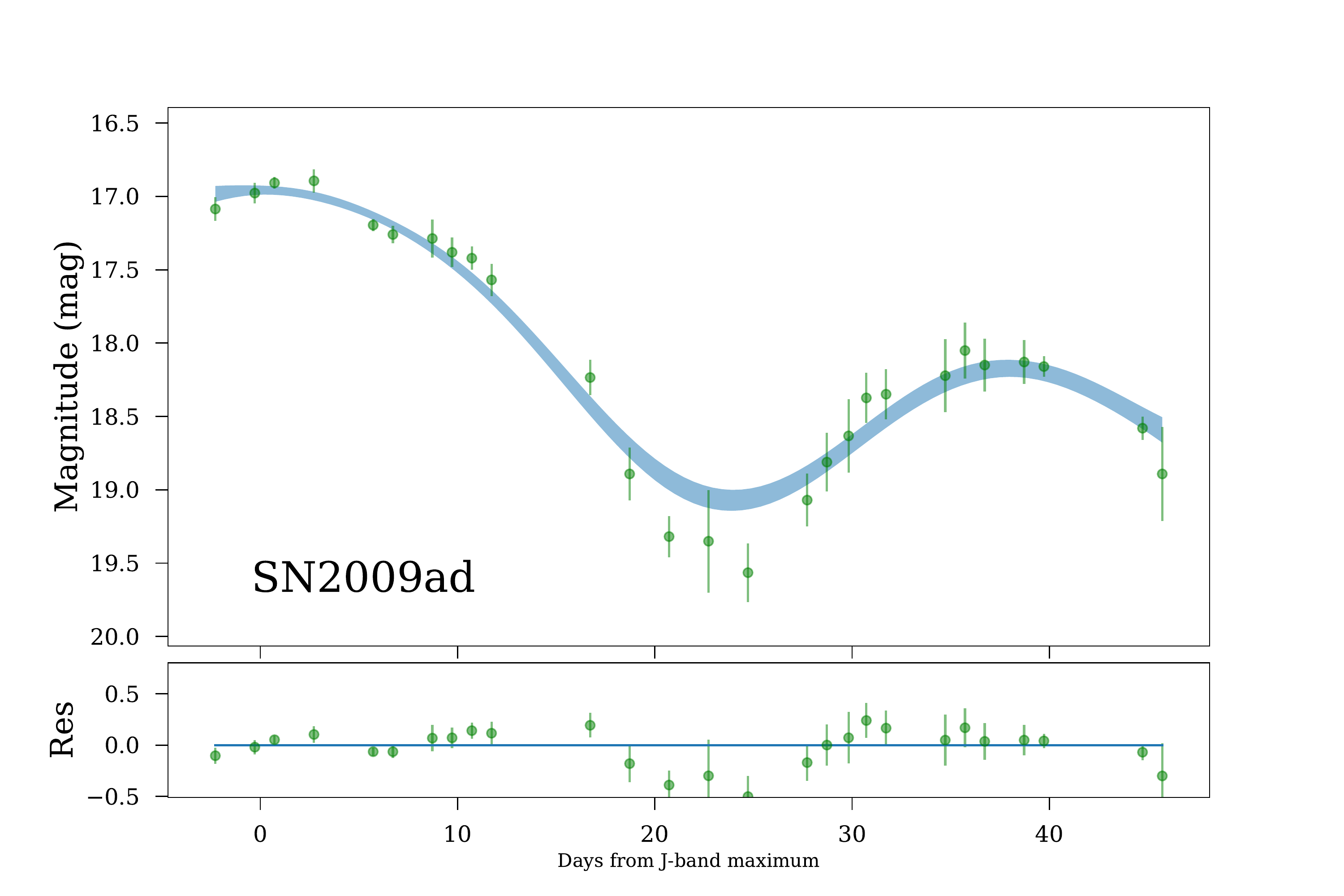}
\includegraphics[width=.2\textwidth]{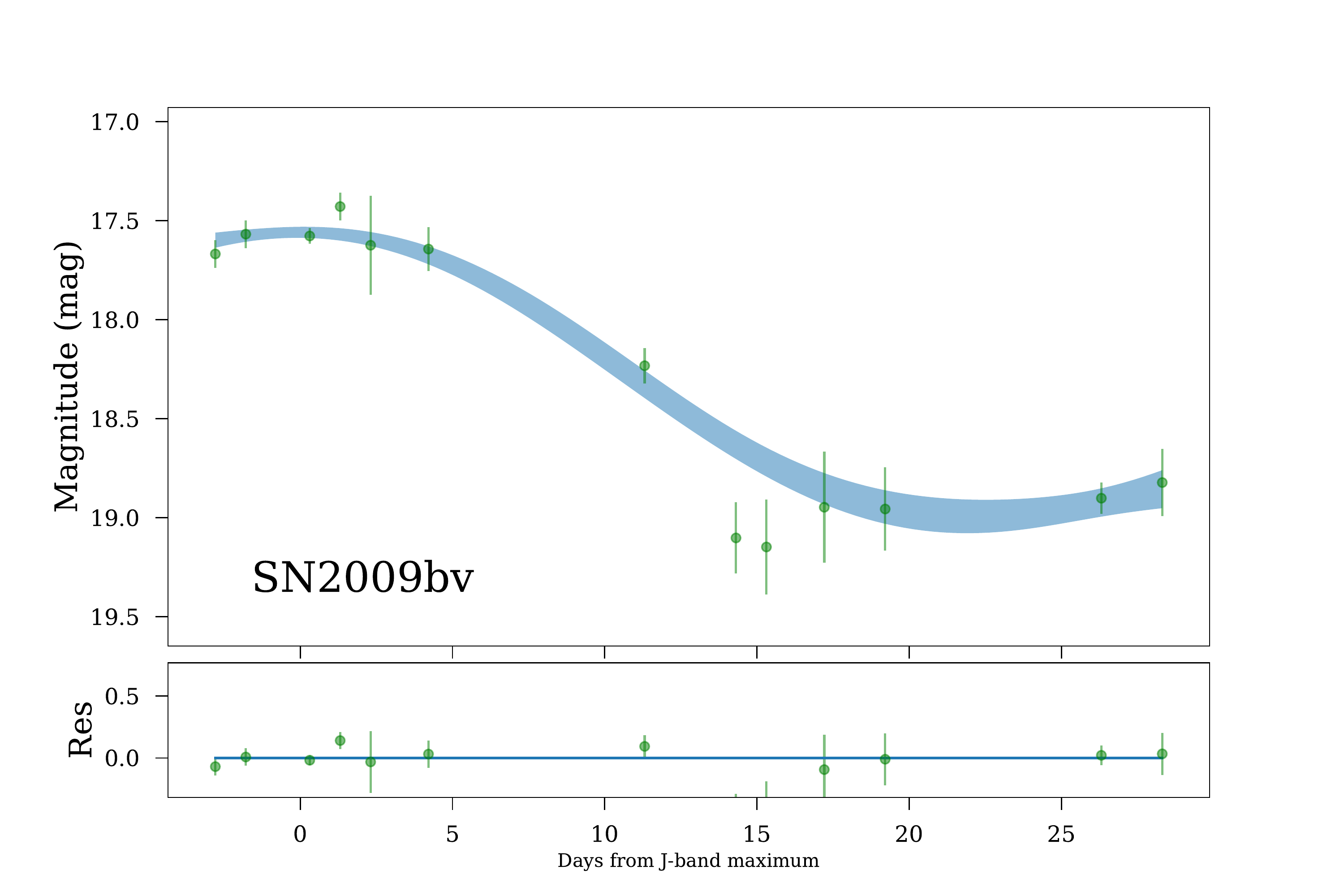}
\includegraphics[width=.2\textwidth]{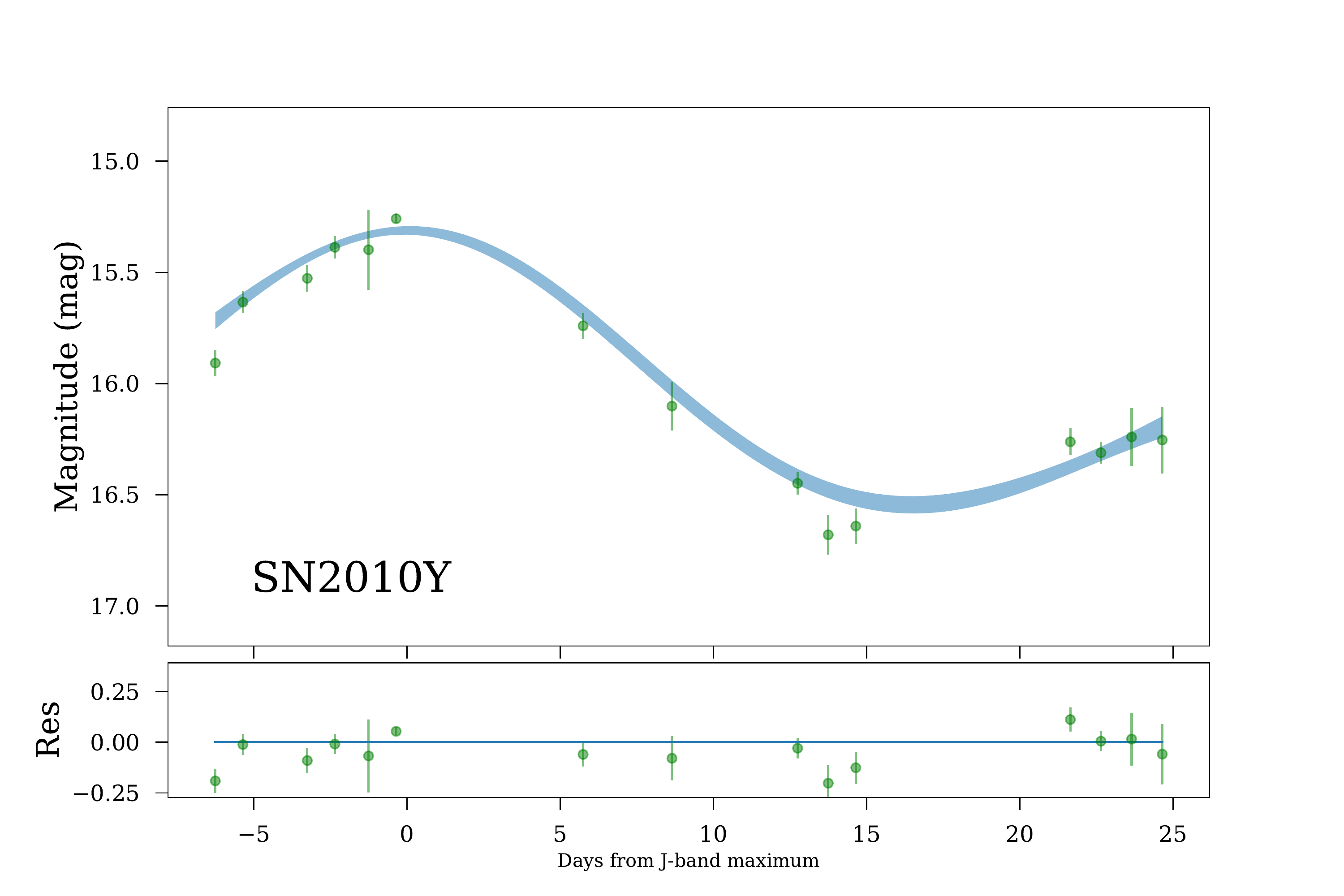}
\includegraphics[width=.2\textwidth]{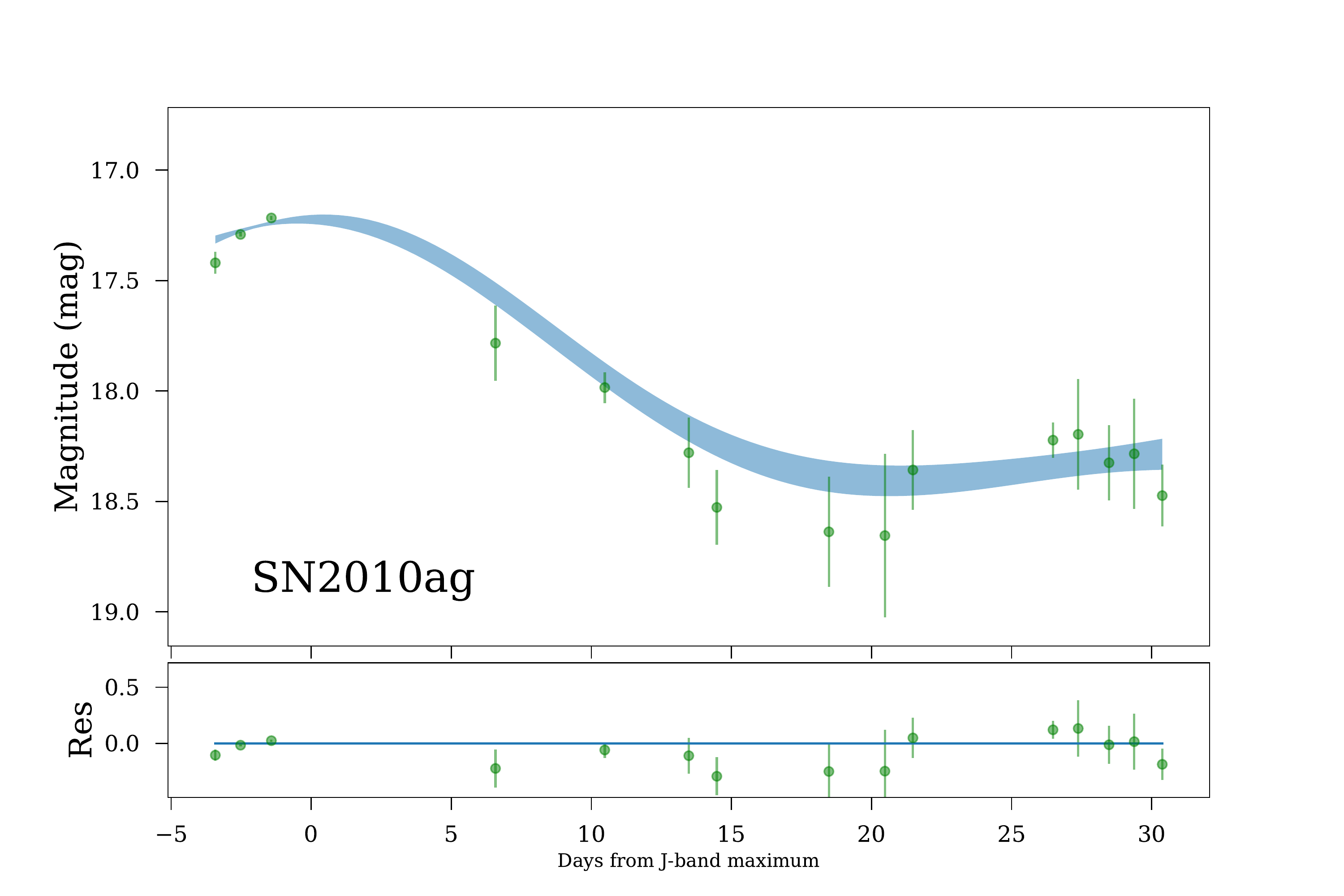}
\includegraphics[width=.2\textwidth]{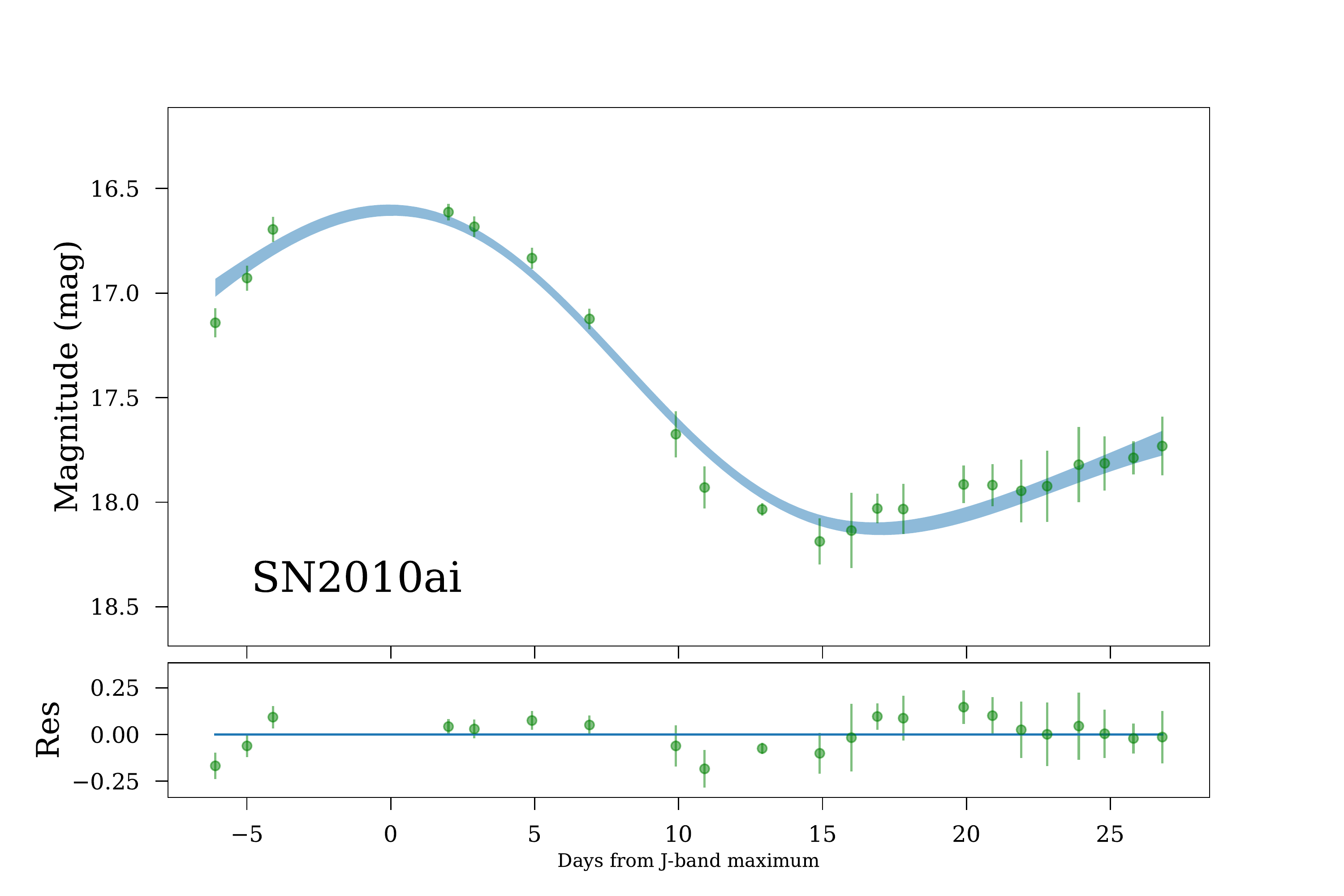}
\includegraphics[width=.2\textwidth]{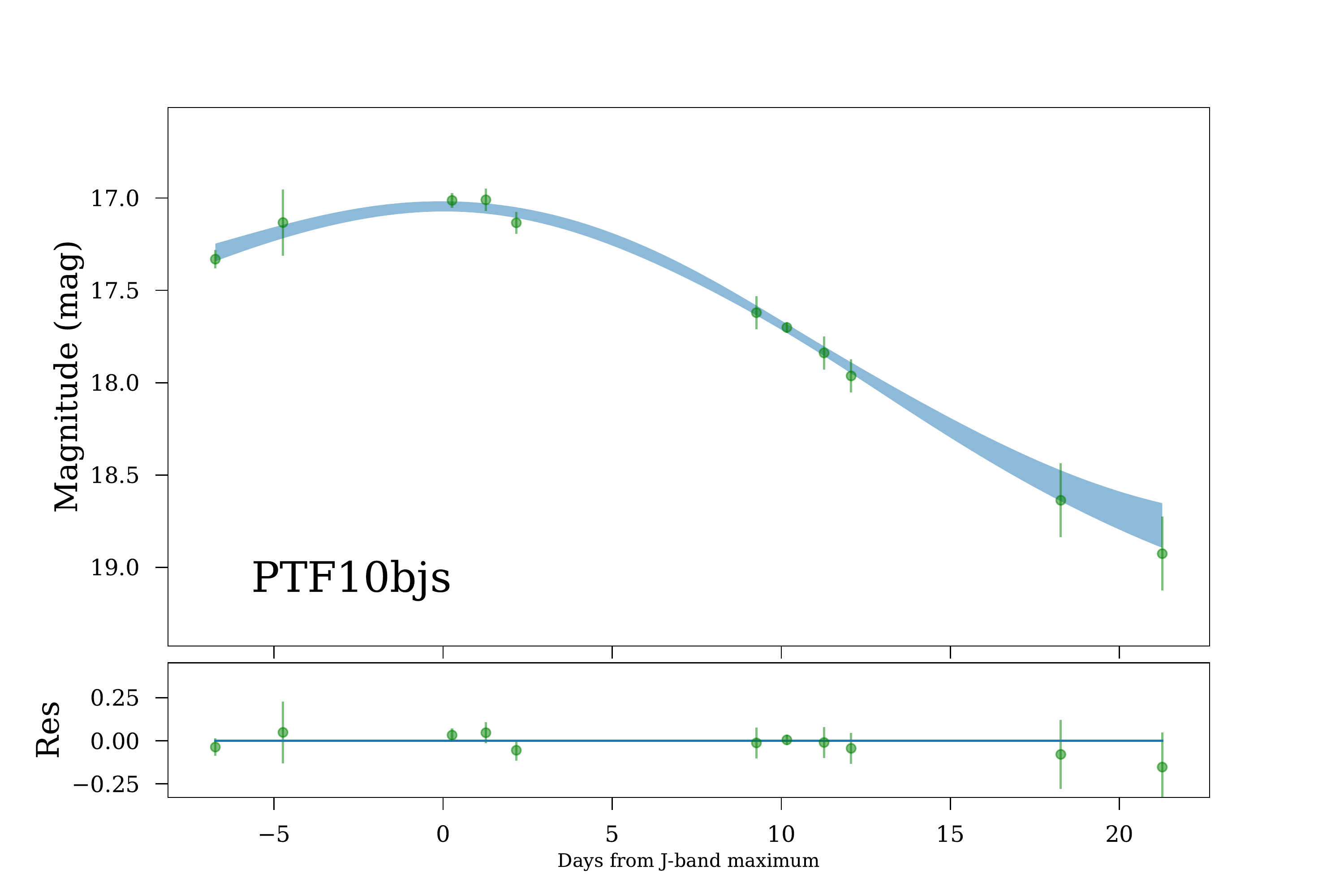}
\includegraphics[width=.2\textwidth]{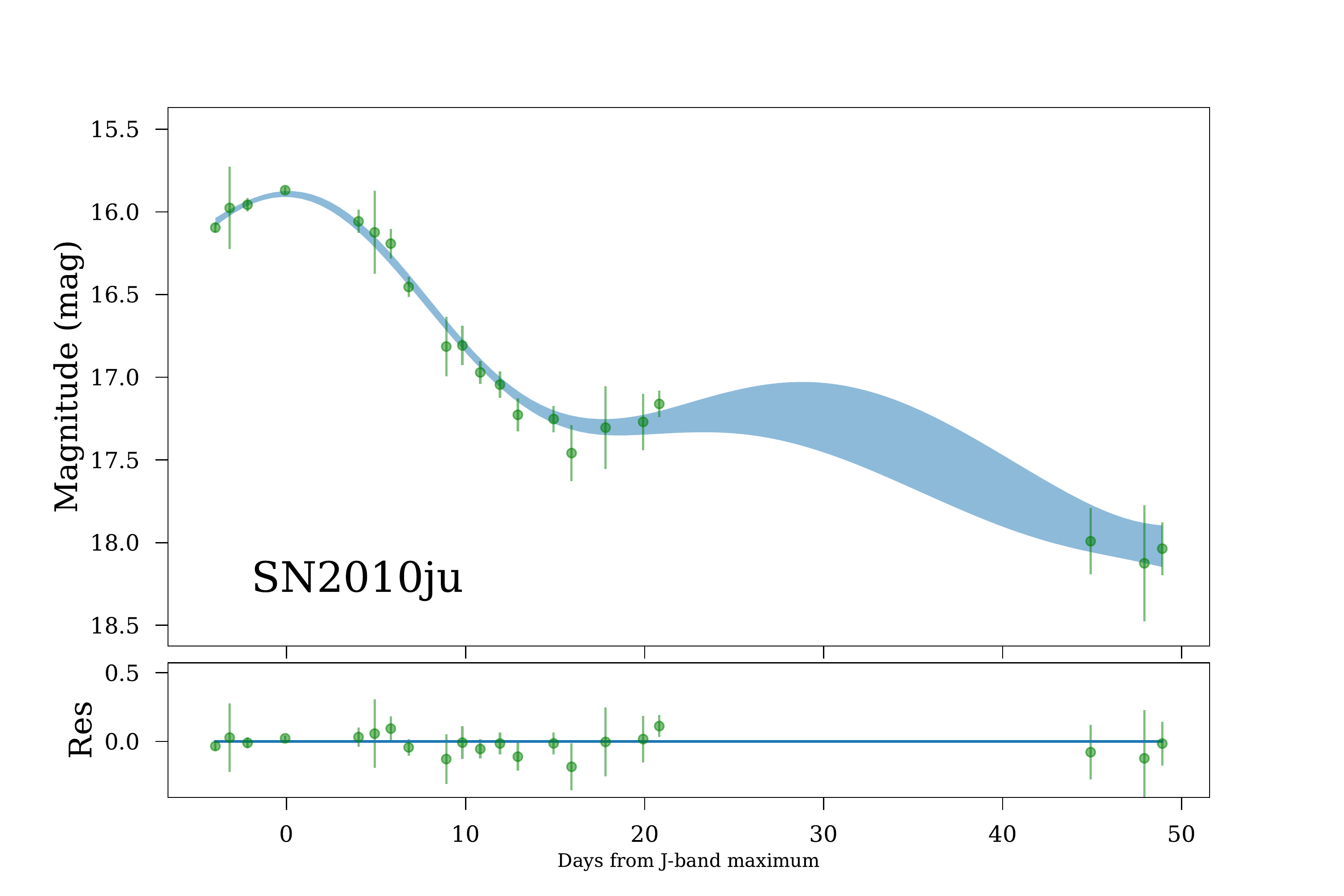}
\includegraphics[width=.2\textwidth]{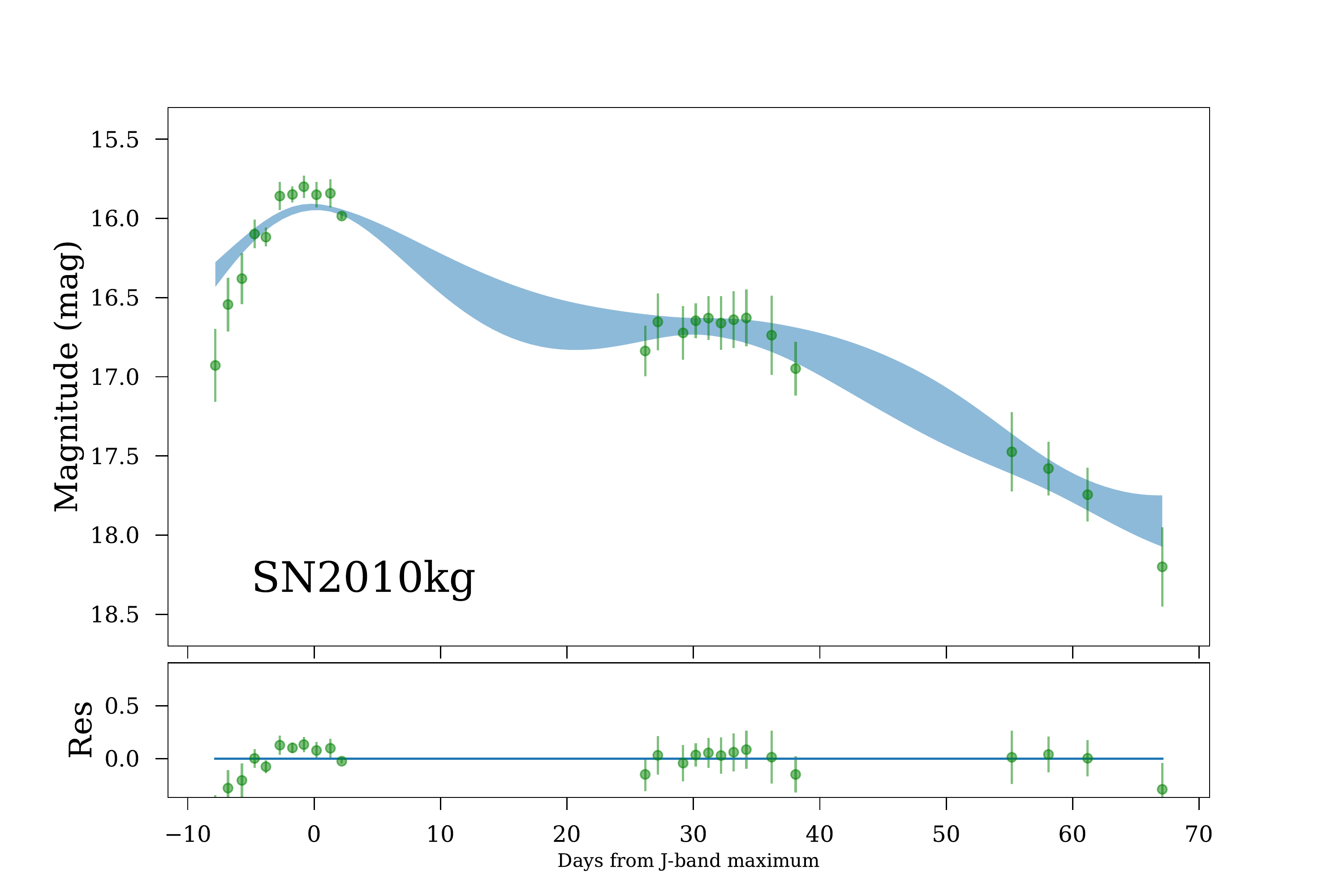}
\includegraphics[width=.2\textwidth]{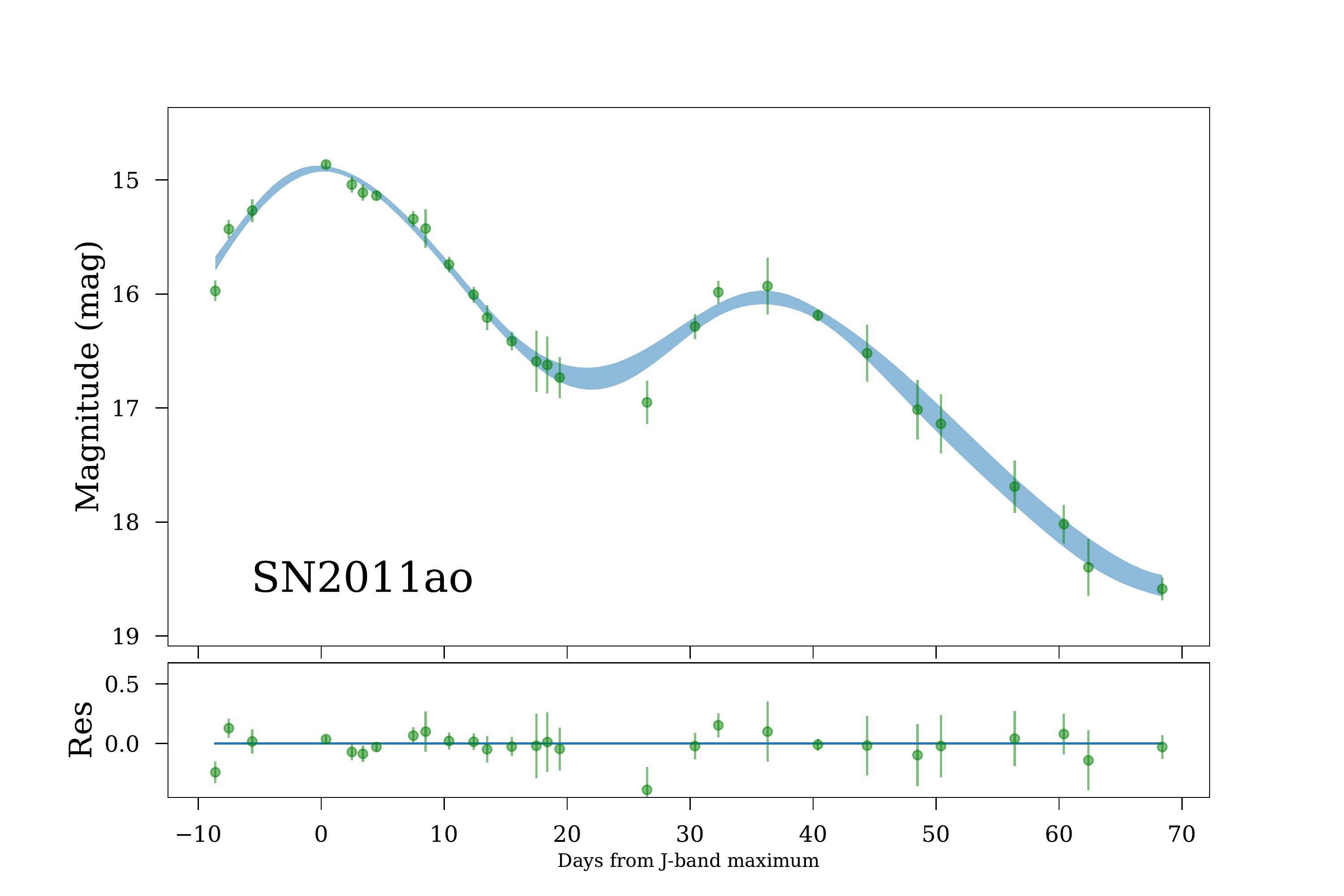}
\includegraphics[width=.2\textwidth]{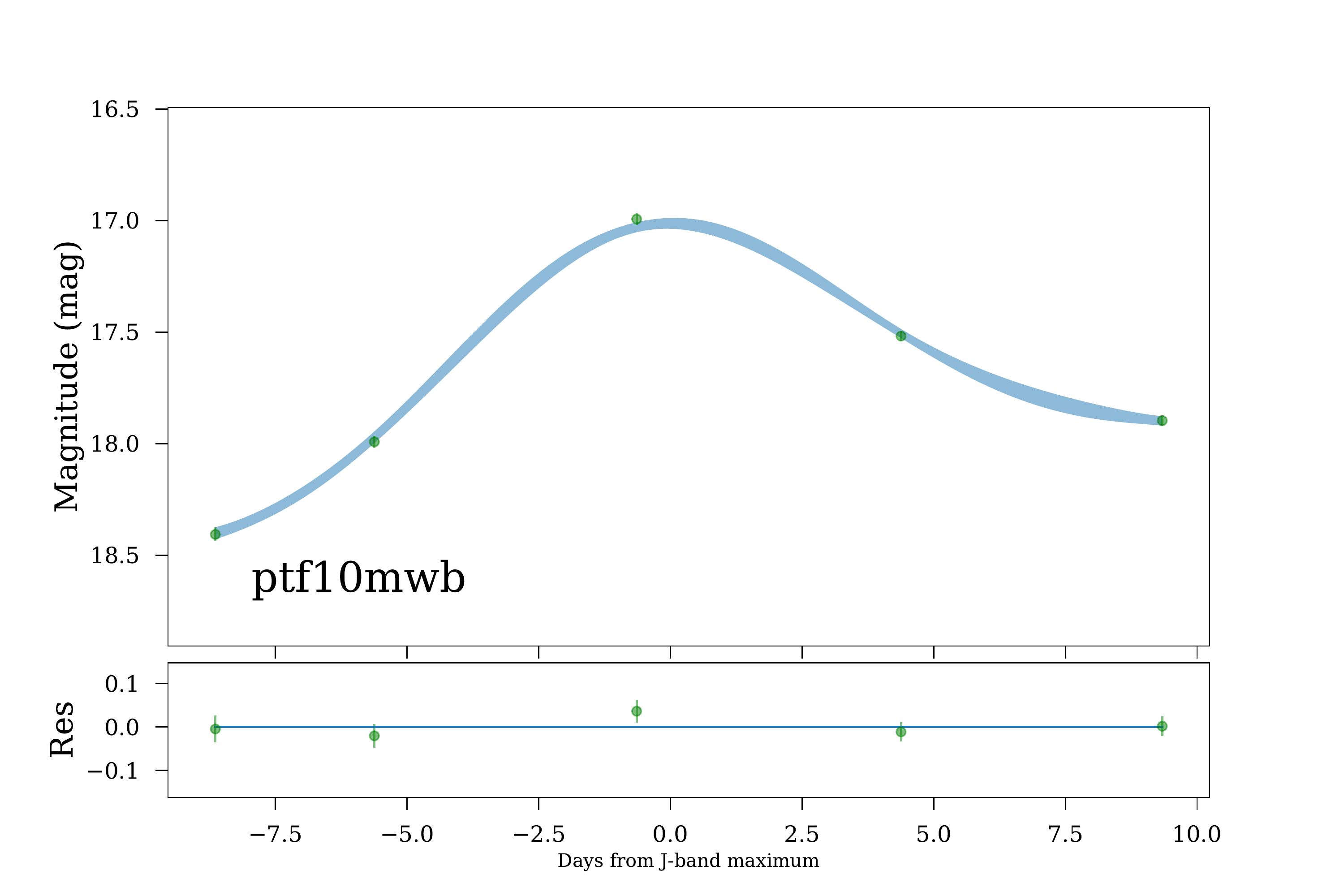}
\includegraphics[width=.2\textwidth]{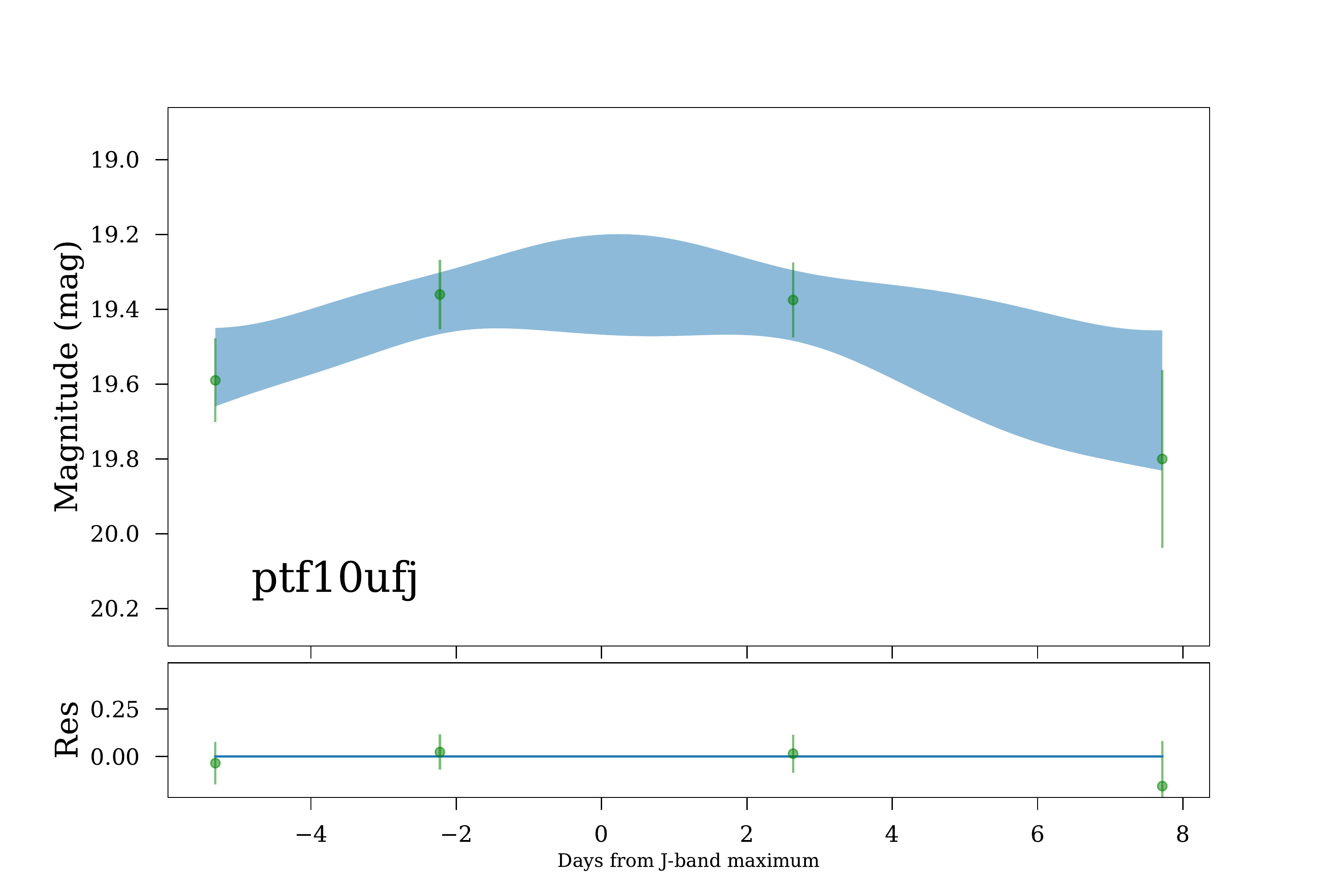}
\caption{Same as Figure~\ref{fig:gpfits_calib}, but for the Hubble flow sample.}
\label{fig:gpfits_hflow}
\end{figure}

In section~\ref{sec:analysis}, we described the light curve fitting
methodology for the SNe in our sample. In Figure~\ref{fig:gpfits_calib} we
plot the light curves of the 9 SNe in our calibration sample along with the
Gaussian process fits to derive the peak magnitude. The same is plotted for
the Hubble-flow sample in Figure~\ref{fig:gpfits_hflow}.

\end{document}

%% file: table_calibrators.tex
\begin{tabular}{|c|c|cc|cc|cc|c|c|c|}
\hline
Supernova & Host Galaxy & $m_J$ & $\sigma_{\rm fit}$ & 
 $\mu_{\rm Ceph}$ & $\sigma_{\rm Ceph}$  & 
 $M_J$ & $\sigma_M$  & MW $A_J$ & $K_J$ & SN $J$-band   \\
 & & (mag) & (mag) & (mag) & (mag) & (mag) & (mag) & (mag) & (mag) & Photometry Reference \\
\hline
SN 2001el & NGC 1448  & 12.837 & 0.022 & 31.311 & 0.045 & $-18.474$ & 0.050 & 0.010 & $-0.011$ & \citet{Krisciunas2004} \\
SN 2002fk & NGC 1309  & 13.749 & 0.010 & 32.523 & 0.055 & $-18.774$ & 0.056 & 0.028 & $-0.020$ & \citet{Cartier2014} \\
SN 2003du & UGC 9391  & 14.325 & 0.056 & 32.919 & 0.063 & $-18.594$ & 0.084 & 0.007 & $-0.015$ & \citet{Stanishev2007} \\
SN 2005cf & NGC 5917  & 13.791 & 0.025 & 32.263 & 0.102 & $-18.472$ & 0.105 & 0.068 & $-0.019$ & \citet{Wang2009} \\
SN 2007af & NGC 5584  & 13.446 & 0.003 & 31.786 & 0.046 & $-18.340$ & 0.046 & 0.027 & $-0.017$ & \citet{Contreras2010}; CSP \\
SN 2011by & NGC 3972  & 13.218 & 0.040 & 31.587 & 0.070 & $-18.369$ & 0.081 & 0.010 & $-0.011$ & \citet{Friedman2015}; CfA \\
SN 2011fe & M101      & 10.464 & 0.009 & 29.135 & 0.045 & $-18.671$ & 0.046 & 0.006 & $-0.002$ & \citet{Matheson2012} \\
SN 2012cg & NGC 4424  & 12.285 & 0.017 & 31.080 & 0.292 & $-18.795$ & 0.292 & 0.014 & $-0.005$ & \citet{Marion2016}; CfA \\
SN 2015F  & NGC 2442  & 13.081 & 0.024 & 31.511 & 0.053 & $-18.430$ & 0.058 & 0.142 & $-0.015$ & \citet{Cartier2017} \\
\hline
\end{tabular}

%% file: table_hflow.tex
\begin{tabular}{|c|c|cc|c|cc|c|c|c|}
\hline
Supernova & Host Galaxy & $z_{\rm CMB}$ & $\sigma_z$ & $z_{\rm CMB}$ (flow- & 
$m_J$ & $\sigma_{\rm fit}$ & MW $A_J$ & $K_J$ & Survey \\
 & or (Cluster) & & & corrected) & (mag) & (mag) & (mag) & (mag) &  \\
\hline
SN 2004eo                  & NGC 6928                         &  0.014747 &  0.000070 &  0.015259 & 15.496 & 0.010 &  0.075  & $-0.037$ & CSP  \\
SN 2005M                   & NGC 2930                         &  0.025598 &  0.000083 &  0.025441 & 16.475 & 0.017 &  0.022  & $-0.055$ & CSP  \\
SN 2005el                  & NGC 1819                         &  0.014894 &  0.000017 &  0.015044 & 15.439 & 0.007 &  0.079  & $-0.041$ & CSP+CfA  \\
SN 2005eq                  & MCG $-$01$-$09$-$06              &  0.028370 &  0.000087 &  0.028336 & 16.793 & 0.059 &  0.051  & $-0.074$ & CfA  \\
SN 2005kc                  & NGC 7311                         &  0.013900 &  0.000087 &  0.014468 & 15.390 & 0.008 &  0.092  & $-0.039$ & CSP  \\
SN 2005ki                  & NGC 3332                         &  0.020384 &  0.000087 &  0.019887 & 16.111 & 0.014 &  0.022  & $-0.054$ & CSP  \\
SN 2006ax                  & NGC 3663                         &  0.017969 &  0.000090 &  0.017908 & 15.719 & 0.010 &  0.033  & $-0.048$ & CSP  \\
SN 2006et                  & NGC 232                          &  0.021662 &  0.000163 &  0.022288 & 16.061 & 0.019 &  0.013  & $-0.045$ & CSP  \\
SN 2006hx                  & (Abell 168)                      &  0.043944 &  0.000073 &  0.044533 & 17.779 & 0.077 &  0.021  & $-0.089$ & CSP  \\
SN 2006le                  & UGC 3218                         &  0.017272 &  0.000027 &  0.018374 & 15.935 & 0.010 &  0.284  & $-0.047$ & CfA  \\
SN 2006lf                  & UGC 3108                         &  0.012972 &  0.000027 &  0.012037 & 14.945 & 0.220 &  0.661  & $-0.037$ & CfA  \\
SN 2007S                   & UGC 5378                         &  0.015034 &  0.000087 &  0.015244 & 15.346 & 0.018 &  0.018  & $-0.041$ & CSP  \\
SN 2007as                  & PGC 026840                       &  0.017909 &  0.000460 &  0.018486 & 15.864 & 0.016 &  0.100  & $-0.043$ & CSP  \\
SN 2007ba\tablefootmark{a} & (Abell 2052)                     &  0.036062 &  0.000304 &  0.035843 & 17.714 & 0.030 &  0.026  & $-0.097$ & CSP  \\
SN 2007bd                  & UGC 4455                         &  0.031849 &  0.000163 &  0.031624 & 17.105 & 0.028 &  0.023  & $-0.082$ & CSP  \\
SN 2007ca                  & MCG $-$02$-$34$-$61              &  0.015080 &  0.000070 &  0.014471 & 15.568 & 0.006 &  0.046  & $-0.041$ & CSP  \\
SN 2008bc                  & PGC 90108                        &  0.015718 &  0.000127 &  0.015623 & 15.542 & 0.009 &  0.182  & $-0.043$ & CSP  \\
SN 2008hs\tablefootmark{a} & (Abell 347)                      &  0.017692 &  0.000050 &  0.018054 & 16.357 & 0.040 &  0.040  & $-0.046$ & CfA  \\
SN 2008hv                  & NGC 2765                         &  0.013589 &  0.000100 &  0.013816 & 15.232 & 0.013 &  0.023  & $-0.038$ & CSP+CfA  \\
SN 2009ad                  & UGC 3236                         &  0.028336 &  0.000007 &  0.028587 & 16.880 & 0.031 &  0.077  & $-0.073$ & CfA  \\
SN 2009bv                  & MCG +06$-$29$-$39                &  0.037459 &  0.000083 &  0.038302 & 17.552 & 0.028 &  0.006  & $-0.093$ & CfA  \\
SN 2010Y\tablefootmark{a}  & NGC 3392                         &  0.011224 &  0.000107 &  0.012261 & 15.302 & 0.019 &  0.009  & $-0.032$ & CfA  \\
SN 2010ag                  & UGC 10679                        &  0.033700 &  0.000177 &  0.033461 & 17.202 & 0.019 &  0.021  & $-0.086$ & CfA  \\
SN 2010ai                  & (Coma)                           &  0.023997 &  0.000063 &  0.022102 & 16.597 & 0.027 &  0.007  & $-0.047$ & CfA  \\
PTF10bjs                   & MCG +09$-$21$-$83                &  0.030551 &  0.000080 &  0.030573 & 17.033 & 0.029 &  0.012  & $-0.077$ & CfA  \\
SN 2010ju                  & UGC 3341                         &  0.015347 &  0.000013 &  0.015020 & 15.600 & 0.018 &  0.292  & $-0.042$ & CfA  \\
SN 2010kg                  & NGC 1633                         &  0.016455 &  0.000037 &  0.017021 & 15.822 & 0.019 &  0.106  & $-0.045$ & CfA  \\
PTF10mwb                   & SDSS J171750.05+405252.5         &  0.030878 &  0.000010 &  0.031004 & 16.995 & 0.026 &  0.021  & $-0.080$ & PTF  \\
PTF10ufj                   & 2MASX J02253767+2445579          &  0.076200 &  0.005000 &  0.076676 & 19.298 & 0.079 &  0.080  & $-0.138$ & PTF  \\
SN 2011ao                  & IC 2973                          &  0.011631 &  0.000063 &  0.012164 & 14.885 & 0.028 &  0.014  & $-0.034$ & CfA  \\
\hline
\end{tabular}

%% file: table_diagnostic.tex
\begin{tabular}{|c|c|c|c|c|c|}
\hline
Supernova & $E(B-V)_{\rm host}$ & $\Delta m_{15}(B)$ & Host Galaxy & Morphology & Code \\ 
 & (mag) & (mag) &  & & \\
\hline
SN 2001el & 0.250 & 1.08 & NGC 1448                  & SAcd       & 4.5 \\
SN 2002fk & 0.010 & 1.20 & NGC 1309                  & SA(s)bc    & 3.5 \\
SN 2003du & 0.000 & 1.02 & UGC 9391                  & SBdm       & 5.0 \\
SN 2005cf & 0.090 & 1.10 & NGC 5917                  & Sb         & 3.0 \\
SN 2007af & 0.170 & 1.17 & NGC 5584                  & SAB(rs)cd  & 4.5 \\
SN 2011by & 0.010 & 1.14 & NGC 3972                  & SA(s)bc    & 3.5 \\
SN 2011fe & 0.013 & 1.20 & M101                      & SAB(rs)cd  & 4.5 \\
SN 2012cg & 0.250 & 0.97 & NGC 4424                  & SB(s)a     & 2.0 \\
SN 2015F  & 0.035 & 1.26 & NGC 2442                  & SAB(s)bc   & 3.5 \\
 & & & & & \\
SN 2004eo & 0.128 & 1.41 & NGC 6928                  & SB(s)ab    & 2.5 \\
SN 2005M  & 0.060 & 0.90 & NGC 2930                  & S?         & 5.0 \\
SN 2005el & 0.015 & 1.34 & NGC 1819                  & SB0        & 1.0 \\
SN 2005eq & 0.044 & 0.75 & MCG $-$01$-$09$-$06       & SB(rs)cd?  & 4.5 \\
SN 2005kc & 0.310 & 1.22 & NGC 7311                  & Sab        & 2.5 \\
SN 2005ki & 0.016 & 1.27 & NGC 3332                  & (R)SA0     & 1.0 \\
SN 2006ax & 0.016 & 1.00 & NGC 3663                  & SA(rs)bc   & 3.5 \\
SN 2006et & 0.254 & 0.89 & NGC 232                   & SB(r)a?    & 2.0 \\
SN 2006hx & 0.210 & 1.38 & PGC 73820                 & S0         & 1.0 \\
SN 2006le & 0.049 & 0.87 & UGC 3218                  & SAb        & 3.0 \\
SN 2006lf & 0.020 & 1.36 & UGC 3108                  & S?         & 4.0 \\
SN 2007S  & 0.478 & 0.77 & UGC 5378                  & Sb         & 3.0 \\
SN 2007as & 0.050 & 1.14 & PGC 026840                & SB(rs)c    & 4.0 \\
SN 2007ba & 0.150 & 1.88 & UGC 9798                  & S0/a       & 1.5 \\
SN 2007bd & 0.058 & 1.10 & UGC 4455                  & SB(r)a     & 2.0 \\
SN 2007ca & 0.350 & 0.90 & MCG $-$02$-$34$-$61       & Sc         & 4.0 \\
SN 2008bc & 0.005 & 0.85 & PGC 90108                 & S          & 2.0 \\
SN 2008hs & 0.019 & 2.02 & NGC 910                   & E+         & 0.0 \\
SN 2008hv & 0.074 & 1.25 & NGC 2765                  & S0         & 1.0 \\
SN 2009ad & 0.045 & 1.03 & UGC 3236                  & Sbc        & 3.5 \\
SN 2009bv & 0.076 & 1.00 & MCG +06$-$29$-$39         & S          & 3.0 \\
SN 2010Y  & 0.000 & 1.76 & NGC 3392                  & E?         & 0.0 \\
SN 2010ag & 0.272 & 1.08 & UGC 10679                 & Sb(f)      & 3.0 \\
SN 2010ai & 0.063 & 1.35 & SDSS J125925.04+275948.2  & E          & 0.0 \\
PTF10bjs  & 0.000 & 1.01 & MCG +09$-$21$-$83         & Sb         & 3.0 \\
SN 2010ju & 0.180 & 1.10 & UGC 3341                  & SBab       & 2.5 \\
SN 2010kg & 0.268 & 1.40 & NGC 1633                  & SAB(s)ab   & 2.5 \\
PTF10mwb  & 0.026 & 1.15 & SDSS J171750.05+405252.5  & S(r)c      & 4.0 \\
PTF10ufj  & 0.000 & 1.20 & 2MASX J02253767+2445579   & S0/a       & 1.5 \\
SN 2011ao & 0.029 & 0.90 & IC 2973                   & SB(s)d     & 5.0 \\
\hline
\end{tabular}

%% file: table_results.tex

\begin{tabular}{|l|cc|cc|cccc|}
\hline
Sample & $N_{\rm calib}$ & $\sigma_{\rm calib}$ & $N_{\rm Hflow}$ & 
$\sigma_{\rm Hflow}$ & $H_0$ & $M_J$ & $-5\,a_J$ & $\sigma_{\rm int}$ \\
& & (mag) & & (mag) & (km s$^{-1}$ Mpc$^{-1}$) & (mag) & (mag) & (mag) \\
\hline
\bf{Fiducial} & \bf{9} & \bf{0.160} & \bf{27} & \bf{0.106} & 
$\mathbf{72.78_{-1.57}^{+1.60}}$ & $\mathbf{-18.524_{-0.041}^{+0.041}}$ & 
$\mathbf{-2.834_{-0.023}^{+0.023}}$ & $\mathbf{0.096_{-0.016}^{+0.018}}$ \\
Flow-corrected redshifts &  9 &  0.160 & 27 &  0.115 & 
$73.18_{-1.68}^{+1.71}$ & $-18.523_{-0.044}^{+0.044}$ & 
$-2.845_{-0.025}^{+0.025}$ & $ 0.109_{-0.017}^{+0.020}$ \\
Host $E(B-V) \leq 0.3$ mag &  9 &  0.160 & 24 &  0.106 & 
$72.90_{-1.62}^{+1.61}$ & $-18.523_{-0.042}^{+0.041}$ & 
$-2.837_{-0.024}^{+0.025}$ & $ 0.098_{-0.016}^{+0.019}$ \\
Spirals only (morphology code $\geq 2$) &  9 &  0.160 & 21 &  0.107 & 
$73.05_{-1.73}^{+1.73}$ & $-18.522_{-0.043}^{+0.042}$ & 
$-2.841_{-0.027}^{+0.027}$ & $ 0.104_{-0.018}^{+0.021}$ \\
Milky Way $A_J \leq 0.3$ mag &  9 &  0.160 & 26 &  0.101 & 
$72.73_{-1.56}^{+1.58}$ & $-18.523_{-0.041}^{+0.041}$ & 
$-2.832_{-0.023}^{+0.023}$ & $ 0.096_{-0.016}^{+0.019}$ \\
Low EBV + Spirals + Low MW $A_J$ &  9 &  0.160 & 15 &  0.094 & 
$73.60_{-1.79}^{+1.80}$ & $-18.523_{-0.043}^{+0.043}$ & 
$-2.857_{-0.031}^{+0.031}$ & $ 0.105_{-0.019}^{+0.025}$ \\
Hubble flow $z \geq 0.02$ &  9 &  0.160 & 13 &  0.104 & 
$73.66_{-1.84}^{+1.86}$ & $-18.524_{-0.043}^{+0.043}$ & 
$-2.860_{-0.033}^{+0.034}$ & $ 0.104_{-0.020}^{+0.025}$ \\
Hubble flow $z \geq 0.03$ &  9 &  0.160 &  7 &  0.091 & 
$72.79_{-2.20}^{+2.25}$ & $-18.524_{-0.045}^{+0.045}$ & 
$-2.835_{-0.047}^{+0.049}$ & $ 0.108_{-0.024}^{+0.033}$ \\
Hubble flow $0.01 \leq z \leq 0.05$ &  9 &  0.160 & 26 &  0.099 & 
$72.87_{-1.55}^{+1.59}$ & $-18.524_{-0.041}^{+0.040}$ & 
$-2.837_{-0.023}^{+0.023}$ & $ 0.095_{-0.015}^{+0.018}$ \\
Hubble flow $0.02 \leq z \leq 0.05$ &  9 &  0.160 & 12 &  0.083 & 
$73.93_{-1.83}^{+1.88}$ & $-18.523_{-0.043}^{+0.043}$ & 
$-2.868_{-0.034}^{+0.034}$ & $ 0.104_{-0.020}^{+0.026}$ \\
Strictest overlap\tablefootmark{a}  &  7 &  0.147 &  8 &  0.058 & 
$73.04_{-2.12}^{+2.21}$ & $-18.532_{-0.048}^{+0.049}$ & 
$-2.849_{-0.042}^{+0.042}$ & $ 0.104_{-0.023}^{+0.031}$ \\
Including fast-decliner outliers &  9 &  0.160 & 30 &  0.170 & 
$71.30_{-2.09}^{+2.11}$ & $-18.524_{-0.057}^{+0.057}$ & 
$-2.789_{-0.030}^{+0.030}$ & $ 0.148_{-0.020}^{+0.024}$ \\
Hubble flow CSP only &  9 &  0.160 & 14 &  0.091 & 
$74.09_{-1.87}^{+1.91}$ & $-18.523_{-0.043}^{+0.043}$ & 
$-2.872_{-0.034}^{+0.034}$ & $ 0.105_{-0.020}^{+0.025}$ \\
Hubble flow CfA only &  9 &  0.160 & 13 &  0.094 & 
$71.47_{-1.72}^{+1.80}$ & $-18.523_{-0.041}^{+0.041}$ & 
$-2.794_{-0.034}^{+0.034}$ & $ 0.098_{-0.020}^{+0.025}$ \\
Hubble flow and calibrators CSP only   &  1 &  0.000 & 14 &  0.091 & 
$80.98_{-2.57}^{+2.55}$ & $-18.338_{-0.066}^{+0.065}$ & 
$-2.880_{-0.022}^{+0.023}$ & $ 0.043_{-0.043}^{+0.040}$ \\
Hubble flow and calibrators CfA only  &  2 &  0.213 & 13 &  0.094 & 
$75.92_{-2.82}^{+2.98}$ & $-18.393_{-0.081}^{+0.081}$ & 
$-2.795_{-0.017}^{+0.018}$ & $ 0.000_{-0.000}^{+0.028}$ \\
\citet{Cardona2017} Cepheid distances &  9 &  0.133 & 27 &  0.106 & 
$73.83_{-1.59}^{+1.61}$ & $-18.492_{-0.042}^{+0.042}$ & 
$-2.833_{-0.021}^{+0.022}$ & $ 0.089_{-0.016}^{+0.018}$ \\
\hline
\end{tabular}
\tablefootwide{
\tablefoottext{a}{This is an extremely restrictive cut to make the calibrators and Hubble flow sample as similar as possible: Low EBV (host $E(B-V) \leq 0.3$ mag) + Spirals only + $1.0 \leq \Delta m_{15}(B) \leq 1.2$ + Milky Way $A_J \leq 0.15$ mag.}
}

%% file: table_csp_cfa.tex
\begin{tabular}{|c|cc|c|}
\hline
SN &  $m_J$ (mag) & $\sigma_{\mathrm{fit}}$ (mag) & LC source \\
\hline
SN2005el 	&	15.439 & 0.007	&  CSP+CfA\\
SN2005el    &	15.438 & 0.007	&  CSP only\\
SN2005el	&   15.445 & 0.016	&  CfA only \\
& & & \\
SN2008hv	&	15.232 & 0.013	&  CSP+CfA\\
SN2008hv 	&	15.213 & 0.019	&  CSP only \\
SN2008hv	& 	15.249 & 0.020	&  CfA only \\
\hline
\end{tabular}